\documentclass[12pt]{iopart}

\usepackage{graphicx}
\usepackage{xcolor}
\usepackage{cite}
\usepackage{relsize}
\usepackage{amssymb}
\usepackage{subcaption}
\usepackage{hyperref}
\usepackage{pdfrender}
\usepackage{esint}

\newcommand*{\bkappa}{%
  \textpdfrender{
    TextRenderingMode=FillStroke,
    LineWidth=.4pt, 
  }{\kappa}%
}

\begin{document}

\title[Omnigenous stellarators with improved ideal and kinetic ballooning stability]{Omnigenous stellarators with improved ideal and kinetic ballooning stability}

\author{R. Gaur\footnote{corresponding author, email: rg6256@princeton.edu, rgaur@terpmail.umd.edu}, R. Conlin, D. Dickinson, J. F. Parisi, D. Panici, D. Dudt, P. Kim, K. Unalmis, W. D. Dorland, E. Kolemen}

\address{Department of Mechanical and Aerospace Engineering, Princeton University, Princeton, New Jersey, USA}
\address{Princeton Plasma Physics Laboratory, Princeton, New Jersey, USA}
\address{School of Physics, Engineering, and Technology, University of York, Heslington, York, UK}
\address{Thea Energy}
\address{IREAP, University of Maryland, College Park, USA}
\address{Department of Physics, University of Maryland, College Park, USA}
\ead{rg6256@princeton.edu, rgaur@terpmail.umd.edu}
\vspace{10pt}
\begin{indented}
\item June 2025
\end{indented}

\begin{abstract}
Omnigenity is a property of a magnetic field which ensures confinement of trapped particles. It is a necessary requirement for any high-performance stellarator. After creating an omnigenous equilibrium, one must also ensure reduced transport resulting from kinetic and magnetohydrodynamic (MHD) instabilities. To this end, we leverage the GPU-accelerated DESC optimization suite, which is used to design stable, finite-$\beta$ omnigenous equilibria with poloidal, toroidal, and helical symmetry, achieving Mercier, ideal ballooning, and as a consequence, improved kinetic ballooning stability. We discover stellarators with second stability, a regime of large pressure gradient where an equilibrium becomes ideal ballooning stable, and demonstrate and explore both using theory and gyrokinetic simulations the connection between ideal and kinetic ballooning stability.

\end{abstract}

\vspace{2pc}

\section{Introduction}
Stellarators\cite{spitzer1958stellarator} are toroidal devices that are used for the magnetic confinement of a hot plasma to achieve nuclear fusion. Unlike their toroidally axisymmetric counterparts, tokamaks, stellarators can provide a greater range of design flexibility\cite{landreman2022mapping} that is used to improve their operational properties. The three-dimensional boundary can be designed to reduce dependence on the toroidal plasma current compared to tokamaks and to reduce current-driven instabilities. However, the lack of a continuous toroidal symmetry also increases the design space of possible reactor configurations. Therefore, finding the optimal stellarator from a large space with over a million configurations~\cite{landreman2022mapping} becomes a complicated optimization problem.

To this end, we have developed the~\texttt{DESC} stellarator\cite{dudt2020desc, conlin2023desc, panici2023desc, dudt2023desc} equilibrium and optimization suite which uses a fully-spectral representation of the MHD equilibria with accurate near-axis behavior and automatic-differentiable objective functions. Using these two properties, \texttt{DESC} can accelerate stellarator equilibrium optimization with favorable characteristics such as omnigenity, MHD stability, low neoclassical transport, and coil shape complexity. 
Previously, multiple reactors such as W7-X~\cite{nuhrenberg1995overview}, NCSX~\cite{zarnstorff2001physics}, HSX~\cite{Anderson01041995}, and LHD~\cite{murakami2002demonstration} have used~\texttt{VMEC} for similar multi-objective optimization design studies.
In this work, we shall use a fully spectral representation in~\texttt{DESC} to minimize multiple objectives while maintaining ideal MHD force balance throughout the device volume at each iteration without re-solving for the equilibrium~\cite{conlin2023desc}.

In this paper, we will demonstrate the utility of the~\texttt{DESC} code by optimizing equilibria for various stability properties, with a focus on a gyrokinetic instability known as the Kinetic Ballooning Mode (KBM). We will also explain how the objectives we have used to obtain omnigenous equilibria differ significantly from our previous optimization studies. Specifically, we will focus on finite-$\beta$ stellarators, where $\beta = 2 \mu_0 p/B^2$ is the ratio of plasma pressure to magnetic pressure. Low-$\beta$ stellarators are typically dominated by electrostatic instabilities such as the ion temperature gradient (ITG) mode~\cite{cowley1991considerations, rudakov1961instability} and trapped electron mode (TEM)~\cite{adam1975destabilization, ryter2005experimental}, which drive the most high heat and particle losses, degrading the performance of reactors. However, upon increasing the $\beta$, the electrostatic modes become subdominant, and electromagnetic instabilities can become a problem~\cite{pueschel2010transport, ishizawa2014electromagnetic}. In particular, much work in the literature discusses how a long-wavelength gyrokinetic instability, known as the Kinetic Ballooning Mode (KBM)~\cite{tang1980kinetic}, can potentially cause huge heat and particle losses as a result of turbulent transport. Therefore, reducing the susceptibility of finite-$\beta$ stellarators to the KBM~\cite{ishizawa2014electromagnetic} is paramount.

We develop a set of omnigenous finite-$\beta$ stellarator equilibria with improved KBM stability.
In section 2, we briefly explain the steady-state ideal MHD model and how we can locally vary the gradients associated with an equilibrium to gain insight into the stability properties of an equilibrium. In section 3, we introduce the linear gyrokinetic model and discuss the limit in which it reduces to the infinite-$n$ ideal ballooning equation. We tested our hypothesis by analyzing a finite-beta W7-X by solving the linear gyrokinetic model numerically using the~\texttt{GS2} code equilibrium and demonstrating a strong correlation between the ideal and kinetic ballooning modes. In section 4, we explain the details of the reverse-mode-differentiable ideal ballooning solver implemented in~\texttt{DESC}. In section 5, we explain how we calculate omnigenous equilibria in~\texttt{DESC}.
In section 6, we obtain and present three stellarator configurations with poloidal, toroidal, and helical omnigenity, respectively. We explain the objective functions used to calculate these equilibria and analyze their physical properties, such as the correlation between the ideal and kinetic ballooning stability (using the~\texttt{GS2} code) and $1/\nu$ neoclassical transport. In section 7, we perform $\hat{s}-\alpha_{\mathrm{MHD}}$ and sensitivity analysis of the optimized equilibria and explain when the distance from ideal ballooning marginality is a good proxy for KBM stability. Finally, in section 8, we conclude our work and discuss various directions in which it can be extended.

\section{Ideal MHD equilibrium}
\label{sec:ideal-MHD}
In this section, we briefly explain how we define and calculate an ideal MHD equilibrium in a stellarator. We then explain how to locally vary the gradients of that equilibrium. 

A divergence-free magnetic field $\bi{B}$ can be written in the Clebsch form~\cite{d2012flux}
\begin{equation}
    \bi{B} =   \nabla \psi \times \nabla\alpha,
    \label{eqn:Div-free-B2}
\end{equation}
We will focus on solutions whose magnetic field lines lie on closed nested toroidal surfaces, known as flux surfaces. We label these surfaces using the enclosed toroidal flux $\psi$. On each flux surface, the lines of constant $\alpha$ coincide with the magnetic field lines. Thus, $\alpha$ is known as the field line label. We define $\alpha = \theta - \iota (\zeta-\zeta_0)$, where $\theta$ is the $\mathrm{PEST}$\cite{grimm1983ideal} straight field line angle, $\zeta$ is the generalized cylindrical toroidal angle, $\zeta_0$ is a constant, and 
\begin{equation}
    \iota = \frac{\bi{B}\cdot \bi{\nabla}\theta}{\bi{B}\cdot \bi{\nabla}\zeta},
\end{equation}
is the pitch of the magnetic field lines on a flux surface, known as the rotational transform. Using the Clebsch form of the magnetic field, we solve the ideal MHD force balance equation
\begin{equation}
    \bi{j} \times \bi{B} = \bi{\nabla} p,
    \label{eqn:ideal-MHD-force-balance}
\end{equation}
where the plasma current $\bi{j} = (\bi{\nabla} \times \bi{B})/\mu_0$, from Ampere's law, $p$ is the plasma pressure, and $\mu_0$ is the vacuum magnetic permeability. Unlike in an axisymmetric case, for stellarators, we have to solve~\eref{eqn:ideal-MHD-force-balance} as an optimization problem. We achieve this with the $\texttt{DESC}$~\cite{dudt2020desc, panici2023desc, conlin2023desc, dudt2023desc} stellarator optimization suite. $\texttt{DESC}$ can simultaneously solve an equilibrium while optimizing multiple objectives such as MHD stability, quasisymmetry, and many more. In this paper, we use some of these metrics to generate various stellarator equilibria. The details of the~\texttt{DESC} package are briefly provided in~\ref{app:DESC-appendix}. In the following sections, we explain how to utilize~\texttt{DESC} to optimize stellarator equilibria for various favorable properties.

\subsection{Locally varying the gradients of pressure and rotational transform}
\label{subsec:Greene-Chance}
In this section, we will briefly explain the method of local equilibrium variation and how we use it to analyze stability in stellarators.

To better understand the stability properties of stellarator equilibria, we can locally vary the gradients of the pressure and rotational transform and obtain a family of local equilibria that are in ideal MHD force balance. We then analyze the stability properties of these equilibria. This gives us an idea of how these equilibria would behave if the pressure and rotational transform profiles were modified. This type of analysis is also known as a $\hat{s}$-$\alpha_{\mathrm{MHD}}$ analysis. According to convention, we normalize the gradients to two parameters: the magnetic shear
\begin{equation}
    \hat{s} = -\frac{\rho}{\iota} \frac{d \iota}{d \rho},
\end{equation}
and the normalized pressure gradient 
\begin{equation}
    \alpha_{\mathrm{MHD}} = -\frac{\mu_0}{B_{\mathrm{N}}^2}\frac{dp}{d\rho},
\end{equation}
respectively, where $\rho = \sqrt{\psi/\psi_{\mathrm{b}}}$ is the normalized radius, $B_N = 2 \psi_{\rm{b}}/a_{\rm{N}}^2$ and $a_{\rm{N}} = \sqrt{A_{\mathrm{b}}/\pi}$ are the effective minor radius, and $A_{\mathrm{b}}$ is the toroidally-averaged cross-section area of the boundary.  In the context of stability, the $\hat{s}-\alpha_{\mathrm{MHD}}$ analysis has been used in both tokamaks~\cite{greene1981second} and stellarators~\cite{hegna-nakajima, hudson2004marginal}. The details and formulation of an $\hat{s}-\alpha_{\mathrm{MHD}}$ analysis with~\texttt{DESC} are described in~\ref{app:s-alpha-appendix}. 

Note that the idea of locally varying the gradients of an equilibrium is only applicable to instabilities that are localized to a flux surface. In this work, we apply this technique to the infinite-$n$ ideal ballooning mode and the $\delta \! f$ gyrokinetic model, both of which are models used to study small-scale instabilities localized to a flux surface.

Using the method of local equilibrium variation with the~\texttt{DESC}, we will vary the gradients $\hat{s}$ and $\alpha_{\mathrm{MHD}}$ on different flux surfaces and scan the MHD stability of a family of equilibria. Using this technique, in the next section, we will analyze the kinetic stability of the same equilibria with the gyrokinetic solver~\texttt{GS2} and explain the relation between the KBM and the infinite-$n$, ideal ballooning mode.

\section{Analyzing local ideal MHD and gyrokinetic stability in stellarators}
\label{sec:GK-W7X}
In this section, we introduce the linear $\delta\! f$ gyrokinetic model, explain the limit in which it can be simplified to obtain the KBM equations, and demonstrate how the KBM may be related to the infinite-$n$ ideal ballooning mode. We further explore this theoretical connection by numerically solving linear gyrokinetic equations using the~\texttt{GS2} code and performing a scan of the maximum KBM growth rate by locally varying the pressure gradient and rotational transform for a typical equilibrium for the W7-X stellarator.

We take the linearized, $\delta \! f$ gyrokinetic model from Abel \textit{et al.}~\cite{abel2013multiscale, antonsen1980kinetic, frieman1982nge} and write a simplified set of equations describing the evolution of the gyrokinetic distribution function in the guiding-center coordinate system $(\bi{R}_s, E_s, \mu_s, t)$
\begin{equation}
    h_s(\bi{R}_s, E_s, \mu_s, t) = \frac{Z_s e \varphi(\bi{r}, t)  F_{0s}}{T_s} + \delta\!f_s(\bi{R}_s, E_s, \mu_s, t),
\end{equation}
the fluctuations of the electrostatic potential $\varphi$, parallel component $\delta\! A_{\parallel}$ of the magnetic vector potential, and the parallel fluctuation of the magnetic field strength $\delta\! B_{\parallel} = \bi{b}\cdot (\bi{\nabla} \times \delta \bi{A}_{\perp})$. The full set of simplified, linear equations
\begin{eqnarray}
\frac{\partial h_s}{\partial t} + (w_{\parallel} \bi{b}+ \bi{v}_{Ds})\cdot \frac{\partial h_s}{\partial \bi{R}_s} = \frac{Z_s e F_{0s}}{T_s}\frac{\partial \left\langle \varphi  - \bi{w} \cdot \delta \bi{A}/c \right\rangle_{\bi{R}_s}}{\partial t} - \bi{V}_{E}\cdot \bi{\nabla} F_{0s},
\label{eqn:electrostatic-GK-equation}
\end{eqnarray}
\begin{eqnarray}
\sum_{s} \frac{(Z_s e)^2 \varphi}{T_s} = \sum_s Z_s e\int d^3\bi{w}\, \left\langle h_{s}\right\rangle_{\bi{r}}, \quad \tau = \frac{T_{\mathrm{e}}}{T_{\mathrm{i}}}\label{eqn:Poisson's-equation},
\end{eqnarray}
\begin{eqnarray}
    -\nabla_{\perp}^2 \delta A_{\parallel} = \frac{4\pi}{c} \sum_s Z_s e \int d^3\bi{w}\, w_{\parallel} \left\langle h_s\right\rangle_{\bi{r}},
    \label{eqn:Parallel-Ampere's-Law}
\end{eqnarray}
\begin{eqnarray}
    \nabla_{\perp}^2 \frac{\delta B_{\parallel}B}{4\pi} = -\nabla_{\perp} \nabla_{\perp} \bi{:}   \sum_s \int d^3\bi{w}\,  \left\langle  m_s \bi{w}_{\perp} \bi{w}_{\perp} h_s\right\rangle_{\bi{r}},
    \label{eqn:Perpendicular-Ampere's-Law}
\end{eqnarray}
where $\bi{v}_{Ds}$ is the magnetic drift velocity and $\bi{V}_E$ is the $\bi{E}\times\bi{B}$ velocity defined as
\begin{eqnarray}
\bi{v}_{Ds} &= \frac{w_{\parallel}^2}{\Omega_s} \bi{b}\times (\bi{b}\cdot \bi{\nabla}\bi{b}) + \frac{w_{\perp}^2}{2\,\Omega_s} \frac{\bi{b}\times \bi{\nabla} B}{B},
\label{eqn:magnetic-drift-velocity}
\end{eqnarray}
\begin{eqnarray}
\bi{V}_{E} &= \frac{c}{B} \bi{b}\times \left\langle \bi{\nabla} \varphi \right\rangle_{\bi{R}_s} - \frac{1}{B} \bi{b}\times \left\langle \bi{\nabla}( \bi{w}\cdot\delta\! \bi{A}) \right\rangle_{\bi{R}_s},
\label{eqn:E-cross-B-drift-velocity}
\end{eqnarray}
and the subscript $s$ is used to define these quantities for different species. The gyrofrequncy of a species is $\Omega_s = Z_s e B/(m_s c)$, $Z_s$ being the charge of the species. For this study, we choose a hydrogen plasma made of ions and electrons, with $Z_{\mathrm{i}} = 1, Z_{\mathrm{e}}=-1$. Both species have the same temperature, which corresponds to $\tau = 1$.

To simplify the model further, we use a normal model ansatz where all the fluctuating quantities are assumed to be periodic perpendicular to the field line. This allows us to write the fluctuating quantities as a Fourier series.
\numparts
\begin{eqnarray}
h_s &= \exp(i\omega t) \sum_{k} h_{k_{\perp}, s}(\zeta, E_s, \hat{\lambda}, \sigma, t)\exp(i \bi{k}_{\perp}\cdot \bi{R}_s ),  \\
X &= \exp(i\omega t)\sum_{k} X_{k_{\perp}}(\zeta, t)\exp(i \bi{k}_{\perp}\cdot \bi{r} ), \quad X = \{\varphi, \delta\!A_{\parallel}, \delta\!B_{\parallel}\}
\label{eqn:normal-mode-ansatz}
\end{eqnarray}
\endnumparts
where $\omega$ is the complex frequency of the mode, $E_s = m_s w^2/2$, is the particle energy, $\mu_s = m_s w_{\perp}^2/(2 B)$ is its magnetic moment, $\hat{\lambda} = \mu/E$ is the pitch angle, $\sigma = w_{\parallel}/|w_{\parallel}|$ is the streaming direction of a particle, and $\bi{k}_{\perp} = (k_y \bi{\nabla}y + k_x \bi{\nabla}x), x = dx/d\psi (\psi-\psi_0), y = d\psi/dx (\alpha-\alpha_0)$. Using this ansatz and dropping the subscript $k_{\perp}$, we can rewrite the linearized $\delta\! f$ gyrokinetic model
\begin{eqnarray}
    \fl i \left(\omega - \omega_{Ds}\right)h_{s} &+ (\bi{b}\cdot \bi{\nabla}\zeta)  w_{\parallel} \frac{\partial h_{s}}{\partial \zeta} \nonumber \\
    &= (\omega - {\omega}^{T}_{*,s}) \Bigg[J_0\left(\frac{k_{\perp}w_{\perp}}{\Omega_s}\right) \left(\varphi -\frac{w_{\parallel}\delta\! A_{\parallel}}{c}\right) +J_1\left(\frac{k_{\perp}w_{\perp}}{\Omega_s}\right) \frac{w_{\perp}}{k_{\perp}} \frac{\delta \! B_{\parallel}}{c}\Bigg] F_{0s},
\label{eqn:electrostatic-gyrokinetic-normal-mode}
\end{eqnarray}
\begin{eqnarray}
\sum_{s} \frac{q_s^2 N_s \delta \varphi}{T_{s}} =  \sum_{s} q_s \int d^3\bi{v} \, J_{0s}h_{s},
\label{eqn:Poisson's-equation-Fourier}
\end{eqnarray}
\begin{eqnarray}
    k_{\perp}^2 \delta \! A_{\parallel} = \sum_s \frac{4 \pi q_s}{c} \int d^3\bi{v} \, v_{\parallel} J_{0s} h_{s},
    \label{eqn:Parallel-Ampere's-Law-Fourier}
\end{eqnarray}
\begin{eqnarray}
    \frac{B}{4\pi} \delta \! B_{\parallel} =  -\sum_s \int d^3\bi{v} \, v_{\perp}^2  \frac{m_{s} J_{1s} \Omega_{s}}{k_{\perp} v_{\perp}} h_{s}.  
\label{eqn:Perpedicular-Ampere's-Law-Fourier}
\end{eqnarray}
Here,  $\omega_{Ds} = \bi{k}_{\perp}\cdot \bi{v}_{Ds}$, is the magnetic drift frequency, and $J_0(k_{\perp}\rho_s)$ and $J_1(k_{\perp}\rho_s)$ are the zeroth- and first-order cylindrical Bessel functions, respectively. The magnetic drift frequency can be further written as 
\begin{eqnarray}
   \omega_{Ds} = E_s(2(1-\hat{\lambda}B) \omega_{\kappa} + \hat{\lambda} B \omega_B), \\
\label{eqn:Particle-drift-frequency}
\end{eqnarray}
where $\omega_{\kappa} = 1/\Omega_s \left[(\bi{b} \times (\bi{b}\cdot\bi{\nabla}\bi{b})) \cdot \bi{k}_{\perp}\right], \omega_{B} = 1/\Omega_s \left[(\bi{b} \times \bi{\nabla}B) \cdot (\bi{k}_{\perp}/B)\right]$ are the curvature and grad-B components, respectively. The normalized temperature and density gradient scale lengths
\begin{eqnarray}
\frac{a_{\rm{N}}}{L_{\mathrm{T}s}} = - \frac{d\log(T_s)}{d\rho},\quad  \frac{a_{\rm{N}}}{L_{\mathrm{n}_s}} = - \frac{d\log(n_s)}{d\rho}, \quad \eta_s = \frac{L_{\mathrm{n}_s}}{L_{\mathrm{T}_s}}, 
\label{eqn:tprim-fprim-definitions}
\end{eqnarray}
$\rho = \sqrt{s} = \sqrt{\psi/\psi_b}$. The instability driving term
\begin{equation}
    \omega^{T}_{*,s} = \omega_{*, s} \left[1 + \eta_s\left(\frac{E_s}{T_s} - \frac{3}{2}\right)\right],
\end{equation}
where
\begin{eqnarray}
    \omega_{*,s} = \frac{T_s}{Z_s e B}\left[(\bi{b} \times \bi{k}_{\perp})\cdot \nabla \log{n_{s}}\right] = k_y \rho_s B \frac{w_{\mathrm{th}, s}}{L_{\mathrm{n}, s}},
    \label{eqn:diamgnetic-frequency}
\end{eqnarray}
is the diamagnetic drift frequency. We have now fully defined the linear gyrokinetic system as an eigenvalue problem. These equations can be further simplified under appropriate limits as described in~\ref{app:GK-KBM} to obtain the infinite-$n$ ideal ballooning equation,
\begin{eqnarray}
    \fl \bi{B}\cdot \bi{\nabla} \left( \frac{|\bi{\nabla}\alpha|^2}{B^2}\bi{B} \cdot \bi{\nabla} \hat{X} \right) + \frac{2}{B^2} \frac{dp}{d\psi} \left[\bi{B} \times (\bi{b}\cdot \bi{\nabla}\bi{b}) \cdot \bi{\nabla} \alpha \right] \hat{X} = \lambda B_{\rm{N}}^2 \frac{|\bi{\nabla}\alpha|^2}{B^2} \hat{X}, \, \lambda = -\frac{a_{\rm{N}}^2}{v_{\rm{A}}^2} \omega^2
    \label{eqn:ideal-ballooning-equation}
\end{eqnarray}
where $\hat{X}$ is the eigenfunction and $\omega$ is the complex frequency of the KBM, and $v_{\rm{A}} = B_{\rm{N}}/\sqrt{4\pi n_i m_i}$ is the Alfv\`{e}n speed and $a_{\rm{N}}$ is the effective minor radius defined in section~\ref{subsec:Greene-Chance} subject to the boundary conditions
\begin{eqnarray}
    \lim_{\zeta \rightarrow \pm \infty} X(\theta) = 0
\end{eqnarray}
where $\lambda$ is the eigenvalue and $\hat{X}$ is the eigenfunction. An equilibrium with $\lambda > 0$ implies an unstable mode, while $\lambda < 0$ implies stability. For each flux surface, equation~\eref{eqn:ideal-ballooning-equation} is solved on multiple field lines $\alpha$ for multiple values of the ballooning parameter $\zeta_0$. For each surface, we choose the maximum $\lambda$ from the $\alpha-\zeta_0$ grid.

As shown in~\ref{app:GK-KBM}, the ideal ballooning mode is closely related to the KBM in the large aspect ratio, high-$\beta$ ordering. Hence, we can use the ideal ballooning problem as a proxy for Kinetic Ballooning Modes and predict their behavior and optimize stellarators for linear ideal and kinetic ballooning instabilities.
After optimization, we also compare the diffusion coefficient estimate calculated using the mixing-length argument 
\begin{eqnarray}
    D_{\mathrm{mix}} \equiv \frac{\gamma_{\mathrm{KBM}} a_{\mathrm{N}}}{v_{\mathrm{th, i}} \langle (k_{\perp} \rho_{\mathrm{i}})^2 \rangle_{\psi}}
\end{eqnarray}
due to the KBM, between the initial and optimized equilibria. The operator $\langle \rangle_{\psi}$ is a flux surface average, defined as 
\begin{eqnarray}
    \langle X \rangle_{\psi} = \frac{\int \int d\theta d\zeta \, \sqrt{g} X}{\int \int d\theta d\zeta \, \sqrt{g}}.
    \label{eqn:FSA}
\end{eqnarray}
We calculate the mixing-length diffusion coefficient because it is often a better estimate of the nonlinear transport than the growth rate.

\subsection{Distance from ideal ballooning marginality as a proxy for KBM stability}
\begin{figure}
    \centering
    \begin{subfigure}[b]{0.243\textwidth}
    \centering
        \includegraphics[width=\textwidth, trim={2mm 2mm 1mm 4mm}, clip]{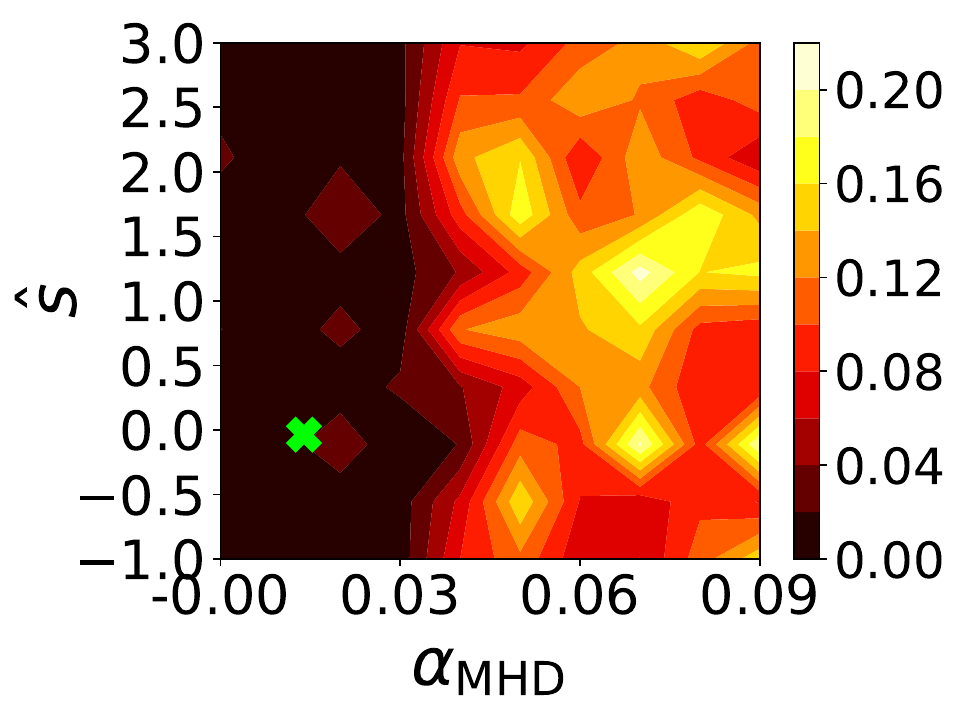}
    \end{subfigure}
    \begin{subfigure}[b]{0.243\textwidth}
        \centering
        \includegraphics[width=\textwidth, trim={2mm 2mm 2mm 4mm}, clip]{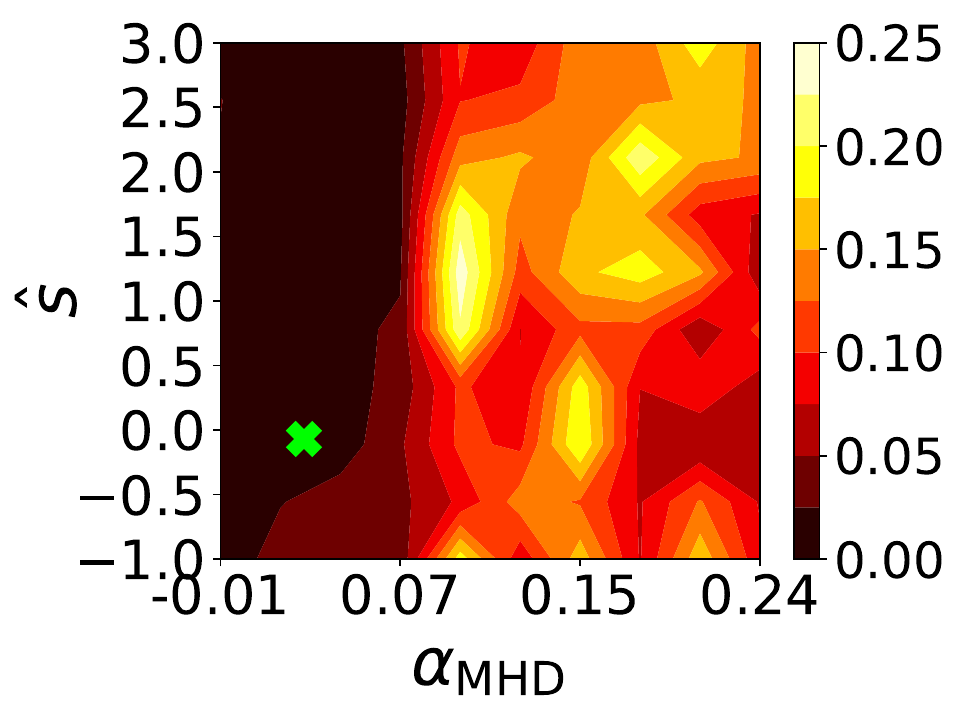}
    \end{subfigure}
    \begin{subfigure}[b]{0.243\textwidth}
        \centering
        \includegraphics[width=\textwidth, trim={2mm 2mm 3mm 4mm}, clip]{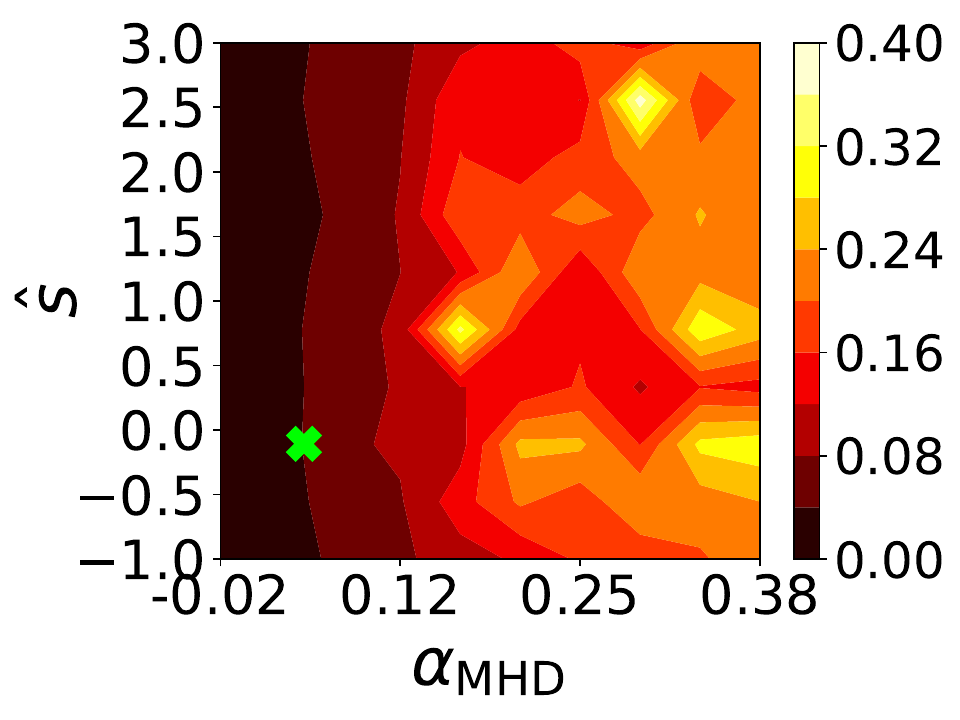}
    \end{subfigure}
    \begin{subfigure}[b]{0.243\textwidth}
        \centering
        \includegraphics[width=\textwidth, trim={2mm 2mm 8mm 4mm}, clip]{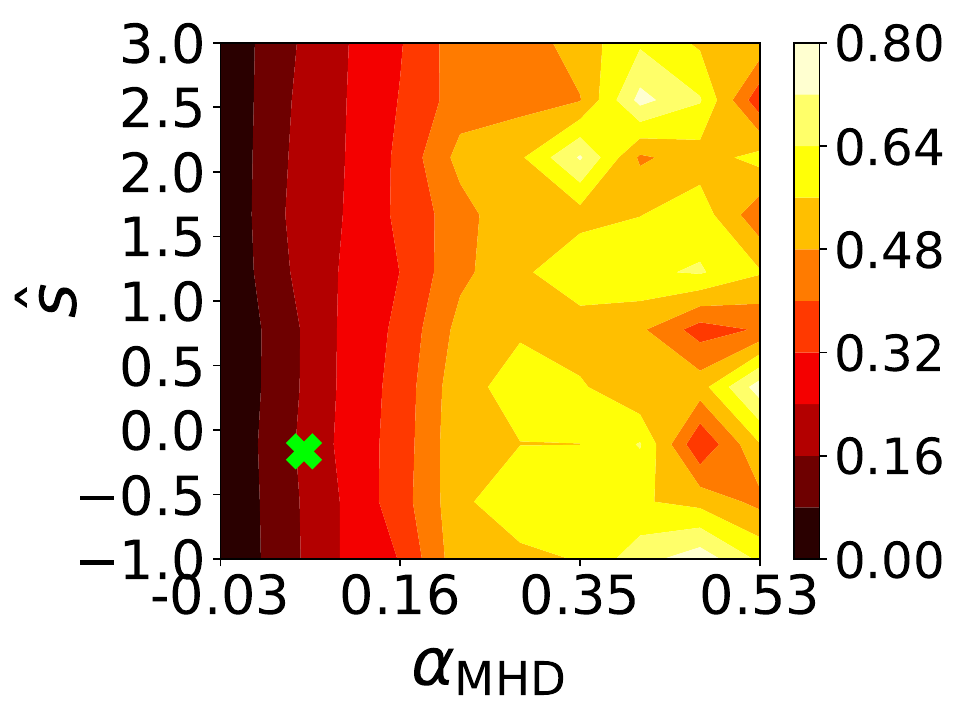}
    \end{subfigure}\\
    
    \begin{subfigure}[b]{0.243\textwidth}
    \centering
        \includegraphics[width=\textwidth, trim={2mm 2mm 8mm 4mm}, clip]{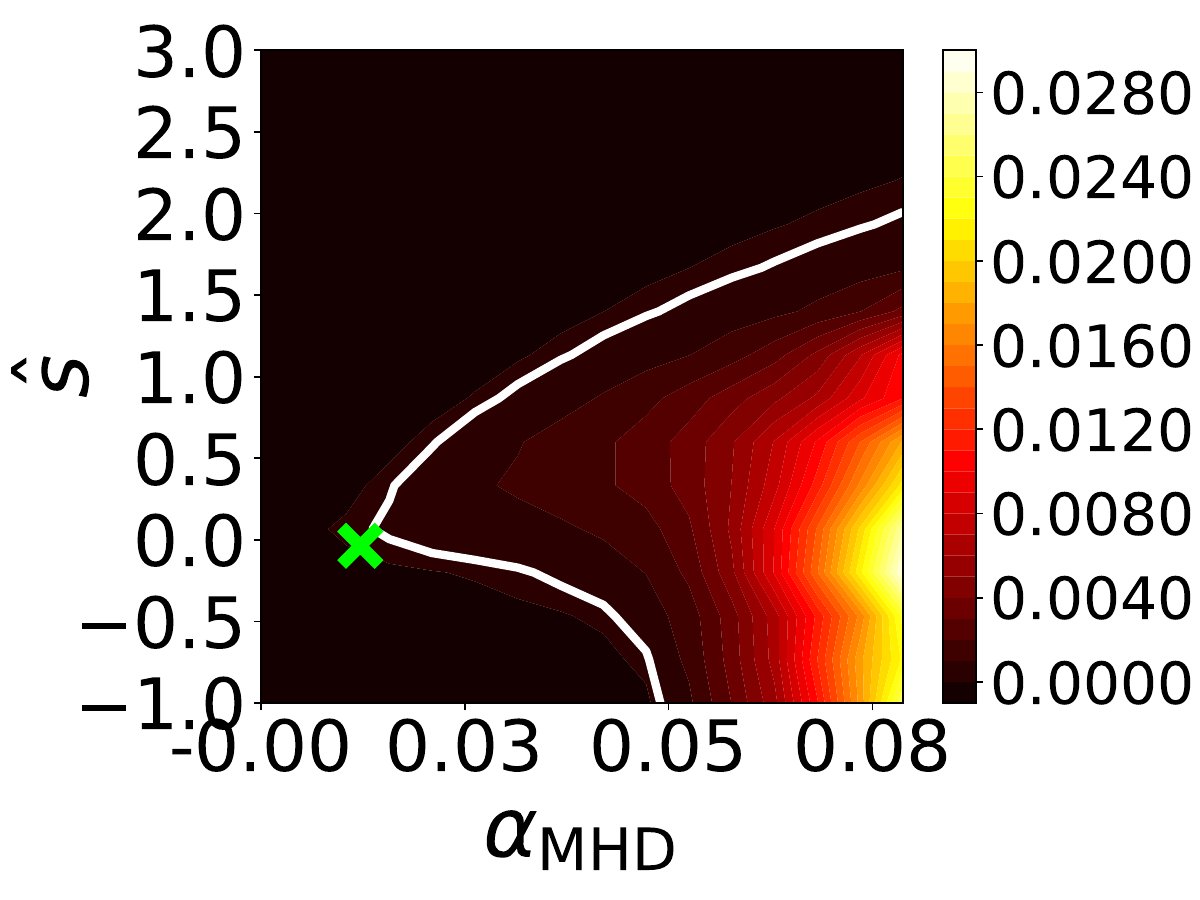}
        \caption{$\rho=0.35$}
    \end{subfigure}
    \begin{subfigure}[b]{0.243\textwidth}
        \centering
        \includegraphics[width=\textwidth, trim={2mm 2mm 13mm 4mm}, clip]{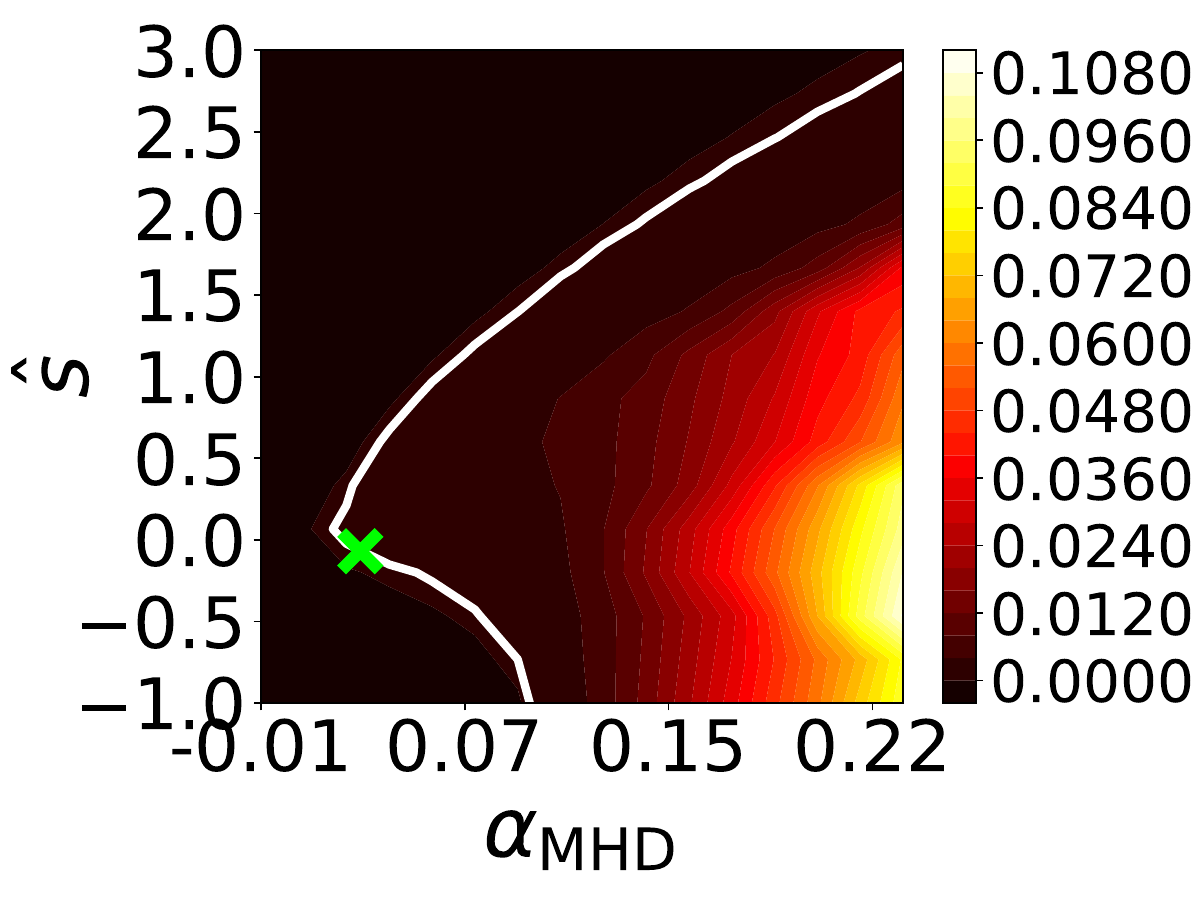}
        \caption{$\rho=0.65$}
    \end{subfigure}
    \begin{subfigure}[b]{0.243\textwidth}
        \centering
        \includegraphics[width=\textwidth, trim={2mm 2mm 13mm 4mm}, clip]{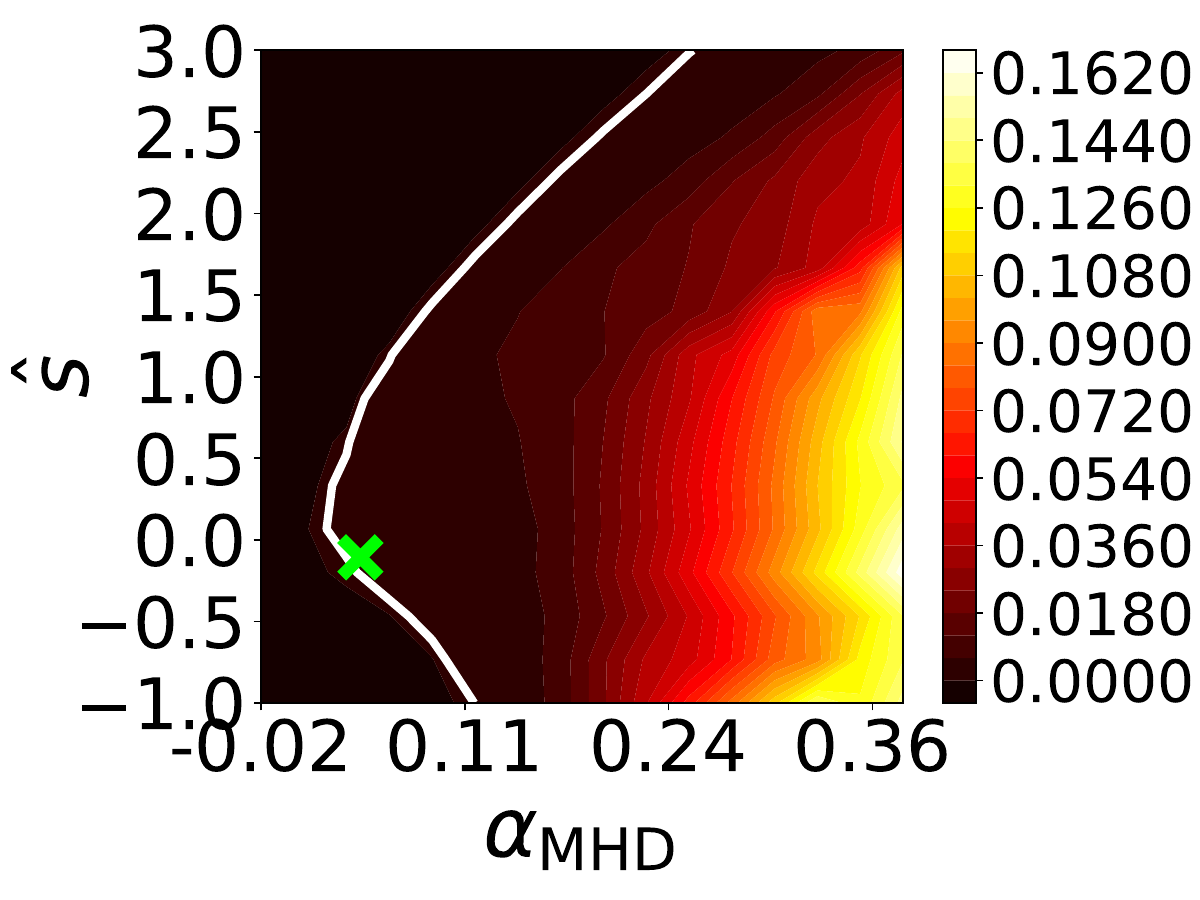}
        \caption{$\rho=0.8$}
    \end{subfigure}
    \begin{subfigure}[b]{0.243\textwidth}
        \centering
        \includegraphics[width=\textwidth, trim={2mm 2mm 13mm 4mm}, clip]{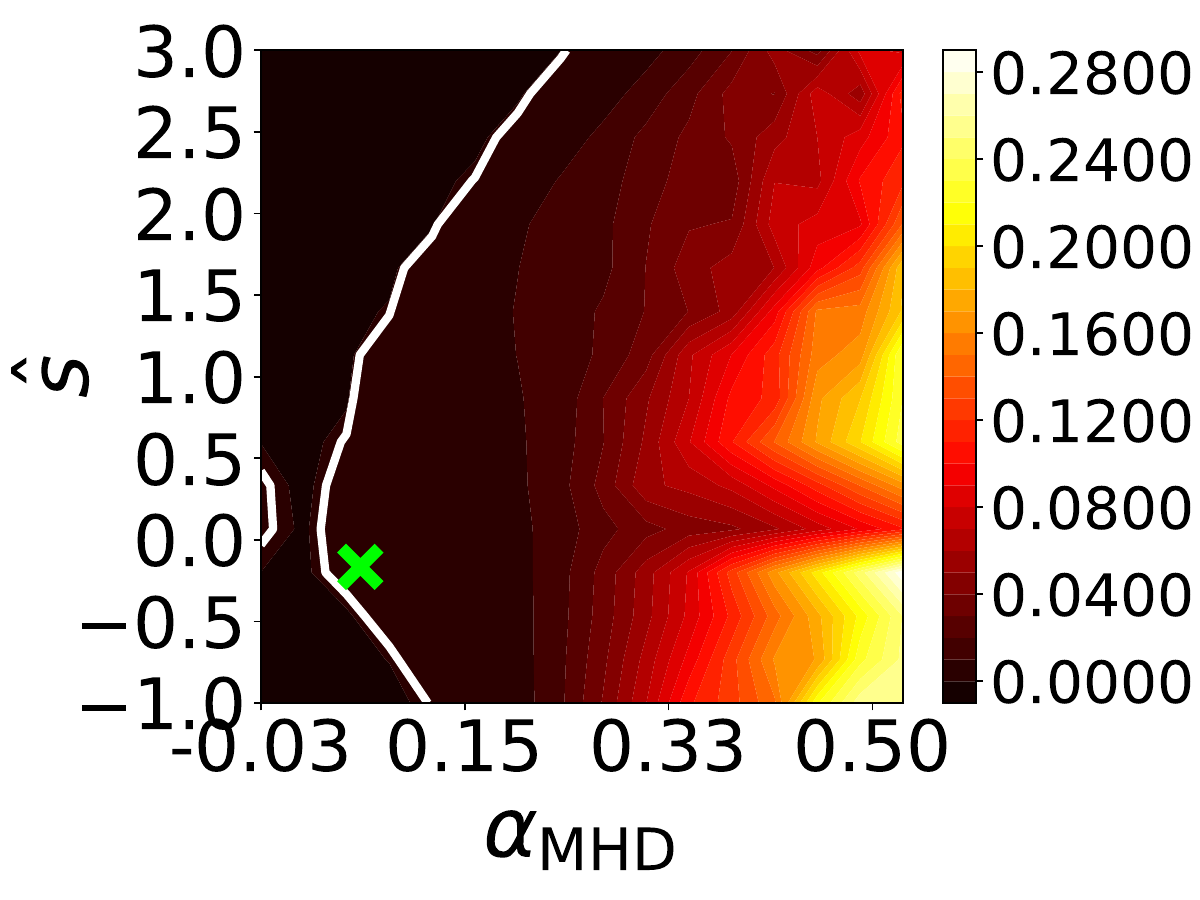}
        \caption{$\rho=0.97$}
    \end{subfigure}
    \caption{An $\hat{s}-\alpha_{\mathrm{MHD}}$ calculation of the maximum KBM growth rate (top row) and the ideal ballooning growth rate (bottom row) on four flux surfaces from core (left) to edge (right) of a W7-X stellarator equilibrium. The crosses mark the original values of $\hat{s}$ and $\alpha_{\mathrm{MHD}}$ whereas the white line is the marginal stability contour. The region enclosed by the contour is ballooning unstable and the region outside is ballooning stable. There is a strong correlation between the ideal-ballooning unstable region and the regions with a high KBM growth rate, that we shall use to optimize equilibria against KBMs.}
    \label{fig:W7X-high-beta-s-alpha}
\end{figure}

To demonstrate the connection between the ideal and kinetic ballooning modes, we perform a $\hat{s}-\alpha_{\mathrm{MHD}}$ analysis for a finite-$\beta$ W7-X equilibrium by numerically solving the linear gyrokinetic equation using the \texttt{GS2}~\cite{kotschenreuther1995quantitative, gs2ref, dorlandETG, jenko2000electron} solver and present our results in Figure~\ref{fig:W7X-high-beta-s-alpha}. We vary $\hat{s}$ and $\alpha_{\rm{MHD}}$ of the equilibrium on four flux surfaces and, for each pair of values, calculate the maximum KBM growth rate in the $\hat{s}-\alpha_{\rm{MHD}}$ space. We then compare it with the ideal ballooning stable and unstable regions in $\hat{s}-\alpha_{\mathrm{MHD}}$ space. The details of these gyrokinetic calculations and the process used to extract the maximum KBM growth rate are explained in~\ref{app:mode-filter}

The most important feature in Figure~\ref{fig:W7X-high-beta-s-alpha} is the strong correlation between the ideal and kinetic ballooning modes. We observe that the KBM growth rate decreases as we move farther away left (outside) of the contour of marginal stability. Hence, the distance from ideal-ballooning marginal stability can be used as a proxy for improving KBM stability. Increasing the distance from ideal-ballooning marginal stability is equivalent to obtaining equilibria with a large negative ideal ballooning eigenvalue $\lambda$.

\begin{figure}
    \centering
    \begin{subfigure}[b]{0.4\textwidth}
    \centering
        \includegraphics[width=\textwidth]{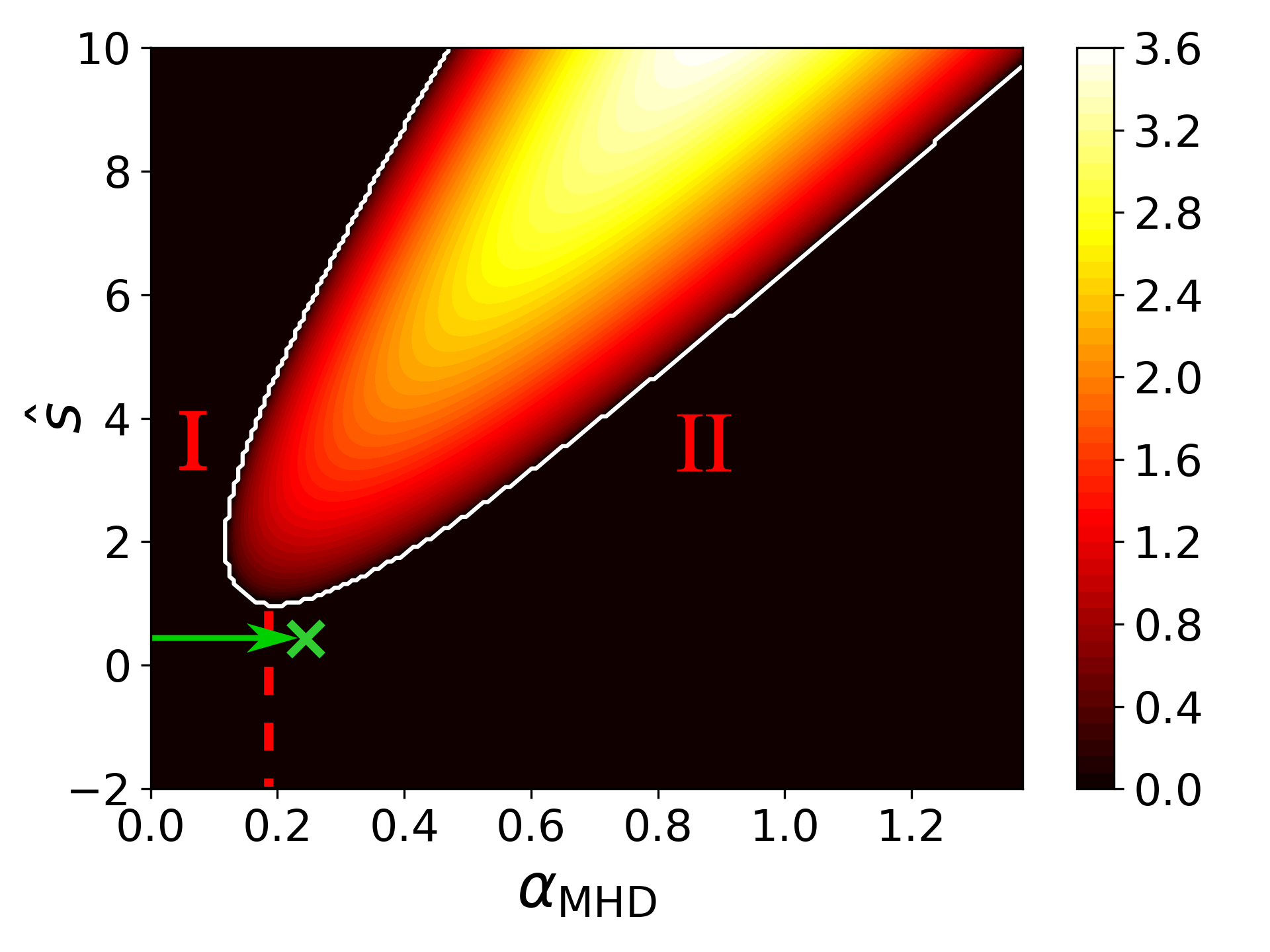}
        \caption{Second stability accessible}
    \end{subfigure}
    \qquad \qquad  
    \begin{subfigure}[b]{0.4\textwidth}
        \centering
        \vspace*{-8mm}
        \includegraphics[width=\textwidth, trim={0mm 0mm 0 0mm}, clip]{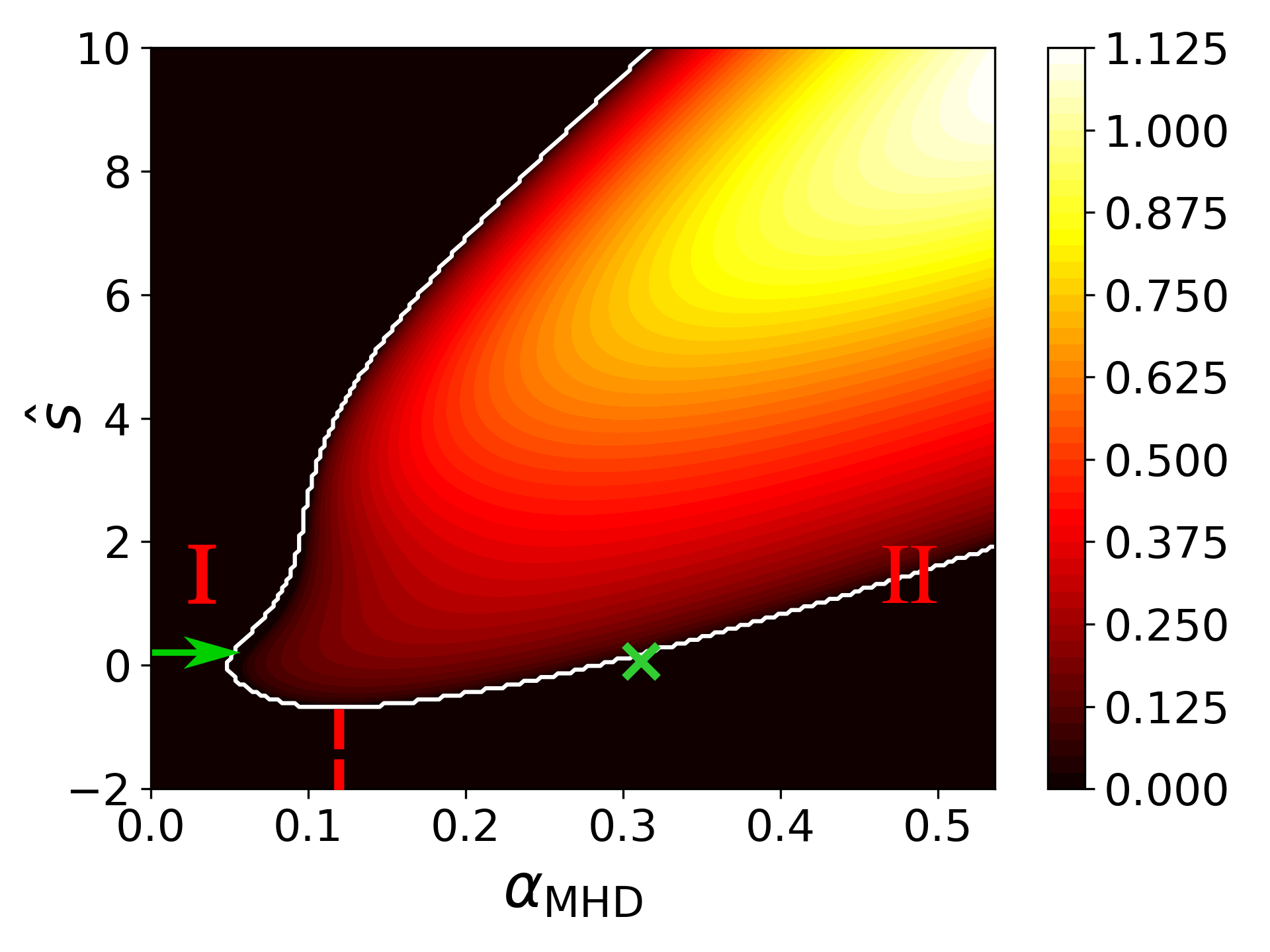}
        \caption{Second stability inaccessible}
    \end{subfigure}
    \caption{$\hat{s}-\alpha_{\mathrm{MHD}}$ diagram plotting the ideal ballooning growth rate, taken from figure 5$\textit{(e)}$(left) and  6$\textit{(c)}$(right) of Gaur \etal ~\cite{gaur2023microstability} for a modified DIII-D-like tokamak equilibrium. The dotted red line marks the region between the first stability (marked as $\mathrm{I}$) and the second stability (marked as $\mathrm{II}$). The cross marks the nominal gradient on the flux surface. The green arrow marks the path that an equilibrium may take to access that state starting from a vacuum state. In contrast to tokamaks, stellarators, in general, typically are not expected to exhibit a second-stable region or a potential path to such a region as shown in figure~\ref{fig:W7X-high-beta-s-alpha}.}
    \label{fig:second-stability}
\end{figure}

The second point concerns the stability against the ballooning mode at large $\alpha_{\mathrm{MHD}}$. In tokamaks, increasing the pressure gradient can make the equilibrium stable against the ideal ballooning as explained by figure~\ref{fig:second-stability}. This is equivalent to the green crosses in figure~\ref{fig:W7X-high-beta-s-alpha} moving from stable to unstable and back to stable regions. This stable region corresponding to large values of $\alpha_{\mathrm{MHD}}$ is called the second stability region. However, second stability generally does not exist in stellarators as has been shown by Hegna and Hudson~\cite{hegna2001loss}. From Figure~\ref{fig:W7X-high-beta-s-alpha}, for the range of $\hat{s}$ and $\alpha_{\mathrm{MHD}}$ it is clear that for this W7X configuration, a region of second stability does not exist at any radius.

\section{Reverse-mode-differentiable infinite-$n$ ideal ballooning solver in~\texttt{DESC}}
\label{sec:ballooning}
In this section, we briefly explain the ballooning solver implemented in the~\texttt{DESC} optimizer and how we leverage automatic differentiation to speed up the gradient calculation. We also benchmark our solver against existing solvers.

Extending the work in\cite{gaur2023adjoint}, we develop a reverse-mode differentiable, fast, and accurate solver for the infinite-$n$ ideal ballooning equation in~\texttt{DESC}. In a package that facilitates automatic differentiation, reverse mode gradients allow us to calculate gradients of physical quantities, which in this case is the maximum eigenvalue $\lambda_{\mathrm{max}}$ in~\eref{eqn:ballooning-matrix-equation}, by automatically applying the chain rule between the final quantity and the input parameters of the problem $\bi{p}$. This speeds up the calculation $\partial \lambda/\partial \bi{p}$ compared to the forward-mode gradient calculation by a factor of the number of parameters $N_{\bi{p}}$, similar to an adjoint method. However, we can obtain reverse mode gradients without solving an adjoint equation and using explicit formulae as used in~\cite{gaur2023adjoint}. In that sense, this technique is much more modular and powerful than adjoint methods.\footnote{One has to carefully define these types of problems in ~\texttt{DESC} as not all objectives are reverse-mode differentiable. This is especially true for objectives involving time-dependent nonlinear dynamics.}

\begin{figure}
    \centering
    \begin{subfigure}[b]{0.4\textwidth}
    \centering
        \includegraphics[width=\textwidth]{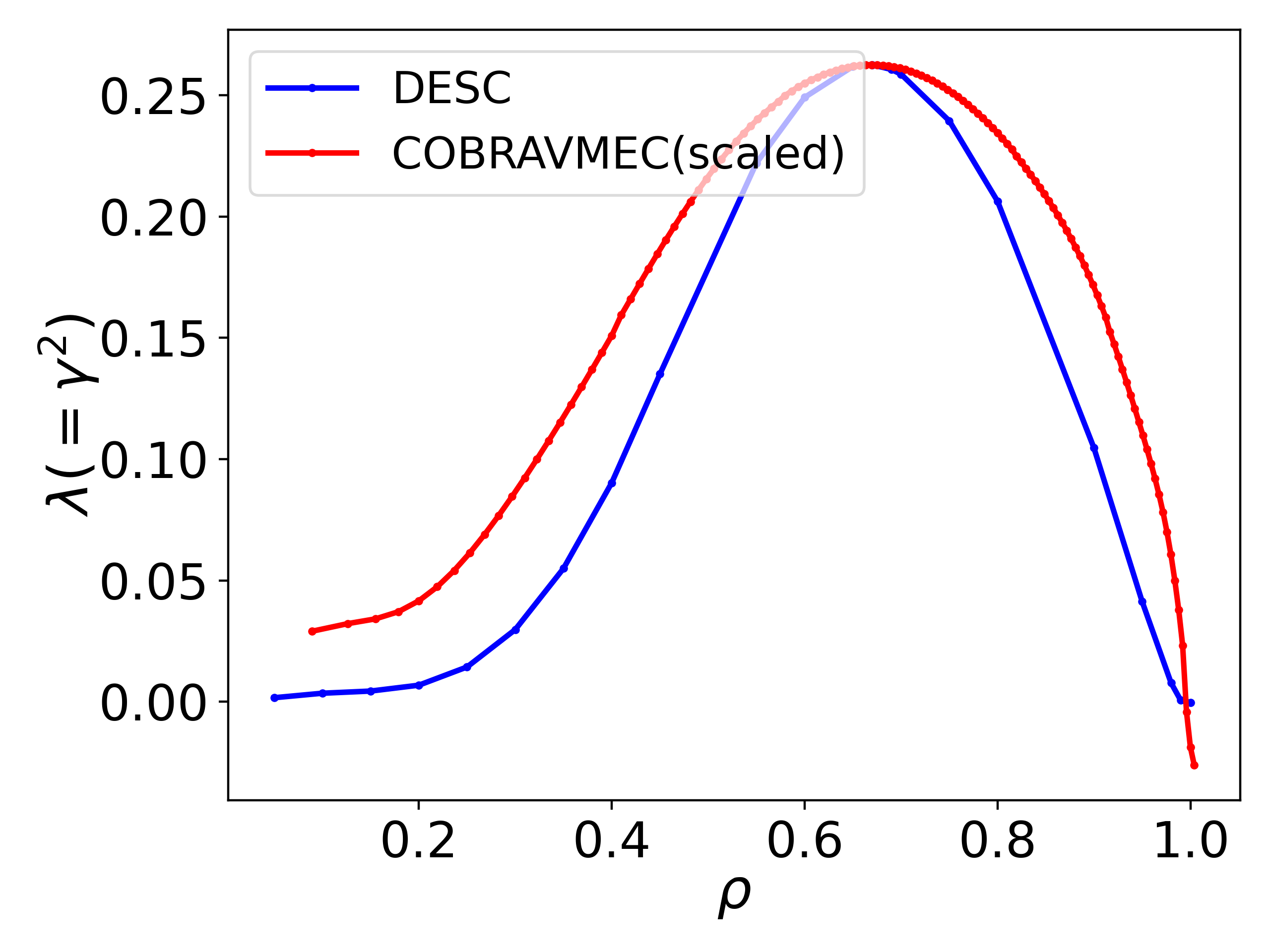}
        \caption{Comparison with~\texttt{COBRAVMEC}}
    \end{subfigure}
    \quad 
    \begin{subfigure}[b]{0.4\textwidth}
        \centering
        \includegraphics[width=\textwidth]{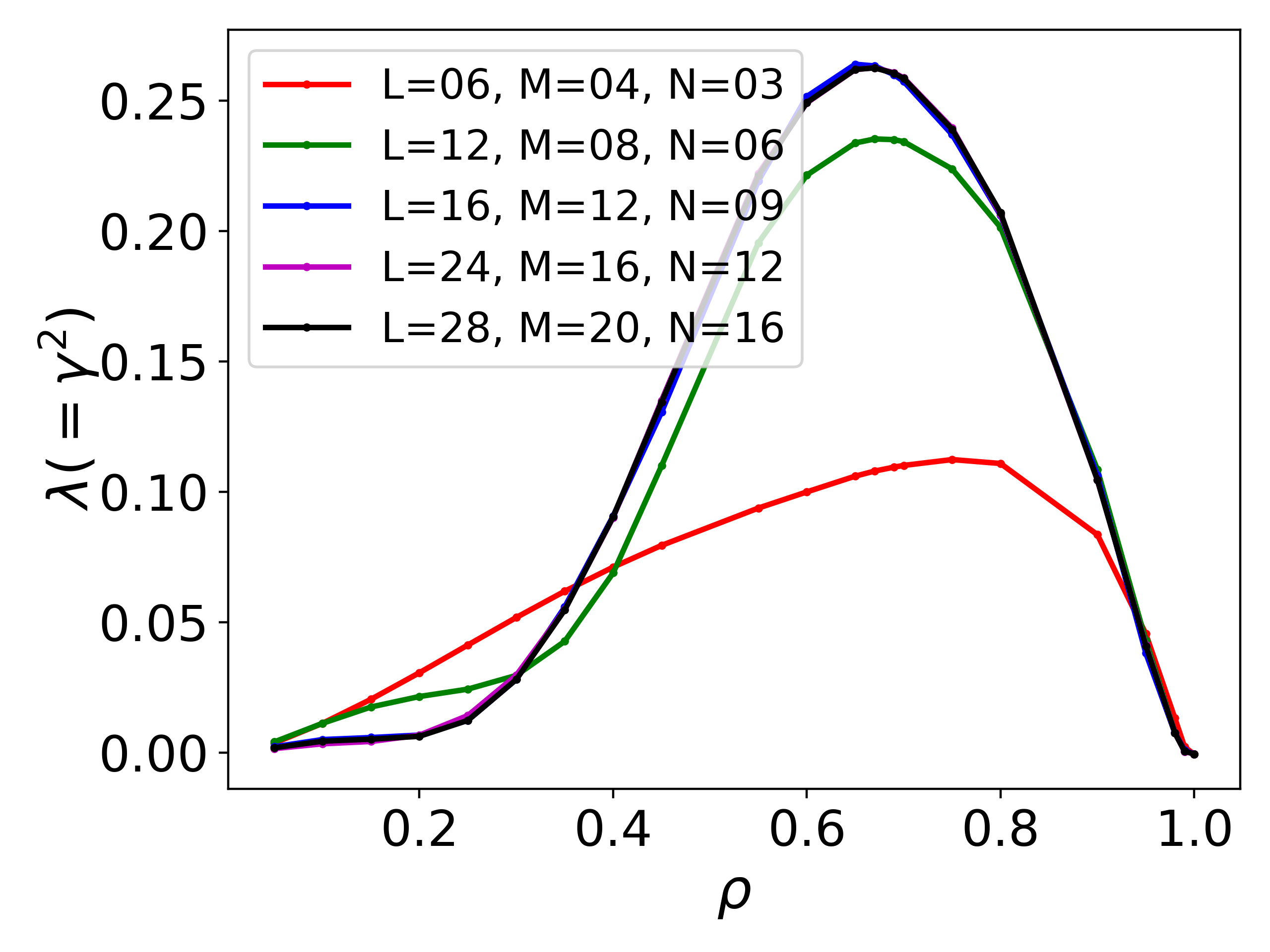}
        \caption{Convergence study in~\texttt{DESC}}
    \end{subfigure}
    \caption{Comparison of the~\texttt{DESC} ballooning solver with~\texttt{COBRAVMEC} shown in $(a)$. Due to different normalizations, the~\texttt{COBRAVMEC} output has been scaled by the ratio $\lambda_{\mathrm{max}, \texttt{DESC}}/\lambda_{\mathrm{max}, \texttt{COBRAVMEC}}$. In $(b)$, we show a convergence test of the maximum ballooning eigenvalue with increasing equilibrium resolution $L, M, N$. To calculate $\lambda_{\mathrm{max}}$, we scan over $N_{\alpha} = 14$ field lines, and $N_{\zeta_0} = 15$ values of the ballooning parameter $\zeta_0 \in [-\pi/2, \pi/2]$. On each field line, we choose $N_0 = 8 \times 1.5 M \times 1.5 N$ points with $\zeta \in [-4\pi, 4\pi]$, where $M, N$ are the poloidal and toroidal resolution of a~\texttt{DESC} equilibrium, respectively. Since the pressure gradient $dp/d\psi = 0$ on the magnetic axis for this equilibrium, $\lambda_{\mathrm{max}} \leq 0$ which is satisfied by the~\texttt{DESC} solver.}
\label{fig:ideal-ballooning-solver-tests}
\end{figure}

To solve the infinite-$n$ ideal ballooning equation, we discretize~\eref{eqn:ideal-ballooning-equation} along a field line $\alpha$ using the finite-difference scheme described in~\cite{sanchez2000cobra}
\begin{equation}
    \fl \mathrm{g}_{j+1/2}\frac{(\widehat{X}_{j+1} - \widehat{X}_{j})}{ \Delta \zeta^2} - \mathrm{g}_{j-1/2}\frac{(\widehat{X}_{j} - \widehat{X}_{j-1})}{\Delta \zeta^2} + (\mathrm{c}_j - \widehat{\lambda} \mathrm{f}_j)\widehat{X}_j = 0, \quad j = 0\ldots N-1,
    \label{eqn:discrete-iball}
\end{equation}
where
\numparts
\begin{eqnarray}
    \mathrm{g} = (\bi{b}\cdot \bi{\nabla}_{\rm{N}} \zeta) \frac{| \bi{\nabla}_{\rm{N}} \alpha |^2}{B/B_{\rm{N}}},\\
    \mathrm{c} = \frac{1}{(B/B_{\rm{N}})^2}\frac{d(\mu_0 (p/B_{\rm{N}}^2))}{d\psi_{\rm{N}}} \frac{2}{(\bi{b}\cdot \bi{\nabla}_{\rm{N}}\zeta)} (\bi{b}\times \bkappa_{\rm{N}}) \cdot \bi{\nabla}_{\rm{N}}\alpha, \, \bkappa_{\rm{N}} = (\bi{b}\cdot \bi{\nabla}_{\rm{N}}\bi{b}) \\
    \mathrm{f} = \frac{1}{(\bi{b}\cdot \bi{\nabla}_{\rm{N}}\zeta)} \frac{| \bi{\nabla}_{\rm{N}}\alpha |^2}{(B/B_{\rm{N}})^3},
\end{eqnarray}
\endnumparts
calculated on a uniformly spaced grid with points being separated by $\Delta \zeta$, subject to the boundary conditions
\begin{equation}
 \widehat{X}_0 =  0,\,  \widehat{X}_{N} = 0.   
\end{equation}
The discretized generalized symmetric eigenvalue equation can be written as
\begin{eqnarray}
    \mathcal{A} X = \lambda \mathcal{B}  X
\label{eqn:ballooning-matrix-equation}
\end{eqnarray}
where $\mathcal{A}$ is a symmetric, tridiagonal matrix 
\[
\fl
\mathcal{A}\! =\!
\left[\begin{array}{ccccccc}
\mathrm{c}_{1} - \frac{(\mathrm{g}_{1/2} +  \mathrm{g}_{3/2})}{(\Delta \zeta)^2} \hspace*{-8mm} &  \frac{\mathrm{g}_{3/2}}{(\Delta \zeta)^2} & \hspace*{-4mm}  0 &\hspace*{-3mm} 0 &  \hspace*{-4mm}\ldots & \hspace*{-6mm} 0 & \hspace*{-10mm} 0\\
\frac{\mathrm{g}_{3/2}}{(\Delta \zeta)^2} \hspace*{-7mm} &  \mathrm{c}_{2} - \frac{(\mathrm{g}_{3/2}  + \mathrm{g}_{5/2})}{(\Delta \zeta)^2}  & \hspace*{-2mm} \frac{\mathrm{g}_{5/2}}{(\Delta \zeta)^2} & \hspace*{-3mm} 0  & \hspace*{-4mm} \ldots & \hspace*{-6mm}  0 & \hspace*{-10mm} 0\\
& & & \ddots\\
& &  & & \ddots\\
0 & 0 &\hspace*{-2mm} 0 & \hspace*{-2mm} 0 & \hspace*{-4mm} \ldots &  \hspace*{-4mm}  \mathrm{c}_{N\!-\!2}-\frac{(\mathrm{g}_{N\!-\!5/2}  + \mathrm{g}_{N\!-\!3/2})}{(\Delta \zeta)^2}  & \hspace*{-10mm}  \frac{\mathrm{g}_{N\!-\!3/2}}{(\Delta \zeta)^2}\\
0 & 0 &\hspace*{-2mm} 0 & \hspace*{-2mm} 0 & \hspace*{-4mm} \ldots &  \hspace*{-6mm} \frac{\mathrm{g}_{N\!-\!3/2}}{(\Delta \zeta)^2} & \hspace*{-10mm}  \mathrm{c}_{N\!-\!1}-\frac{(\mathrm{g}_{N\!-\!3/2}  + \mathrm{g}_{N\!-\!1/2})}{(\Delta \zeta)^2} \\
\end{array}\right]
\]
and $\mathcal{B}$ is a symmetric, diagonal matrix
\[
\centering
\mathcal{B}\! = \!
\left[\begin{array}{ccccccc}
\mathrm{f}_{1} \hspace*{-2mm} &  0 & \hspace*{-2mm}  0 &\hspace*{-3mm} 0 &  \hspace*{-4mm}\ldots & \hspace*{-6mm} 0 & \hspace*{-10mm} 0\\
0 \hspace*{-2mm} & \mathrm{f}_{2}  & \hspace*{-2mm} 0 & \hspace*{-3mm} 0  & \hspace*{-4mm} \ldots & \hspace*{-6mm}  0 & \hspace*{-10mm} 0\\
& & & \ddots\\
& &  & & \ddots\\
0 & 0 &\hspace{-2mm} 0 & \hspace*{-2mm} 0 & \hspace*{-3mm} \ldots &  \hspace*{-3mm} \mathrm{f}_{N\!-\!2}  & \hspace*{-2mm}0\\
0 & 0 &\hspace*{-2mm} 0 & \hspace*{-2mm} 0 & \hspace*{-4mm} \ldots &  \hspace*{-3mm} 0 & \hspace*{-2mm}  \mathrm{f}_{N-1}\\
\end{array}\right]
\]
Using DESC, we solve~\eref{eqn:ballooning-matrix-equation} on $N_{\rho} =  6$ flux surfaces with $\rho \in [0.15, 0.95]$. On each flux surface, we solve the ballooning equation on the $N_{\alpha} = 14$ field lines with $\alpha \in [0, \pi]$ and for $N_{\zeta_0} = 15$ values of the ballooning parameter $\zeta_0 \in [-\pi/2, \pi/2]$ to find the maximum eigenvalue $\lambda_{\mathrm{max}} = \mathrm{max} (\lambda_{j, k})\, \, \forall\,  j \in [1, N_{\zeta_0}], k \in [1, N_{\alpha}], j, k \in \mathbb{Z}$. Using this definition, we define the following ballooning objective function for each flux surface
\begin{eqnarray}
    f_{\mathrm{ball}}(\rho) = w_0 \,  \mathrm{ReLU}(\lambda_{\mathrm{max}}(\rho) - \lambda_0) + w_1 \sum_{j=1}^{N_{\zeta_0}} \sum_{k = 1}^{N_{\alpha}} \mathrm{ReLU} (\lambda_{j, k}(\rho) - \lambda_0)
\end{eqnarray}
where $\mathrm{ReLU}$ is the Rectified Linear Unit operator, $i$ is the index of the flux surface, $w_0, w_1$ are constant weights, and $\lambda_0 < 0$ is the desired distance from ideal-ballooning marginality. As a benchmark, we present a comparison of our ideal ballooning solver with~\texttt{COBRAVMEC}~\cite{sanchez2000cobra} and a convergence study with equilibrium resolution in Figure~\ref{fig:ideal-ballooning-solver-tests}. 

\section{Omnigenity}
\label{sec:omnigenity}
Omnigenity is a favorable property of a magnetic field that ensures the radial confinement of the trapped particles in a stellarator. Mathematically, an ideal MHD equilibrium with nested flux surfaces is omnigenous~\cite{hall1975three} if
\begin{eqnarray}
\int_{\zeta_{\mathrm{b}1}}^{\zeta_{\mathrm{b}2}}\frac{d \zeta}{(\bi{b}\cdot \bi{\nabla}\zeta)} (\bi{v}_{Ds} \cdot \bi{\nabla} \psi) = 0
\end{eqnarray}
where the particle drift $\bi{v}_{Ds}$ is the magnetic drift velocity, defined in~\eref{eqn:magnetic-drift-velocity}, $\zeta$ is a field-line following coordinate, $\zeta_{\mathrm{b}1}$ and $\zeta_{\mathrm{b}2}$ are the bounce points of a particle and the integral operation, known as bounce-averaging, denotes the average radial deviation of a particle from a flux surface. For a system with a longitudinal adiabatic invariant,
\begin{eqnarray}
    \mathcal{J}_{\parallel} = \int_{\zeta_{\mathrm{b}1}}^{\zeta_{\mathrm{b}2}}\frac{d \zeta}{(\bi{b}\cdot \bi{\nabla}\zeta)} w_{\parallel},
\end{eqnarray}
the omnigenity condition is equivalent to $\partial_{\alpha} \mathcal{J}_{\parallel} = 0$ implying that the longitudinal adiabatic invariant is a flux function for omnigenous equilibria.

Due to axisymmetry, a continuous toroidal symmetry, tokamaks are inherently omnigenous. Since stellarators do not possess continuous toroidal symmetry, they have to be optimized for omnigenity. Most of the work over the last decade in stellarator optimization has focused on a subset of omnigenity, a hidden symmetry known as quasisymmetry~\cite{garren1991magnetic}. For quasi-symmetric equilibria, the magnetic field strength $B$ on a flux surface can be completely defined by a single angular coordinate, unlike a general stellarator magnetic field $B = B(\theta, \zeta)$ that requires two angular coordinates.

Depending on the topology of the magnetic field, a quasisymmetric configuration can have $B = B(\theta_{\mathrm{B}}), B = B(\theta_{\mathrm{B}} - n_{\mathrm{FP}}\zeta_{\mathrm{B}})$, or $B = B(\zeta_{\mathrm{B}})$, where $\theta_{\mathrm{B}}, \zeta_{\mathrm{B}}$ are Boozer angles, and $n_{\mathrm{FP}}$ is the field period, a positive integer characterizing the discrete toroidal symmetry of a stellarator. In the same order as defined in the previous sentence, these equilibria are called quasiaxisymmetric (QA), quasihelically symmetric (QH), and quasipoloidally symmetric (QP). Analogous to quasisymmetry, Cary and Shasharina\cite{cary1997helical} developed a specialized angular coordinate system in which one can define omnigenity in terms of the magnetic field strength in a manner analogous to quasisymmetry. Since omnigenity is a superset of quasisymmetry, it expands the space of possible omnignous configurations, allowing us to optimize for additional favorable properties, apart from omnigenity. In this work, we optimize stellarators for equilibria with poloidal omnigenity (OP), toroidal omnigenity (OT), and helical omnigenity (OH) along with additional objectives related to MHD and kinetic stability.

In recent years, there have been many stellarator designs with poloidal omgnigenity OP (also known as QI)~\cite{plunk2019direct,goodman2023constructing,jorge2022single, mata2022direct} but toroidal or helical omnigenity has not been explored. In this work, we use the technique from Cary and Shasharina~\cite{cary1997omnigenity} and numerically implemented by Dudt~\textit{et al.}~\cite{dudt2024magnetic} in the~\texttt{DESC} code to find omnigenous stellarator configurations. The theory and implementation are briefly described in the following paragraphs.

\begin{figure}
    \centering
    \begin{subfigure}[b]{0.4\textwidth}
        \centering
        \includegraphics[width=\textwidth]{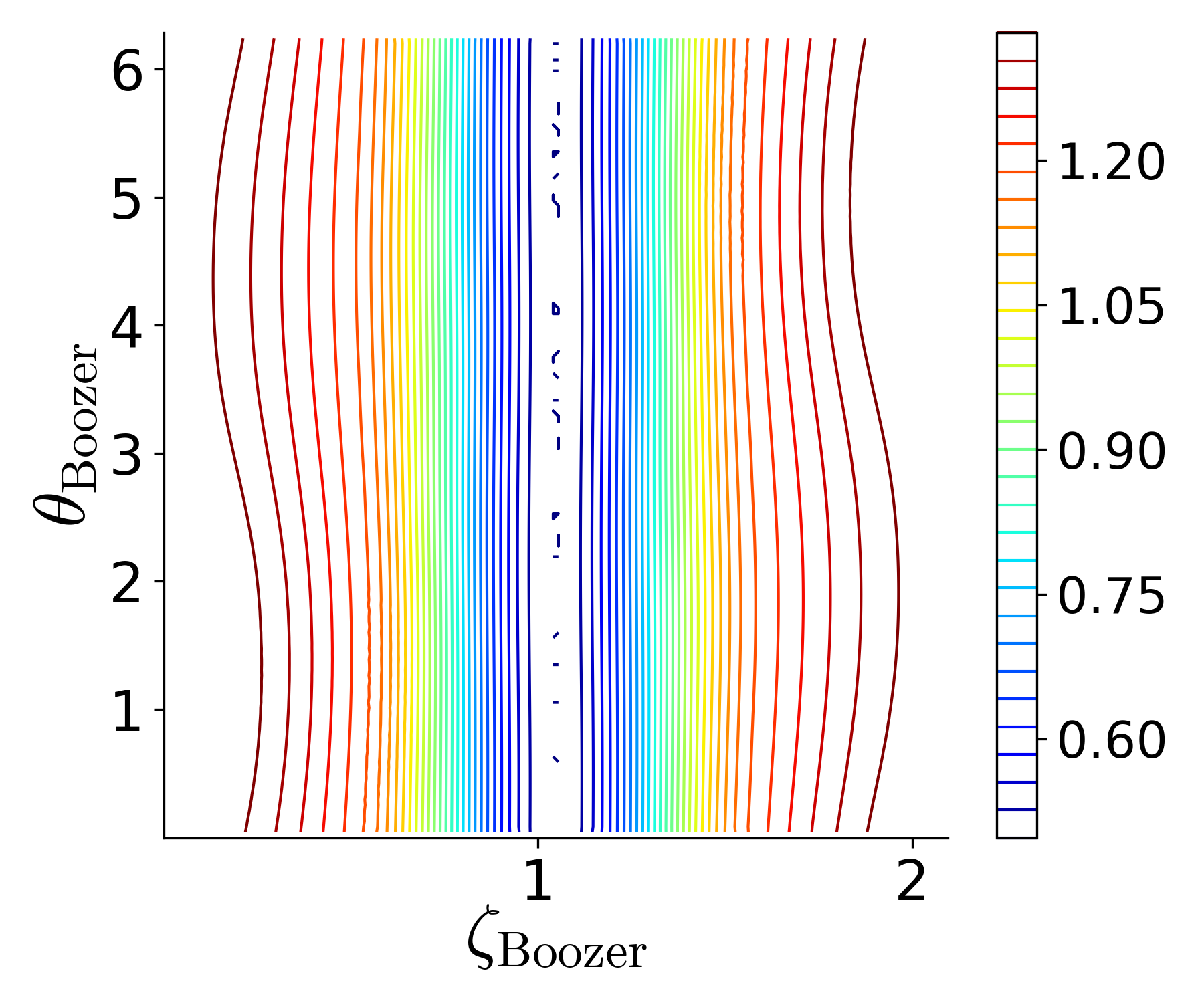}
        \caption{Target $B$}
    \end{subfigure}
    \quad 
    \begin{subfigure}[b]{0.4\textwidth}
    \centering
        \includegraphics[width=\textwidth]{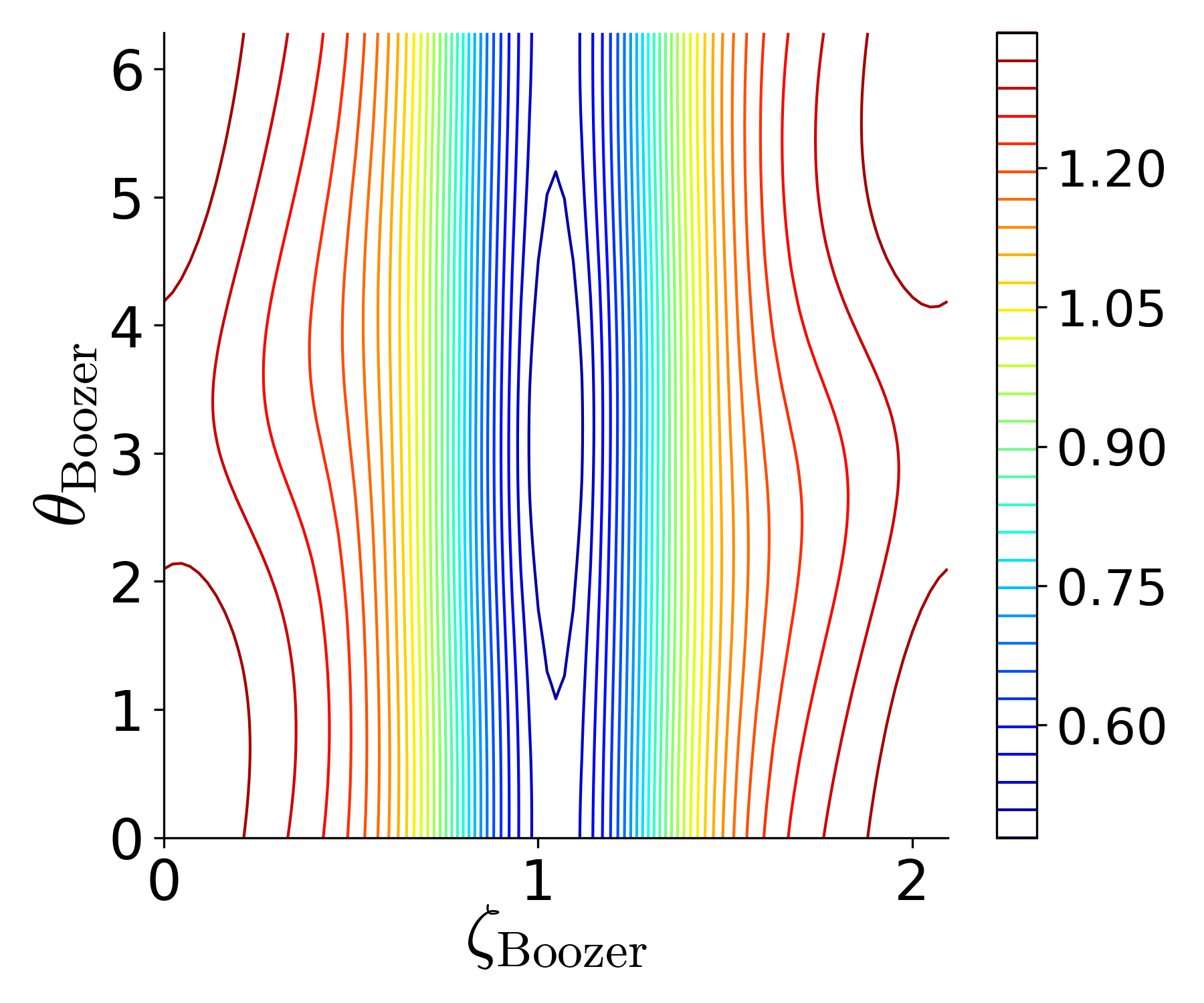}
        \caption{Optimized $B$}
    \end{subfigure}
    \caption{Comparison of a typical poloidally omnigenous magnetic field (in Boozer coordinates) on the flux surface for an optimized configuration. Figure $(a)$ shows the target omnigenous field generated using the Cary-Shasharina prescription, whereas $(b)$ shows the optimized equilibrium field.}
\label{fig:OP-example}
\end{figure}

Cary and Shasharina derived the following conditions that an omnigenous field must satisfy on each flux surface:
\begin{enumerate}
    \item The minimum and maximum magnetic field strength $B_{\mathrm{min}}$ and $B_{\mathrm{max}}$ must be independent of the field line label $\alpha$
    \item The contour of the maximum magnetic field strength must be a closed line in Boozer coordinates 
\end{enumerate}
To implement these omnigenous target magnetic fields, Dudt~\textit{et al.}~\cite{dudt2024magnetic} designs a target omnigenous field profile $B$ in~\texttt{DESC} and penalizes the equilibrium field $B_{\mathrm{eq}}$ through the objective
\begin{eqnarray}
    f_{\mathrm{om}} = \sum_{i} (B_{\mathrm{eq}, i} - B)^2,
    \label{eqn:Omnigenity_error}
\end{eqnarray}
where $B$ is the target omnigenous field and $i$ is the index of the flux surface over which $f_{\mathrm{om}}$ is calculated. By minimizing $f_{\mathrm{om}}$ over multiple flux surfaces, we ensure that the equilibrium field is as close to omnigenous as possible~\footnote{Note that minimizing $f_{\mathrm{om}}$ is a sufficient but not a necessary condition to ensure omnigenity. An equilibrium field could deviate significantly from the target field $B$, \textit{i.e.}, have a large $f_{\mathrm{om}}$ and still be omnigenous}. A typical example showing the equilibrium and target omnigenous field is presented in Figure~\ref{fig:OP-example}.

The Cary-Shasharina prescription differs significantly from the objectives used to optimize poloidal omnigenity by Isaev \etal~\cite{isaev1997plasma} and Subbotin \etal~\cite{subbotin2006integrated}, which are based on Mikhailov \etal~\cite{mikhailov2002improved}, where the equilibrium is optimized for both quasi-poloidal symmetry and omnigenity by enforcing $\mathcal{J}_{\parallel} = \mathcal{J}_{\parallel}(\psi)$. In that sense, our method offers greater flexibility in the optimization process because we are not enforcing quasisymmetry along with omnigenity. To corroborate our point, in addition to the enhanced stability results in the following section, we have displayed the neoclassical transport coefficient $\epsilon_{\mathrm{eff}}^{3/2}$ in figure~\ref{fig:effective-ripple}.

\section{Omnigenous equilibria with improved KBM stability}
\label{sec:Results}
In this section, we explain how we generate omnigenous equilibria with improved stability with~\texttt{DESC}. Using the ideas explained in the previous sections, we first search for omnigenous stellarators with poloidally closed contours. To do that, we use the following objective function
\begin{eqnarray}
 \fl  \mathcal{F} =  w_{A} f_{\mathrm{aspect}}^2 + w_{B} f_{\mathrm{ball}}^2 +  w_{C} f_{\mathrm{curv}}^2 + w_{D} f_{\mathrm{DMerc}}^2 + w_{E} f_{\mathrm{elongation}}^2 + w_{O} f_{\mathrm{om}}^2 ,
\end{eqnarray}
where $f_{x}$ are various objectives on the right side for omnigenity, distance from ideal-ballooning marginality, Mercier stability, boundary curvature, boundary elongation, and aspect ratio, and $w_{A}, w_{B}, w_{C}, w_{D}, w_{E}, w_{O}$ are weights used with each objective function. The exact definitions of these objectives are provided in~\ref{app:objectives-FoMs-defn}. At each iteration of a DESC optimization, $\mathcal{F}$ is minimized while satisfying~\eref{eqn:ideal-MHD-force-balance}. Mathematically,
\begin{equation}
    \min (\mathcal{F}_{\mathrm{stage-one}}), \quad \textrm{s.t.} \, \bi{\nabla}\left(\mu_0 p + \frac{B^2}{2}\right) - \bi{B}\cdot\bi{\nabla}\bi{B} =  0, \, \psi_{\mathrm{b}} = \psi_{\mathrm{b}0}, p = p_{0}(\psi), \iota = \iota_0(\psi)
\end{equation}
which is minimized with respect to a vector of parameters $\bi{p}$ that denote the coefficients representing the shape of the boundary and the field of omnigenity. Note that optimization is performed while maintaining force balance, fixed profiles, and the toroidal flux enclosed by the plasma boundary.

With the objective $\mathcal{F}$ defined, we run~\texttt{DESC} on a single NVIDIA A100 GPU. A single optimization takes less than four hours. Currently, the speed of optimization is limited by the Boozer transformation needed for the omnigenity objective and the ideal-ballooning growth rate calculation. We expect a significant speedup as we vectorize these calculations and move towards~\texttt{DESC} \textit{v1.0}. 

\subsection{Poloidal omnigenity (OP)}
\label{subsec:OP}
\begin{figure}[!h]
    \centering
    \begin{subfigure}[b]{0.32\textwidth}
    \centering
        \includegraphics[width=\textwidth]{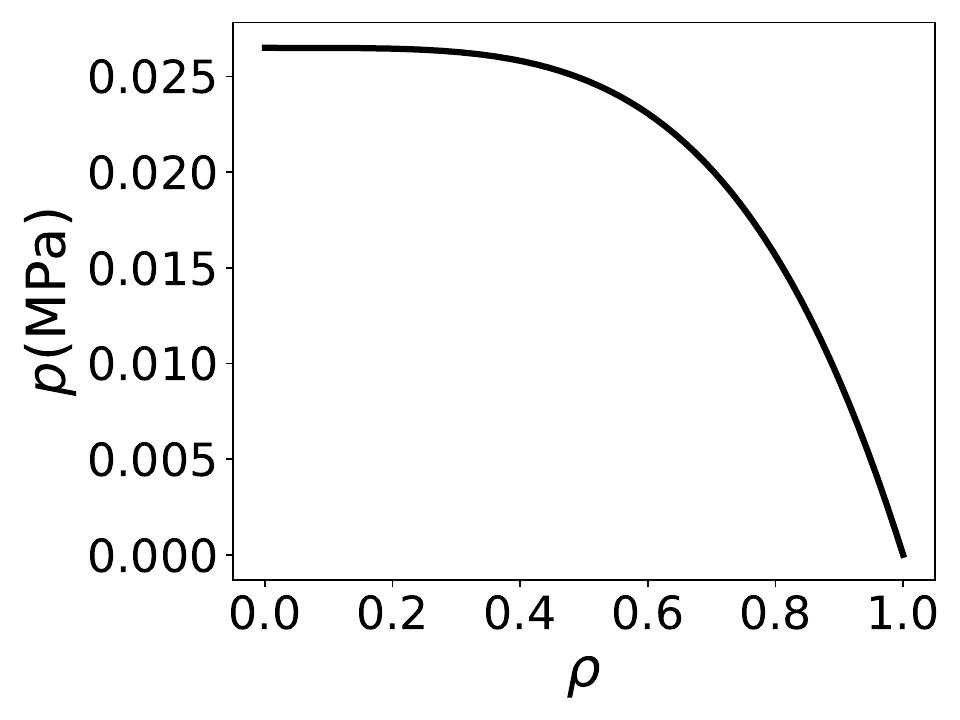}
        \caption{Pressure}
    \end{subfigure}
    \begin{subfigure}[b]{0.32\textwidth}
        \centering
        \includegraphics[width=\textwidth]{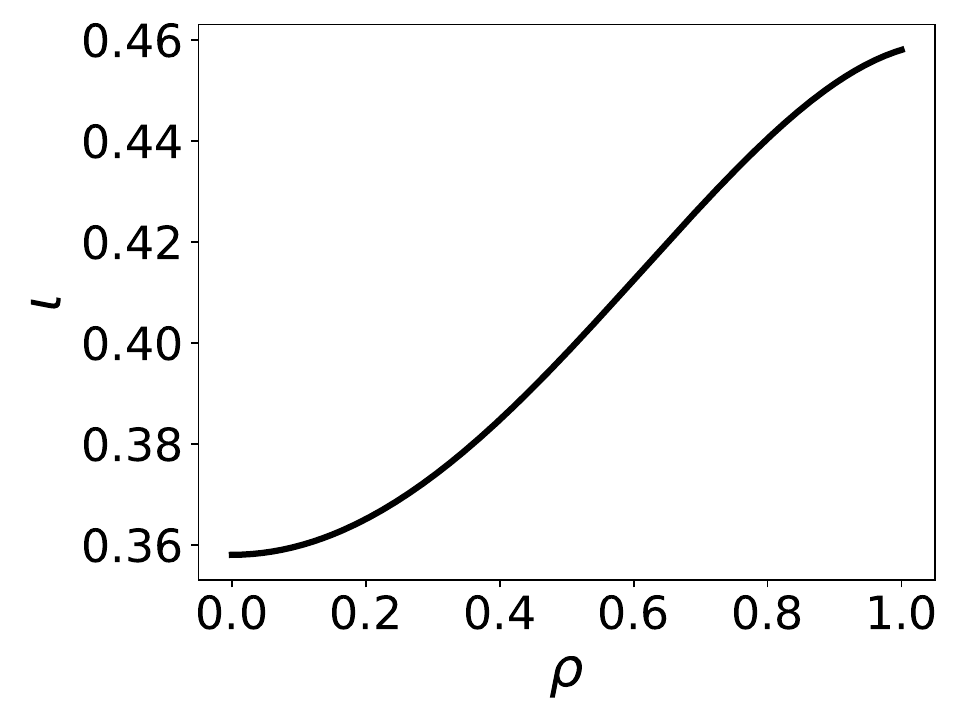}
        \caption{Rotational transform}
    \end{subfigure}
    \begin{subfigure}[b]{0.32\textwidth}
        \centering
        \includegraphics[width=\textwidth, trim={0mm 12mm 0 0}, clip]{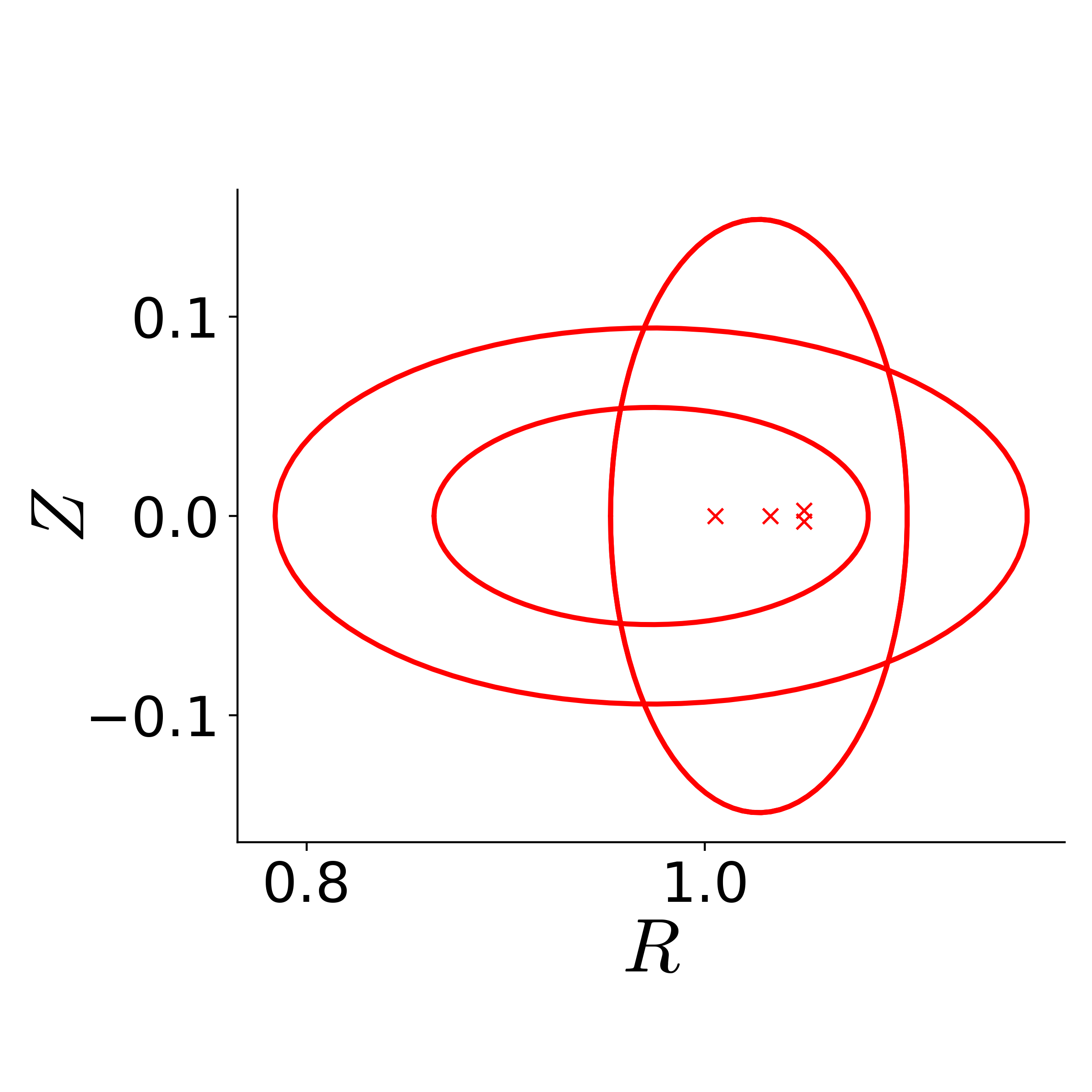}
        \caption{Initial boundary}
    \end{subfigure}
    \caption{Inputs to the optimization module in~\texttt{DESC} for the OP case. Figures $(a)$ and $(b)$ show the various profiles and figure $(c)$ has the boundary cross-section at different toroidal angles for a single field period.}
\label{fig:OP-inputs}
\end{figure}

\begin{figure}
    \centering
    \begin{subfigure}[b]{0.32\textwidth}
    \centering
        \includegraphics[width=\textwidth]{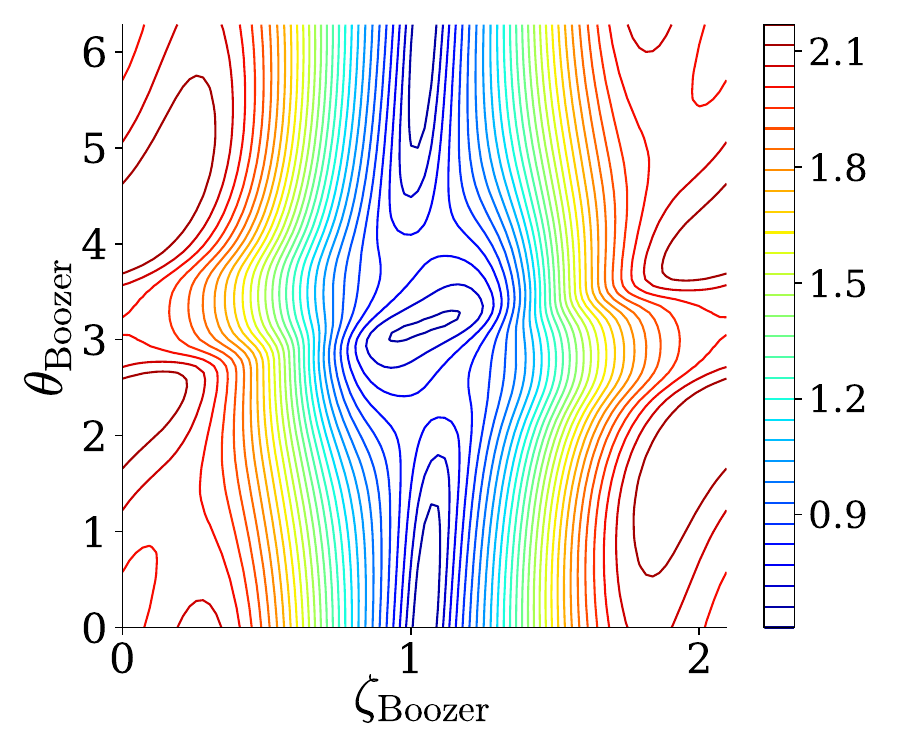}
        \caption{Initial B $(\rho=1)$}
    \end{subfigure}
    \quad 
    \begin{subfigure}[b]{0.32\textwidth}
        \centering
        \includegraphics[width=\textwidth]{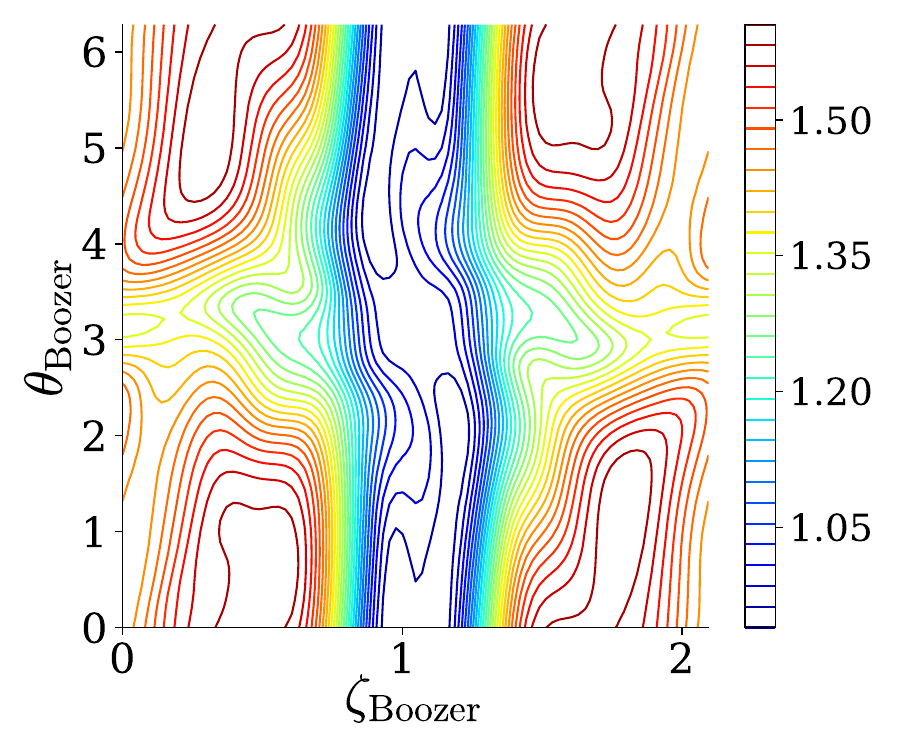}
        \caption{Optimized B $(\rho=1)$}
    \end{subfigure}
    \begin{subfigure}[b]{0.3\textwidth}
        \centering
        \includegraphics[width=\textwidth, trim={0mm 4mm 0 9mm}, clip]{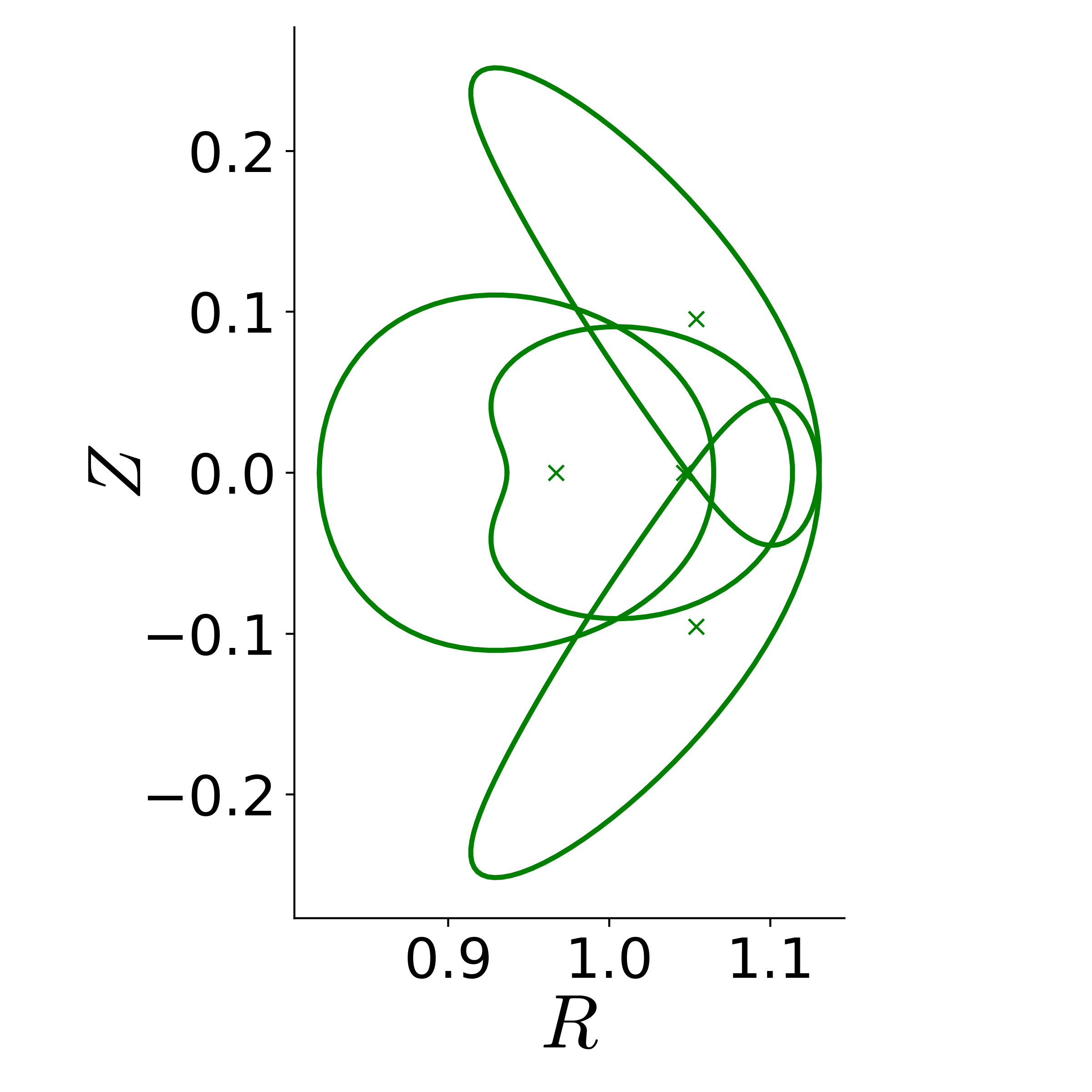}
        \caption{Optimized boundary}
    \end{subfigure}\\
    
    \clearpage
    \hspace*{-6mm}
    \begin{subfigure}[b]{0.325\textwidth}
    \centering
        \includegraphics[width=\textwidth, trim={2mm 2mm 4mm 3mm}, clip]{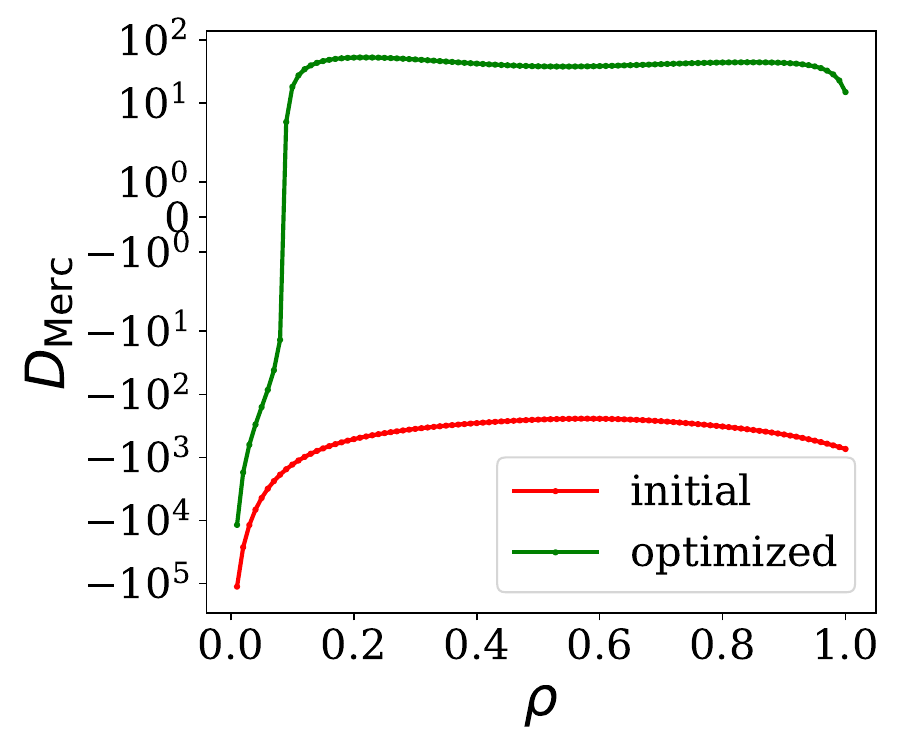}
        \caption{Mercier stability}
    \end{subfigure}
    \quad 
    \begin{subfigure}[b]{0.317\textwidth}
        \centering
        \includegraphics[width=\textwidth, trim={2mm 2mm 4mm 3mm}, clip]{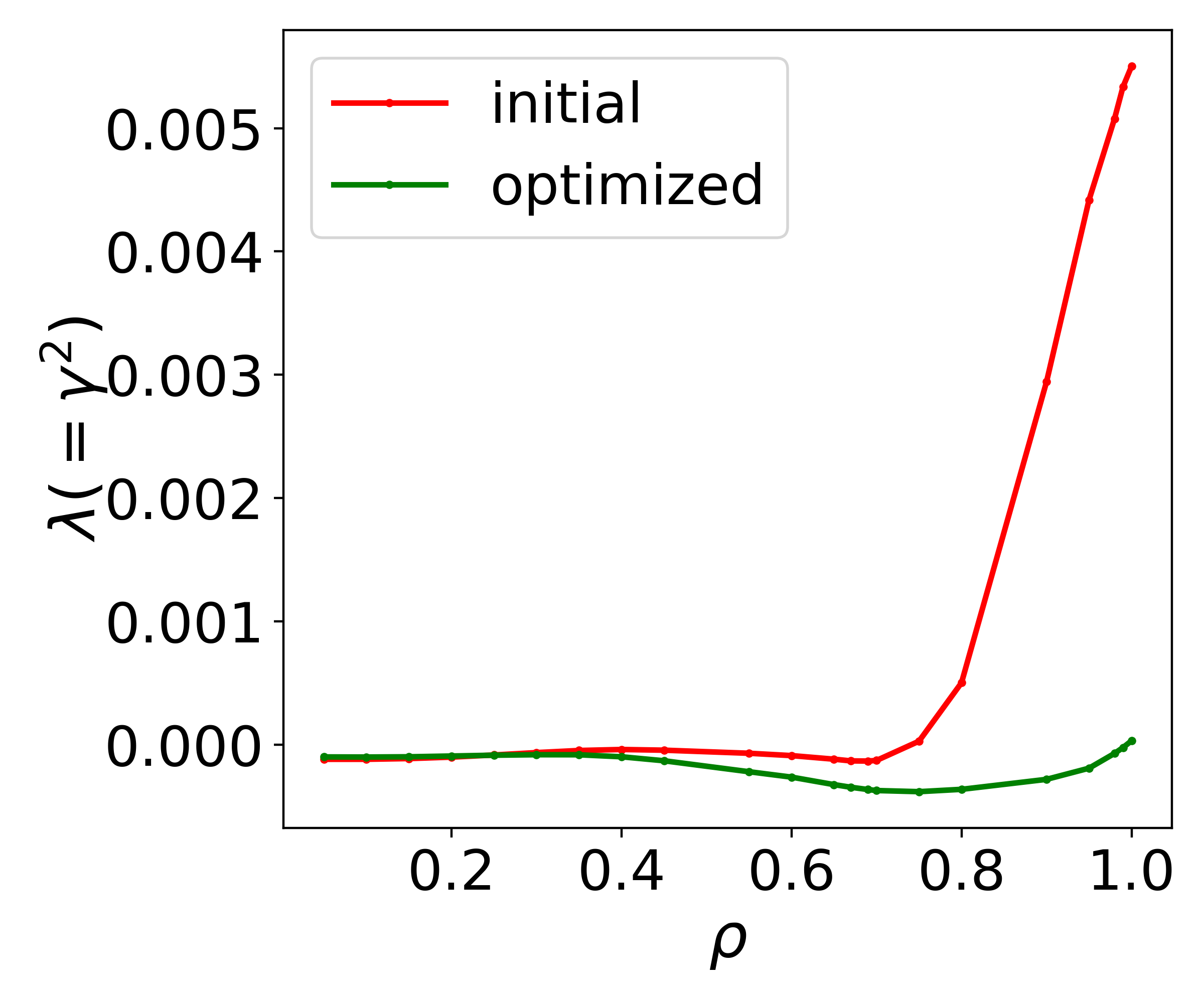}
        \caption{Ballooning stability}
    \end{subfigure}
    \begin{subfigure}[b]{0.322\textwidth}
        \centering
        \includegraphics[width=\textwidth, trim={0mm 0mm 0 0mm}, clip]{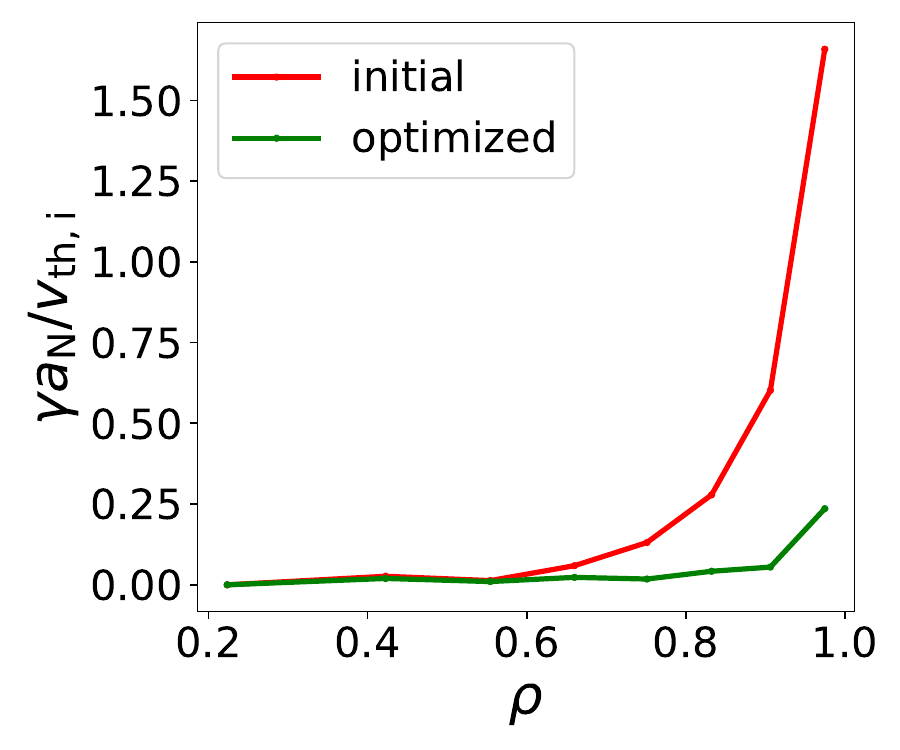}
        \caption{KBM growth rate}
    \end{subfigure}\\
    \clearpage
    \begin{subfigure}[b]{0.35\textwidth}
    \centering
        \includegraphics[width=\textwidth, trim={0mm 0mm 0 4mm}, clip]{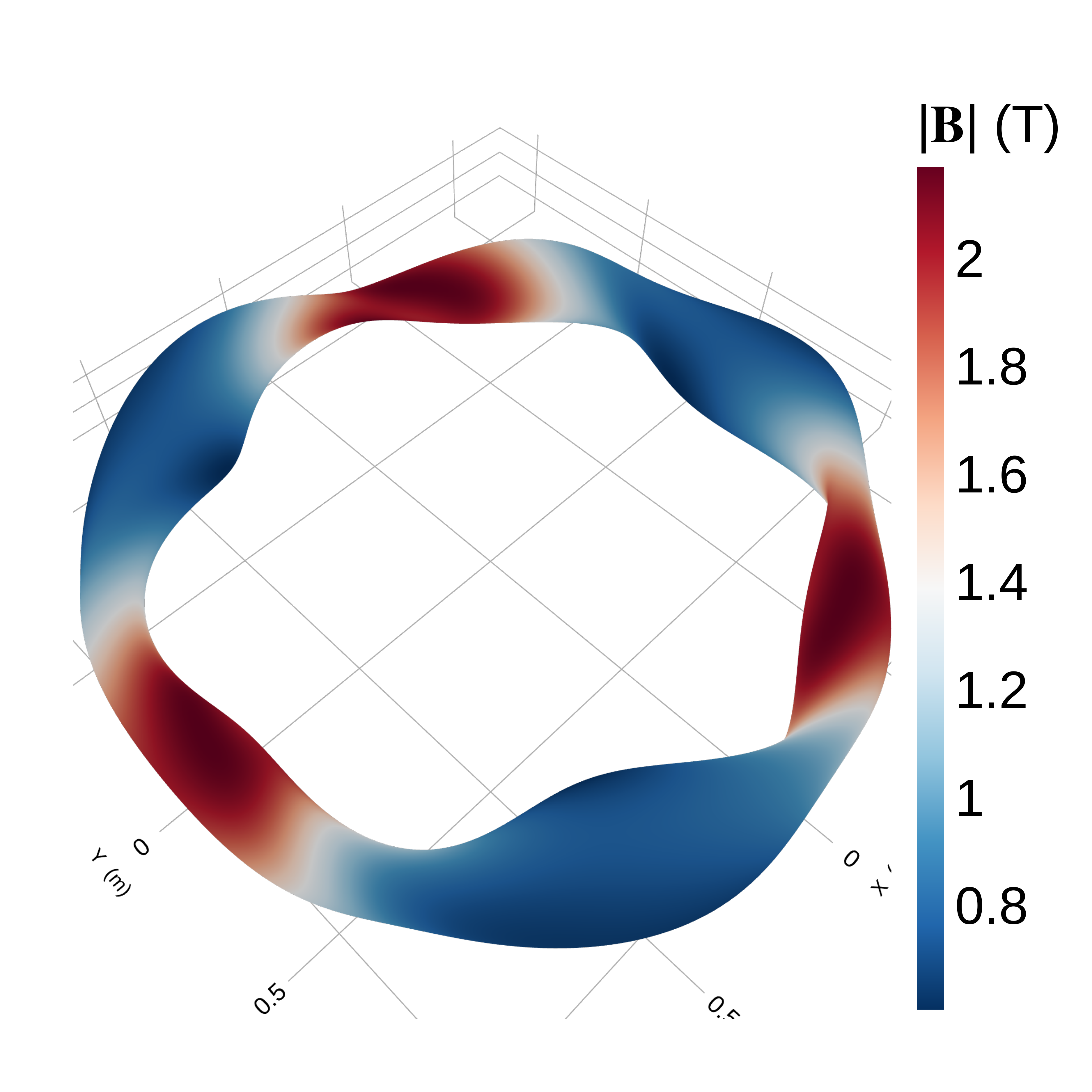}
        \caption{Initial $B$ on boundary}
    \end{subfigure}
    \qquad  \qquad
    \begin{subfigure}[b]{0.35\textwidth}
        \centering
        \includegraphics[width=\textwidth, trim={0mm 0mm 0 4mm}, clip]{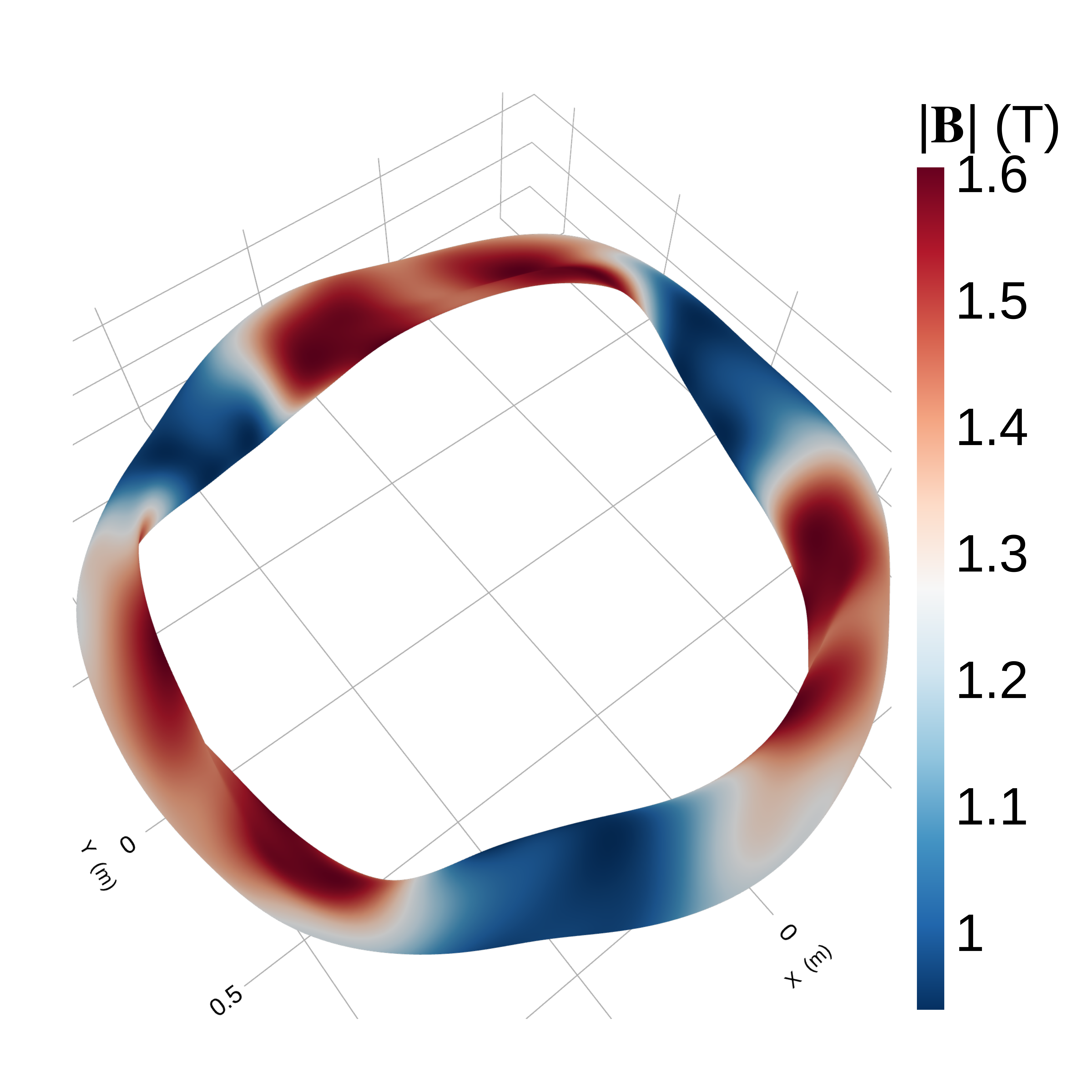}
        \caption{Optimized $B$ on boundary}
    \end{subfigure}  
    \caption{Outputs from the OP stability optimization in~\texttt{DESC}. Figures $(a)$ and $(b)$ show the magnetic field strength $B$ on the boundary surface, respectively; $(c)$ is the optimized boundary cross-section at different toroidal angles for a single field period. Figures $(d), (e), (f)$ compare various instabilities between the initial and optimized equilibria and figures $(g), (h)$ illustrate the magnetic field strength on the plasma boundary.}
\label{fig:OP-outputs1}
\end{figure}

Our objective is to find a poloidally omnigenous (OP) equilibrium with improved stability using~\texttt{DESC}. Implementing the heuristic analytical model provided by Goodman et al.~\cite{goodman2023constructing} and using the~\texttt{DESC} omnigenity module, we have generated a database of $\sim 10^6$ omnigenous equilibria~\cite{Gaur-DPP23}. From this database, we select a finite-$\beta$ equilibrium with $n_\mathrm{FP} =3$, a low omnigenity error, and negative magnetic shear. The pressure profile is $p = p_0 (1-\rho^4)$ with $p_0 = 0.026$ MPa and the rotational transform profile is $\iota = \iota_0 + \iota_2 \rho^2 + \iota_4 \rho^4$ with $\iota_0 = 0.358, \iota_2 = 0.18, \iota_4 = -0.08$. We then optimize this equilibrium for improved stability using~\texttt{DESC} and present the results in figures~\ref{fig:OP-inputs} and~\ref{fig:OP-outputs1}, along with important figures of merit in table~\ref{tab:OP-quantities}. The definitions of various figures of merit are provided in~\ref{app:objectives-FoMs-defn}.
\begin{table}
\caption{Figures of merit of the initial and optimized OP equilibria}
\lineup
\begin{tabular*}{\textwidth}{@{}l*{15}{@{\extracolsep{0pt plus12pt}}l}}
\br                              
Equilibrium & Aspect ratio & \0$\langle \beta \rangle$ &\m OP error & \0$\Psi_{\mathrm{b}} (\mathrm{T-m^2})$ & \0 $ I_{\mathrm{b}} (\mathrm{kA})$ & \0 $D_{\mathrm{mix}}(\rho=0.9)$ \cr 
\mr
initial   &\0\0 9.33   & 0.032 & \0\0\0 $2.02$ & \0\0 $0.037$ & \0\0 -5.3 & \0\0 97.07\cr 
optimized &\0\0 9.82   & 0.029 & \0\0\0 $0.20$ & \0\0 $0.037$ & \0\0 31.9 & \0\0 0.72\cr 
\br
\end{tabular*}
\label{tab:OP-quantities}
\end{table}

Using DESC, we are able to successfully stabilize the initially unstable equilibrium while maintaining reasonable poloidal omnigenity. Due to the curvature objective, the boundary has a minimal ``bean''-like shaping along with a small mirror ratio $(B_{\mathrm{max}} - B_{\mathrm{min}})/(B_{\mathrm{max}} + B_{\mathrm{min}})$, which would simplify coil design. As we stabilize the ideal ballooning mode, we also see a significant reduction in the maximum KBM growth rate and an even greater reduction in the diffusion coefficient $D_{\mathrm{mix}}$ in table~\ref{tab:OP-quantities}\footnote{The significantly large value of the mixing-length diffusion estimate is due to the fact that for ideal-ballooning unstable modes, KBM growth rate peaks around $|k_{\perp}| = 0$ .This is one of the reasons why nonlinear flux tube simulations of KBMs can sometimes fail to saturate, as the most unstable mode depends starts to depend strongly on the domain size.}. Moreover, since this equilibrium has a negative shear, we hypothesize that it will have a reduced turbulent transport compared to that of an equilibrium with a positive shear. Since the rotational transform does not cross any low-order rational values, this equilibrium will not form any low-order magnetic islands. The negative shear will also stabilize any bootstrap-current driven high-order magnetic islands~\cite{hegna1994stability}.

Using~\texttt{DESC} have obtained a lower aspect ratio and better omnigenity than the method used by Isaev \etal~\cite{isaev1997plasma}. In the next section, we generate a toroidally omnigenous equilibrium using the same process. 

\subsection{Toroidal omnigenity (OT)}
\begin{figure}[!h]
    \centering
    \begin{subfigure}[b]{0.32\textwidth}
    \centering
        \includegraphics[width=\textwidth]{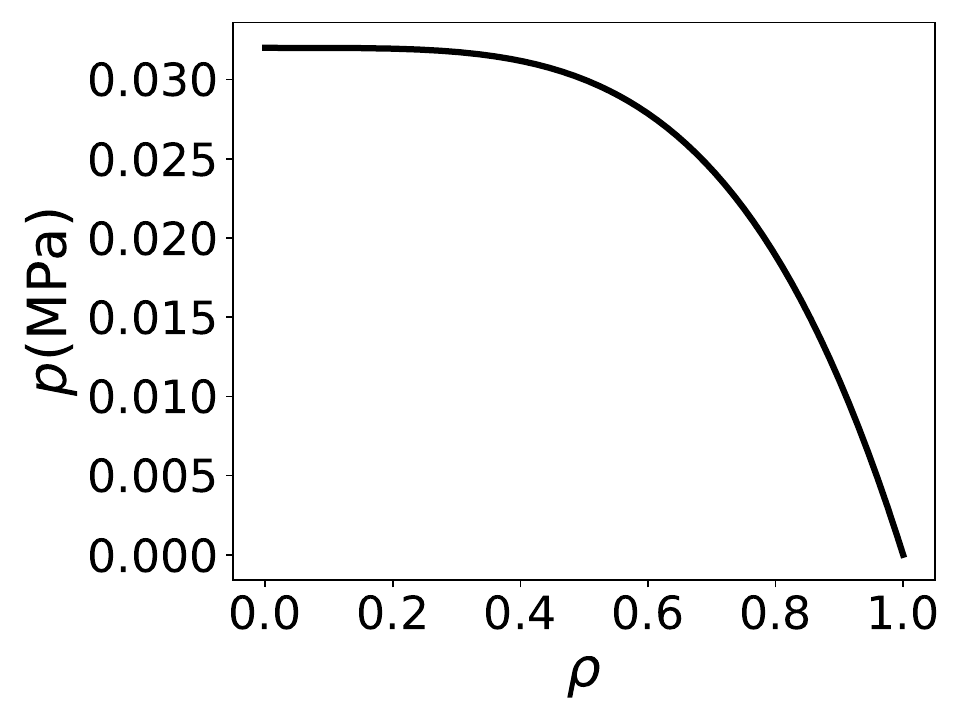}
        \caption{Pressure}
    \end{subfigure}
    \begin{subfigure}[b]{0.32\textwidth}
        \centering
        \includegraphics[width=\textwidth]{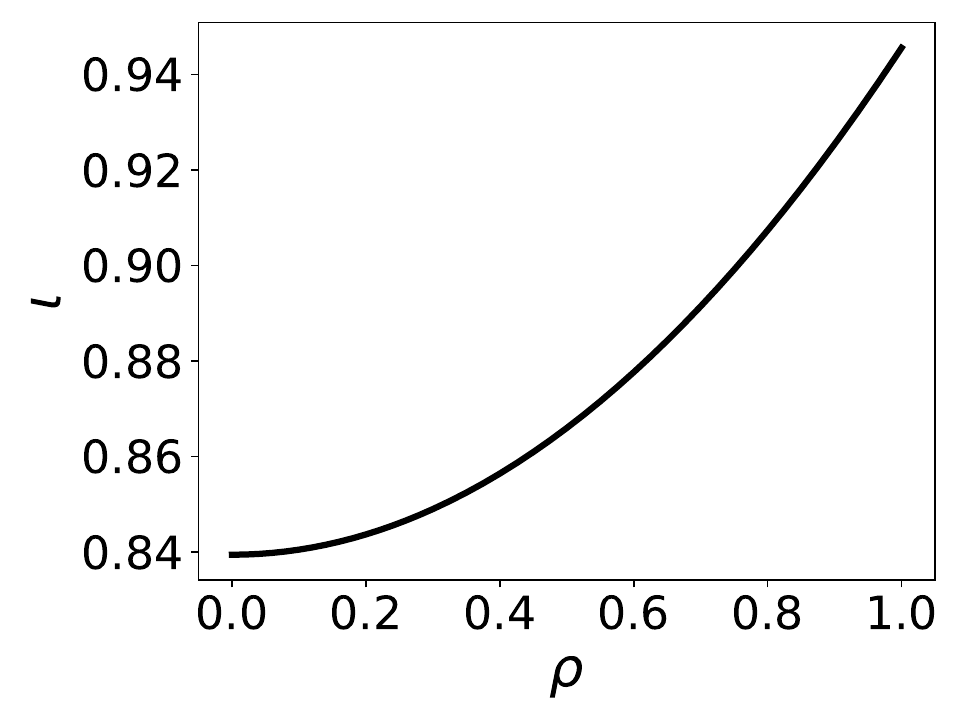}
        \caption{Rotational transform}
    \end{subfigure}
    \begin{subfigure}[b]{0.30\textwidth}
        \centering
        \includegraphics[width=\textwidth, trim={0mm 4mm 0 8mm}, clip]{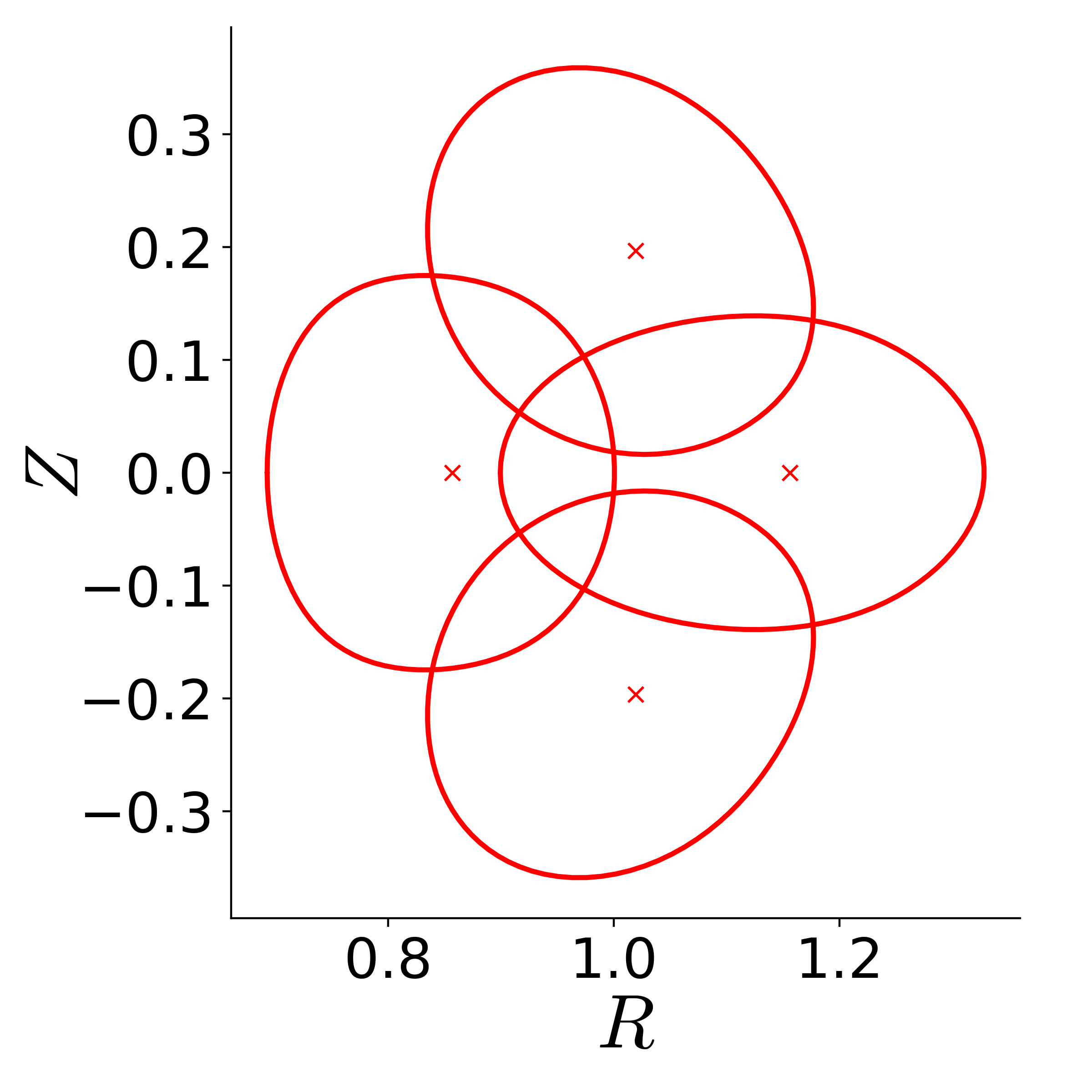}
        \caption{Initial boundary}
    \end{subfigure}
    \caption{Inputs to the optimization module in~\texttt{DESC}. Figures $(a)$ and $(b)$ show the pressure and rotational transform profiles and figure $(c)$ is the initial boundary cross-section at different toroidal angles for a single field period.}
\label{fig:OT-inputs}
\end{figure}

In this section, we will find a toroidally omnigenous (OT) equilibrium with improved stability using DESC. To obtain toroidal omnigenity, we start with an equilibrium that is ballooning-stable and close to quasiaxisymmetry from the omnigenity database. The OT database is generated by first calculating quasiaxisymmetric equilibria from pyQSC~\cite{landreman2019constructing} and then optimizing them for toroidal omnigenity using DESC. The initial equilibrium has a finite-$\beta$ with $n_\mathrm{FP} =1$ and a negative magnetic shear. The pressure profile has the form $p = p_0 (1-\rho^4)$ with $p_0 = 0.031$ MPa and the rotational transform profile is $\iota = \iota_0 + \iota_2 \rho^2$ with $\iota_0 = 0.84, \iota_2 = 0.106$. The input data are presented in figure ~\ref{fig:OT-inputs} and important properties of this equilibrium are given in table~\ref{tab:OT-quantities}.

\begin{figure}
    \centering
    \begin{subfigure}[b]{0.325\textwidth}
    \centering
        \includegraphics[width=\textwidth, trim={2mm 2mm 4mm 6mm}, clip]{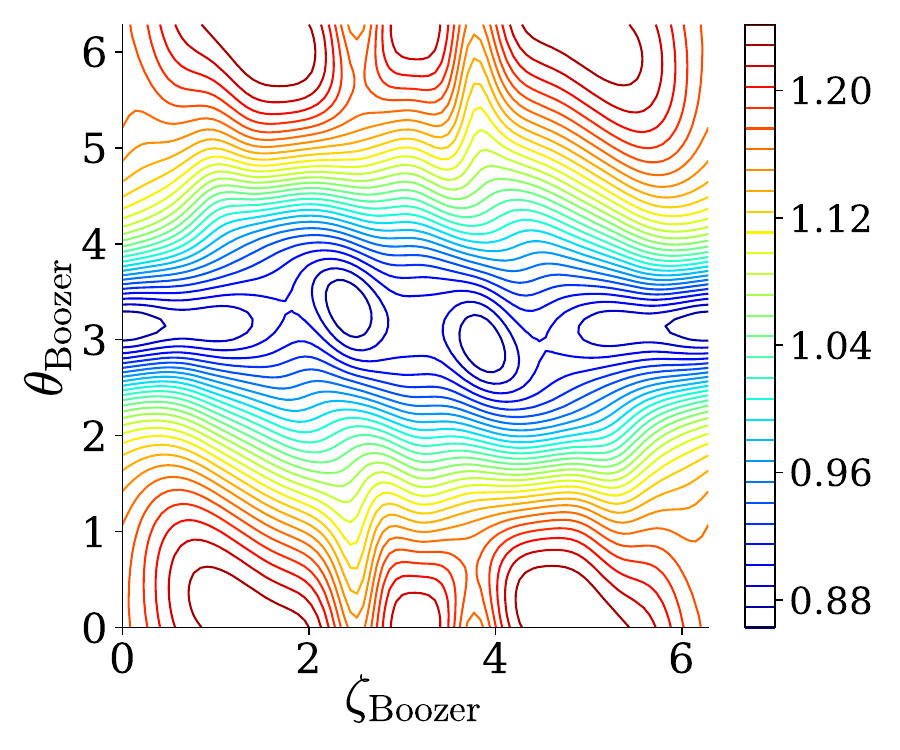}
        \caption{Initial B $(\rho=1)$}
    \end{subfigure}
    \quad 
    \begin{subfigure}[b]{0.325\textwidth}
        \centering
        \includegraphics[width=\textwidth, trim={2mm 2mm 4mm 6mm}, clip]{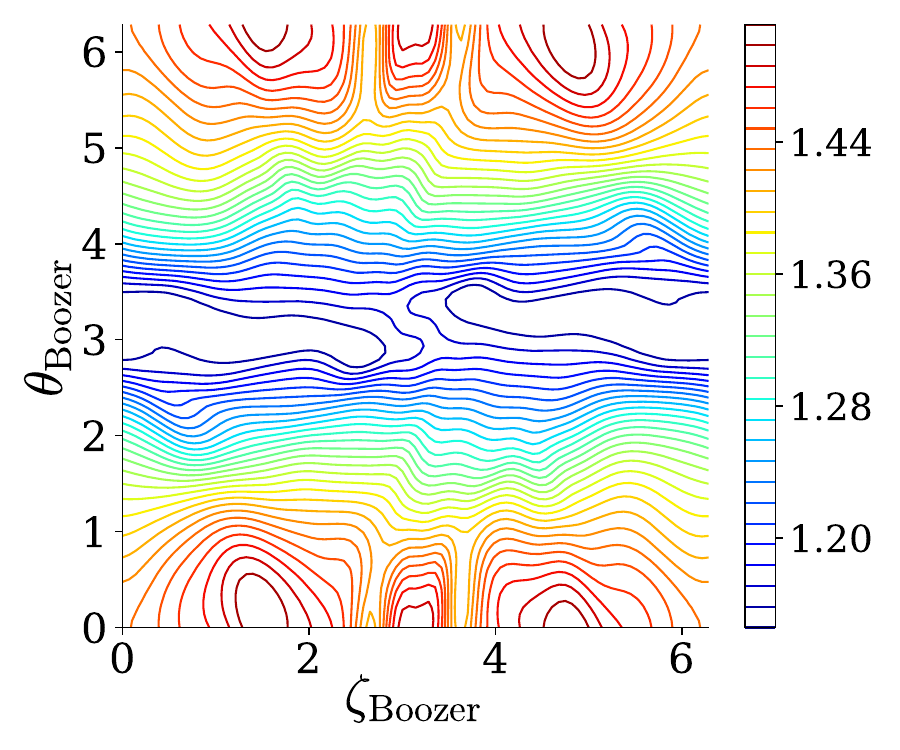}
        \caption{Optimized B $(\rho=1)$}
    \end{subfigure}
    \begin{subfigure}[b]{0.3\textwidth}
        \centering
        \includegraphics[width=\textwidth, trim={0mm 4mm 0 9mm}, clip]{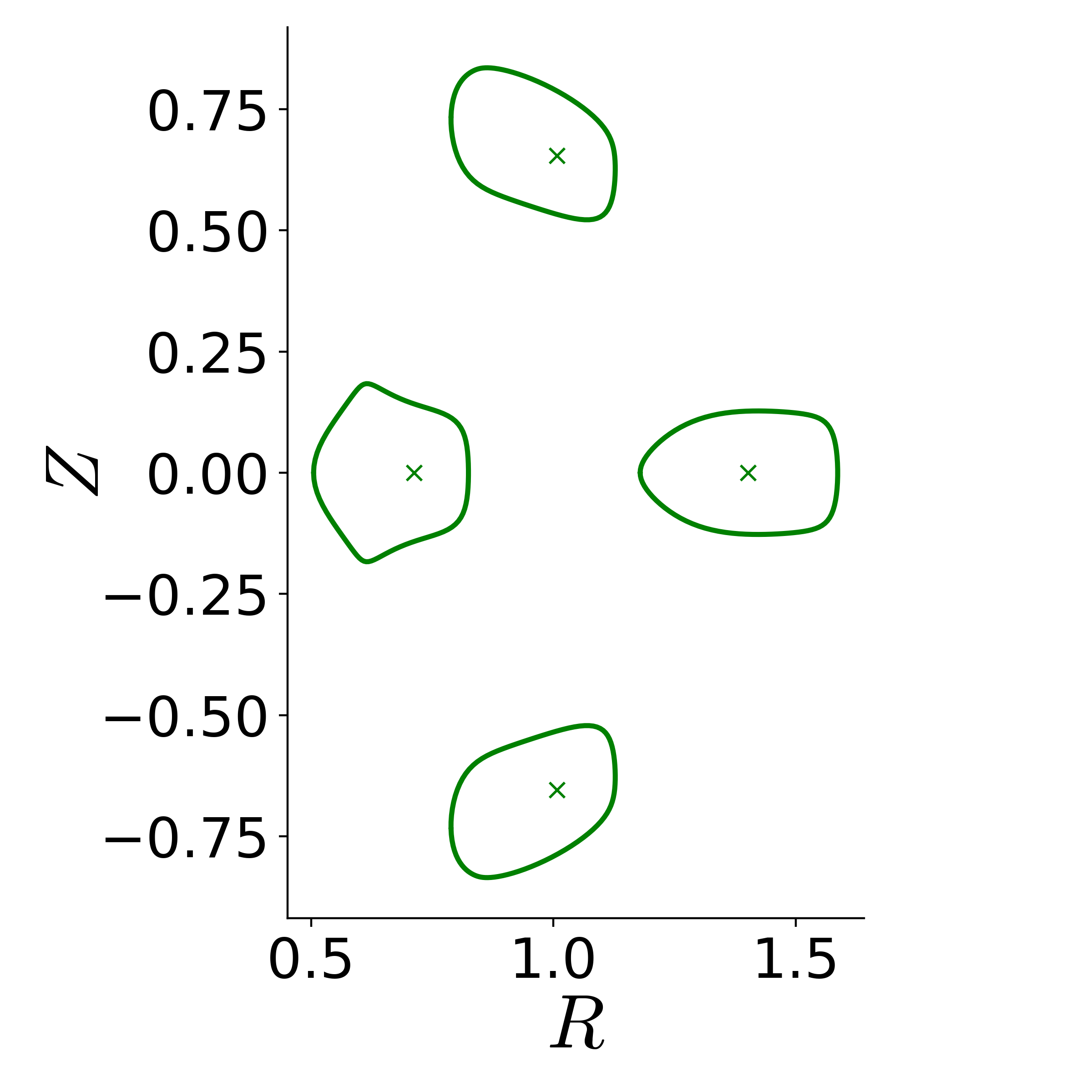}
        \caption{Optimized boundary}
    \end{subfigure}\\

    \clearpage
    \hspace*{-6mm}
    \begin{subfigure}[b]{0.32\textwidth}
    \centering
        \includegraphics[width=\textwidth, trim={0mm 0mm 0 0mm}, clip]{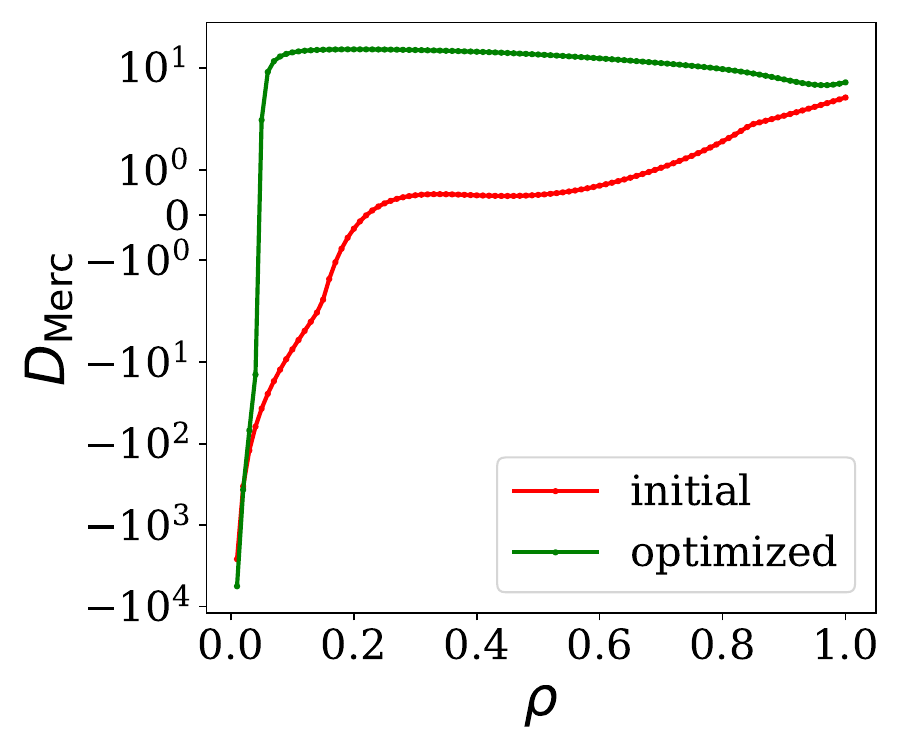}
        \caption{Mercier stability}
    \end{subfigure}
    \quad
    \begin{subfigure}[b]{0.32\textwidth}
        \centering
        \includegraphics[width=\textwidth]{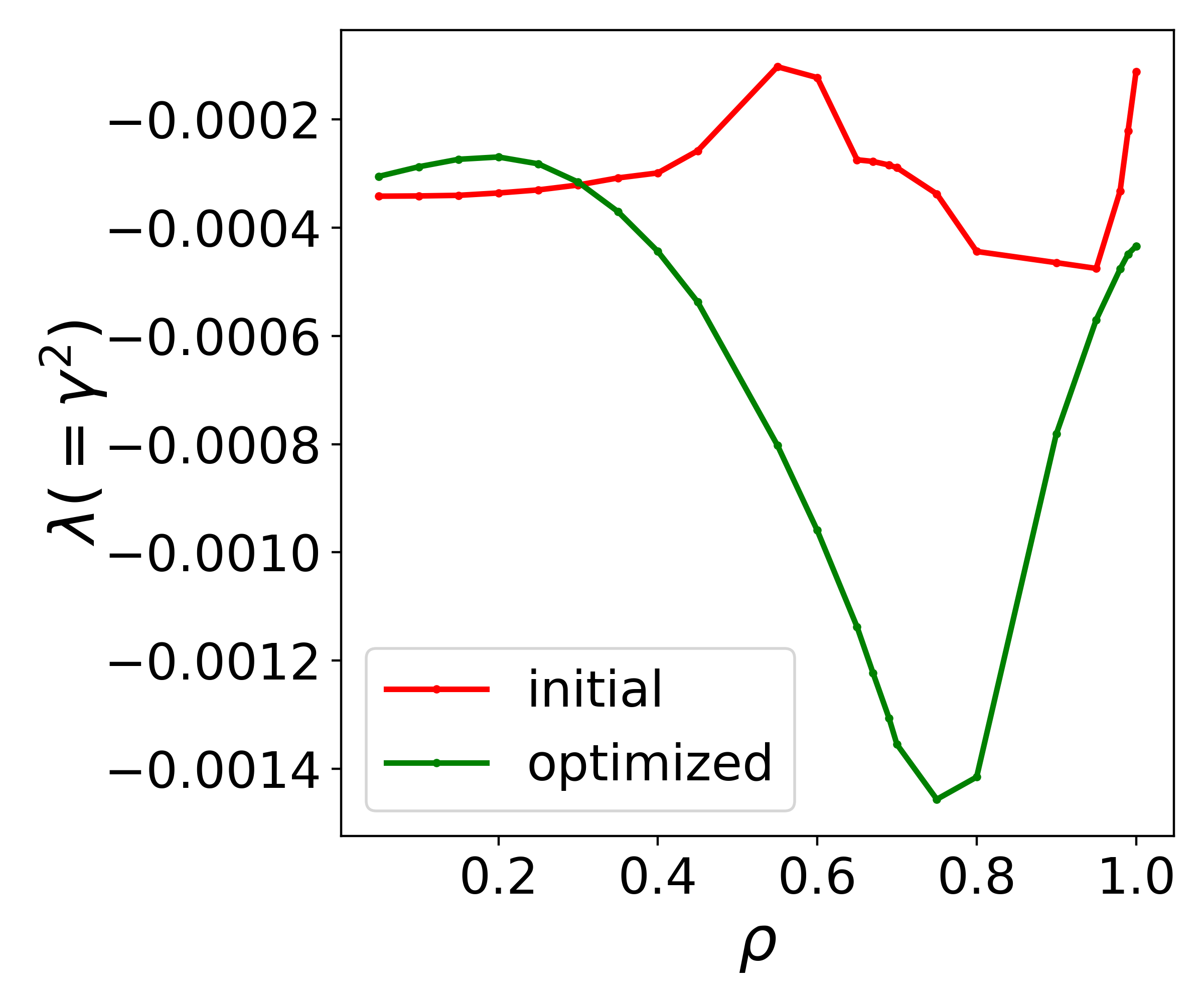}
        \caption{Ballooning growth rate}
    \end{subfigure}
    \begin{subfigure}[b]{0.32\textwidth}
        \centering
        \includegraphics[width=\textwidth, trim={0mm 0mm 0 0mm}, clip]{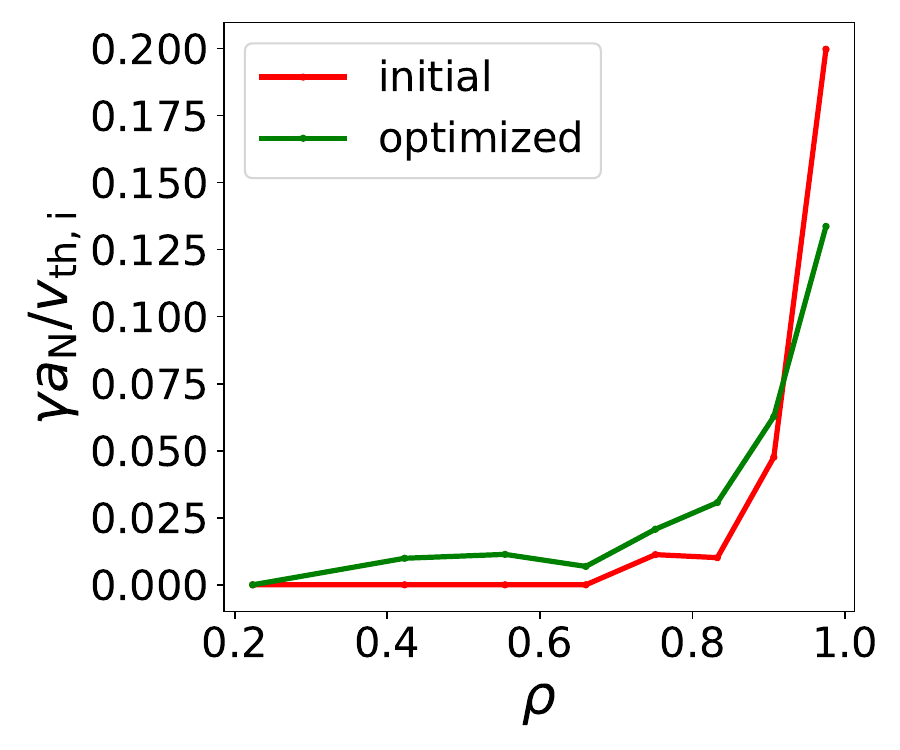}
        \caption{KBM growth rate}
    \end{subfigure}\\

    \clearpage
    \begin{subfigure}[b]{0.35\textwidth}
    \centering
        \includegraphics[width=\textwidth]{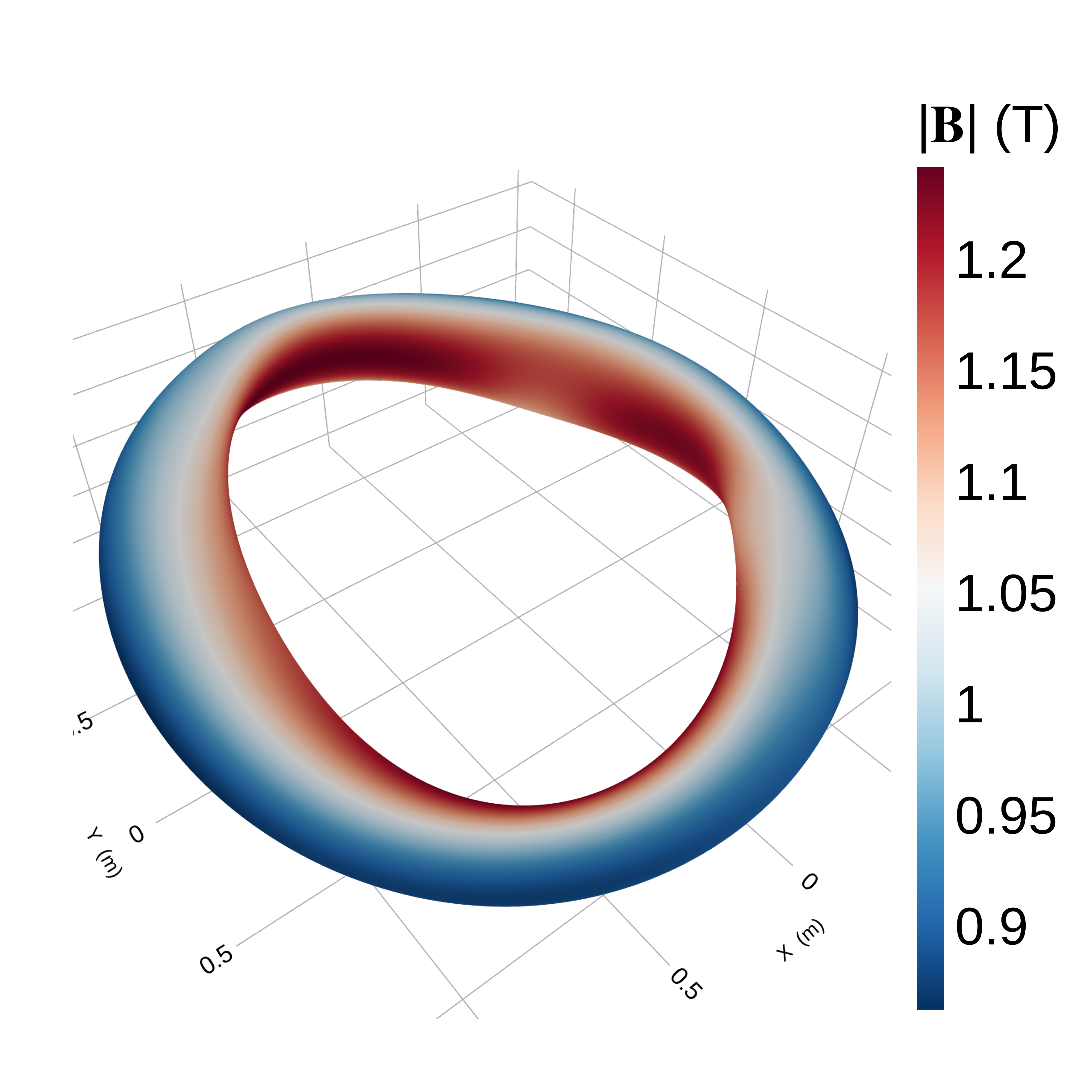}
        \caption{Initial $B$ on boundary}
    \end{subfigure}
    \qquad  \qquad
    \begin{subfigure}[b]{0.35\textwidth}
        \centering
        \includegraphics[width=\textwidth]{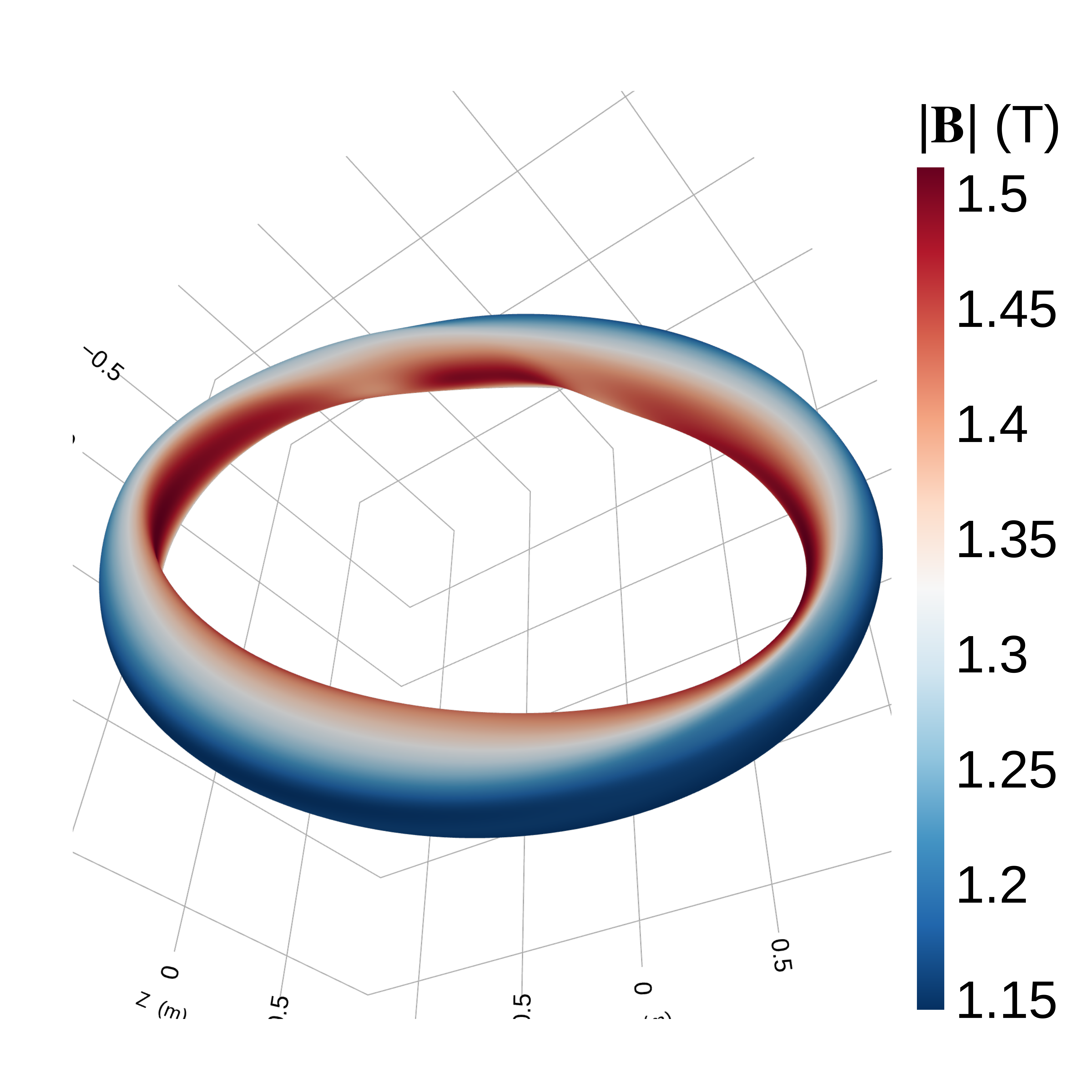}
        \caption{Optimized $B$ on boundary}
    \end{subfigure}
    \caption{Outputs from the OT stability optimization in~\texttt{DESC}. Figures $(a)$ and $(b)$ show the magnetic field strength $B$ on the boundary surface, respectively; $(c)$ is the optimized boundary cross-section at different toroidal angles for a single field period, figures $(d), (e), (f)$ show comparison of various instabilities between the initial and optimized equilibria and figures $(g), (h)$ illustrate the magnetic field strength on the plasma boundary.}
\label{fig:OT-outputs1}
\end{figure}

\begin{table}
\caption{Figures of merit of the initial and optimized OT equilibria}
\lineup
\begin{tabular*}{\textwidth}{@{}l*{15}{@{\extracolsep{0pt plus12pt}}l}}
\br                              
Equilibrium & Aspect ratio & \0$\langle \beta \rangle$ &\m OT error & \0$\Psi_{\mathrm{b}} (\mathrm{T-m^2})$ & \0 $I_{\mathrm{b}} (\mathrm{kA})$  & \0 $D_{\mathrm{mix}}(\rho=0.9)$ \cr 
\mr
initial   &\0\0 5.92   & 0.054 & \0\0\0 $0.22$ & \0\0 $0.087$ & \0\0 156 & \0\0  0.51\cr 
optimized &\0\0 6.25   & 0.033 & \0\0\0 $0.04$ & \0\0 $0.087$ & \0\0 151 & \0\0 0.4\cr 
\br
\end{tabular*}
\label{tab:OT-quantities}
\end{table}
We have successfully obtained a toroidal omnigenous equilibrium with improved stability. The equilibrium becomes Mercier stable for most of the volume without degrading the ballooning stability and omnigenity. We also see a small reduction in the KBM growth rate and the mixing-length diffusion coefficient, especially near the plasma edge. Due to the curvature objective, we can again avoid ``bean''-like shapes in the inboard side of the optimized stellarator. It is important to point out that, much like quasiaxisymmetric equilibria, our results indicate that obtaining low-aspect-ratio OT equilibria is simpler than achieving low-aspect-ratio OP or OH equilibria. Note that, even though the magnetic axis of the optimized configuration appears to have a high torsion from figure~\ref{fig:OT-outputs1}$\mathrm{(c)}$, the 3D figure~\ref{fig:OT-outputs1}$\mathrm{(h)}$ reveals that the stellarator axis remains nearly planar. 

\subsection{Helical omnigenity (OH)}
\begin{figure}[!h]
    \centering
    \begin{subfigure}[b]{0.32\textwidth}
    \centering
        \includegraphics[width=\textwidth]{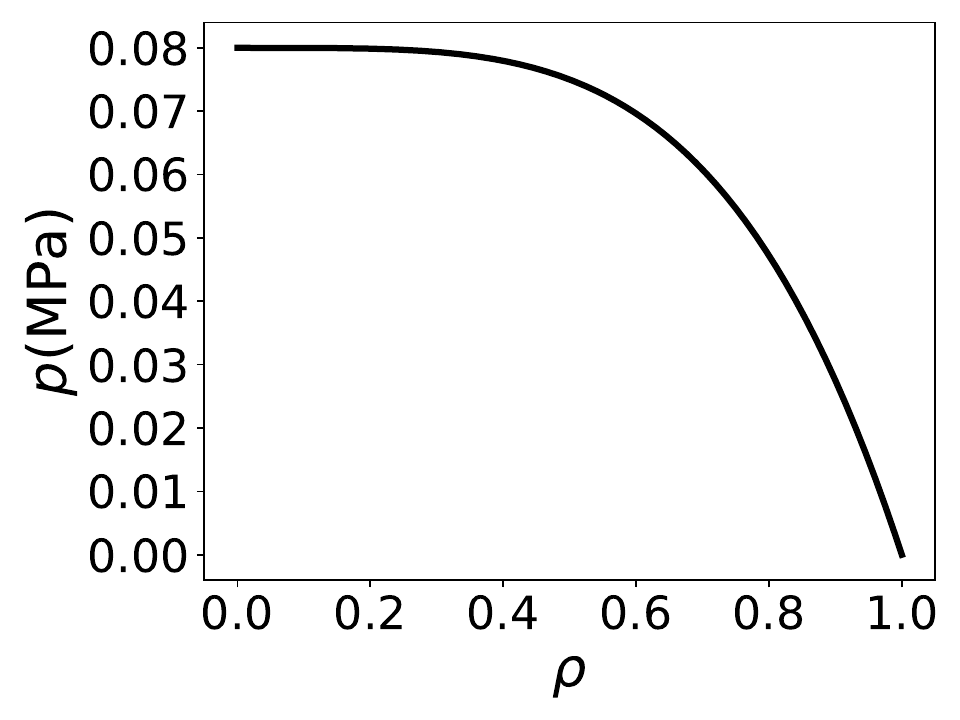}
        \caption{Pressure}
    \end{subfigure}
    \begin{subfigure}[b]{0.32\textwidth}
        \centering
        \includegraphics[width=\textwidth]{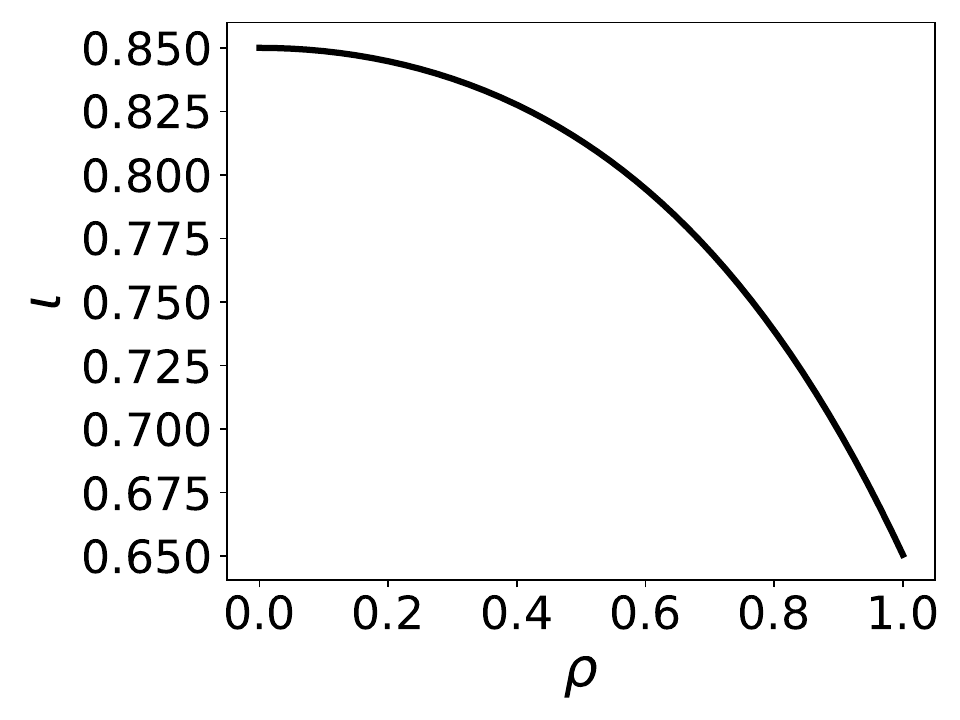}
        \caption{Rotational transform}
    \end{subfigure}
    \begin{subfigure}[b]{0.30\textwidth}
        \centering
        \includegraphics[width=\textwidth, trim={0mm 4mm 0 8mm}, clip]{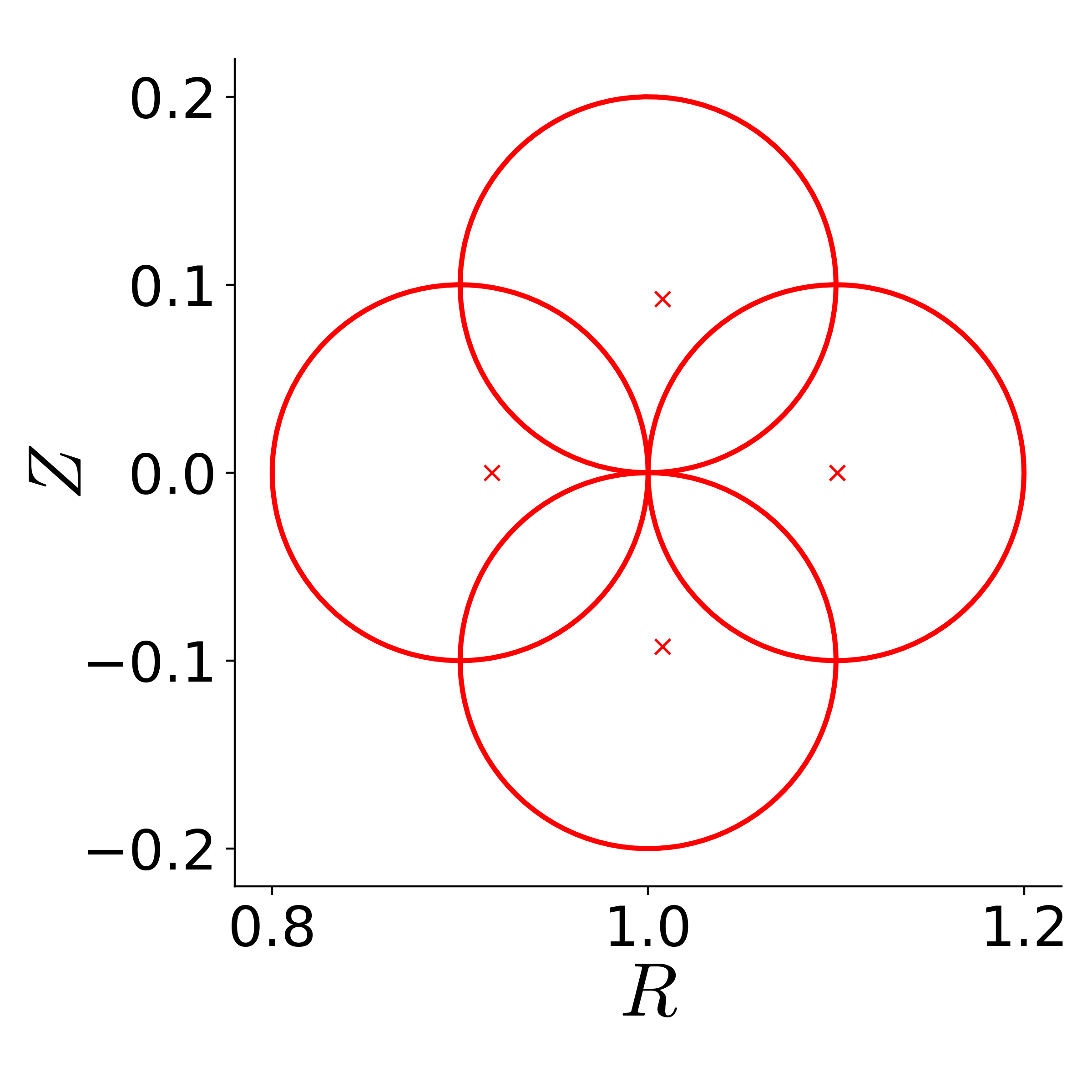}
        \caption{Initial boundary}
    \end{subfigure}
    \caption{Inputs to the optimization module in~\texttt{DESC} for the OH case. Figures $(a)$ and $(b)$ show the pressure and rotational transform profiles and figure $(c)$ is the boundary cross-section at different toroidal angles for a single field period.}
\label{fig:OH-inputs}
\end{figure}

In this section, we will generate a helically omnigenous (OH) equilibrium with improved stability using DESC. To obtain helical omnigenity, we start with a finite-$\beta$ circular vacuum equilibrium with $n_{\mathrm{FP}} = 5$ and a finite magnetic axis torsion using the following boundary parametrization
\numparts
\begin{eqnarray}
    R_{\mathrm{b}} = 1 + 0.1 \cos(\theta) + 0.1 \cos(5 \phi) \\ 
    Z_{\mathrm{b}} =  0.1 \sin(\theta) + 0.1 \sin(5 \phi).
\end{eqnarray}
\endnumparts
The pressure profile has the form $p = p_0 (1-\rho^4)$ with $p_0 = 0.08$ MPa and the rotational transform profile is $\iota = \iota_0 + \iota_2 \rho^2 + \iota_4 \rho^4$ with $\iota_0 = 0.85, \iota_2 = -0.128, \iota_4 = -0.071$. The shape of this equilibrium and input profiles are provided in figure~\ref{fig:OH-inputs} and its characteristic properties are given in table~\ref{tab:OH-quantities}.
\begin{figure}
    \centering
    \begin{subfigure}[b]{0.325\textwidth}
    \centering
        \includegraphics[width=\textwidth, trim={2mm 2mm 4mm 6mm}, clip]{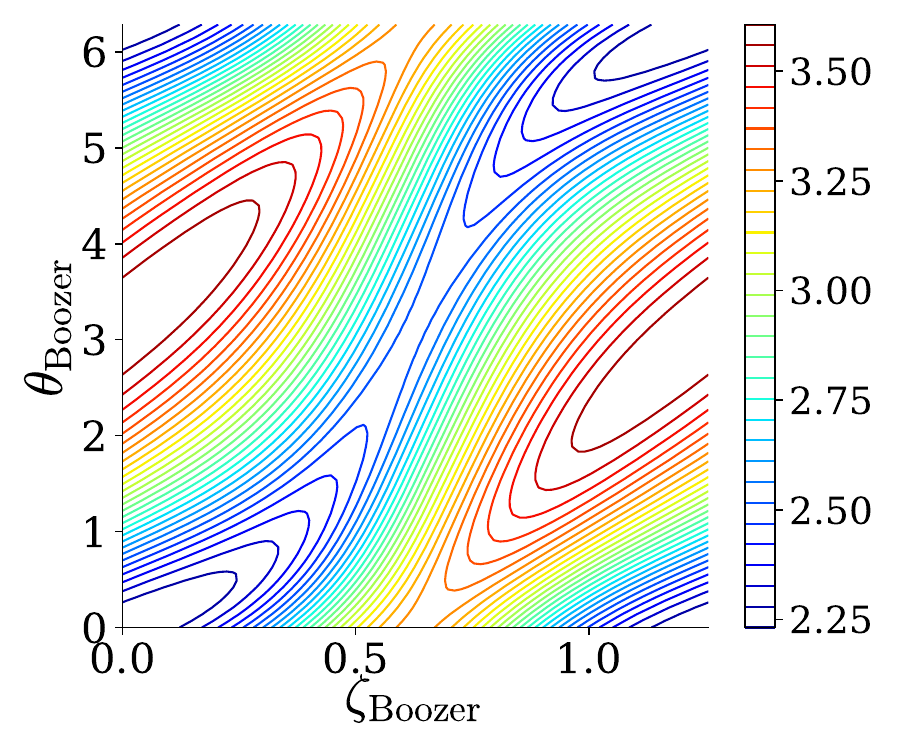}
        \caption{Initial B $(\rho=1)$}
    \end{subfigure}
    \quad 
    \begin{subfigure}[b]{0.325\textwidth}
        \centering
        \includegraphics[width=\textwidth, trim={2mm 2mm 4mm 6mm}, clip]{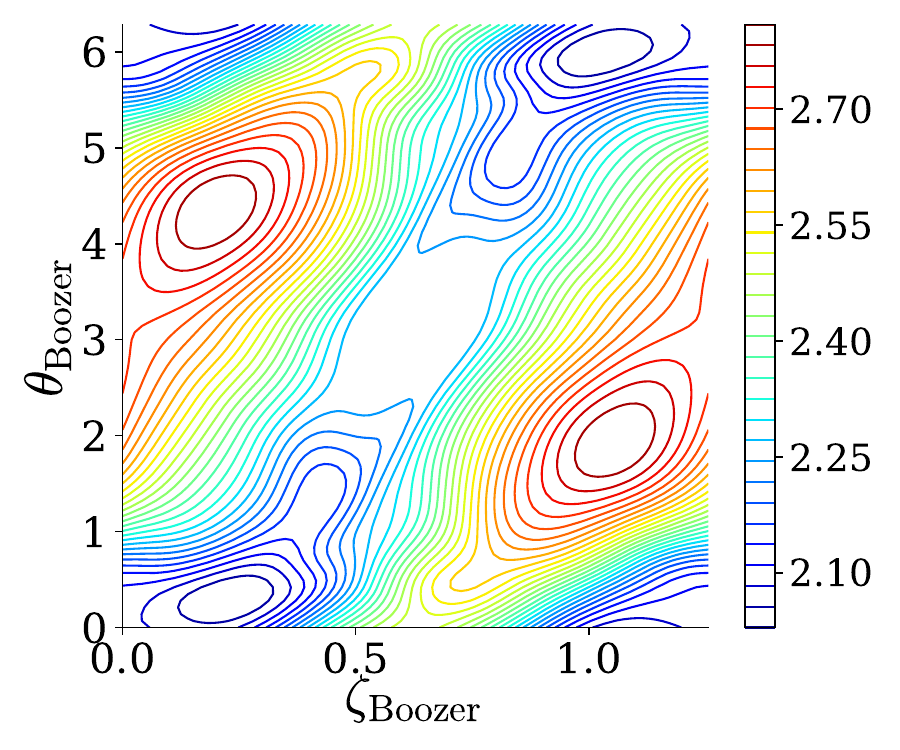}
        \caption{Optimized B $(\rho=1)$}
    \end{subfigure}
    \begin{subfigure}[b]{0.3\textwidth}
        \centering
        \includegraphics[width=\textwidth, trim={0mm 4mm 0 7mm}, clip]{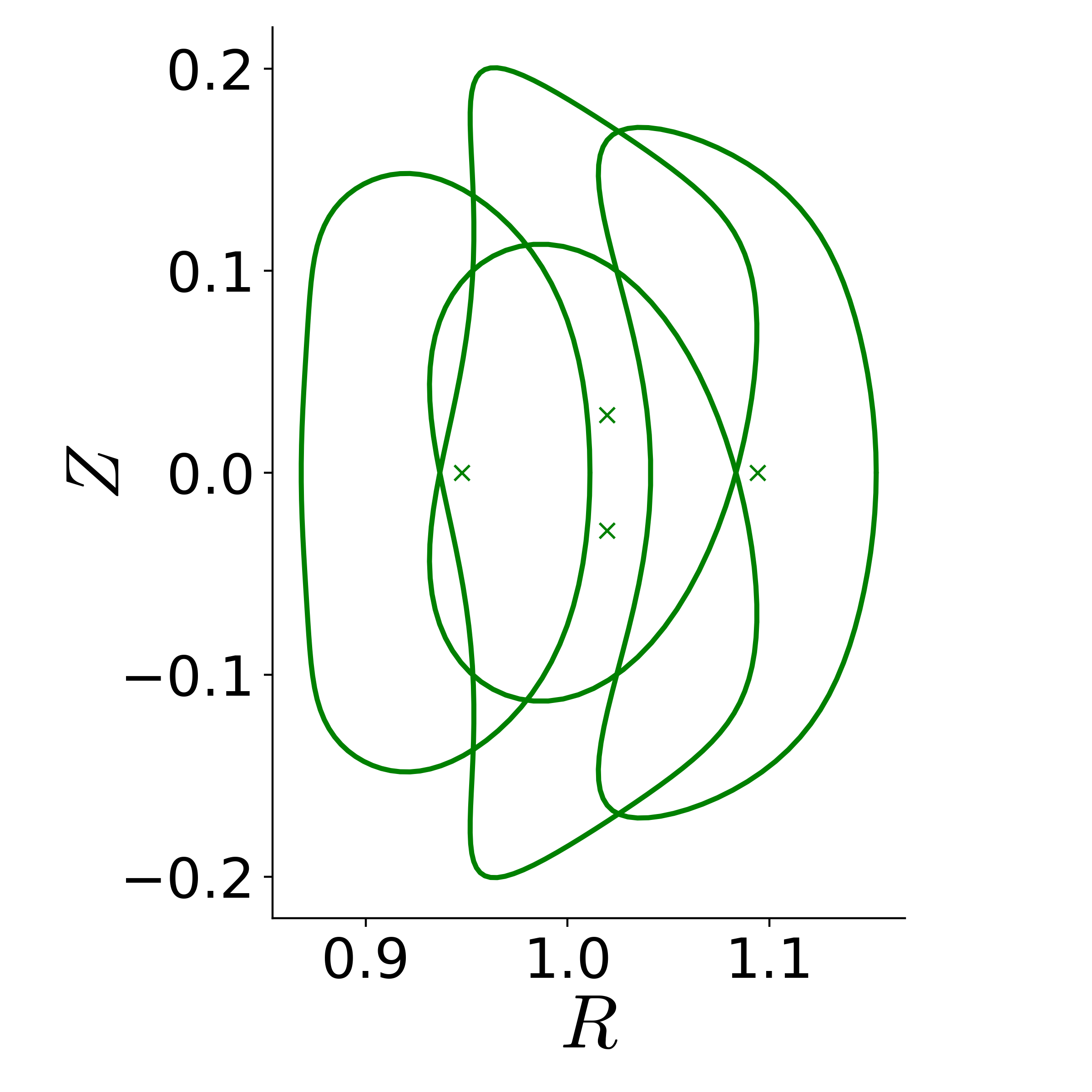}
        \caption{Optimized boundary}
    \end{subfigure}\\
    \clearpage
    \hspace*{-6mm}
    \begin{subfigure}[b]{0.32\textwidth}
    \centering
        \includegraphics[width=\textwidth, trim={0mm 0mm 0 0mm}, clip]{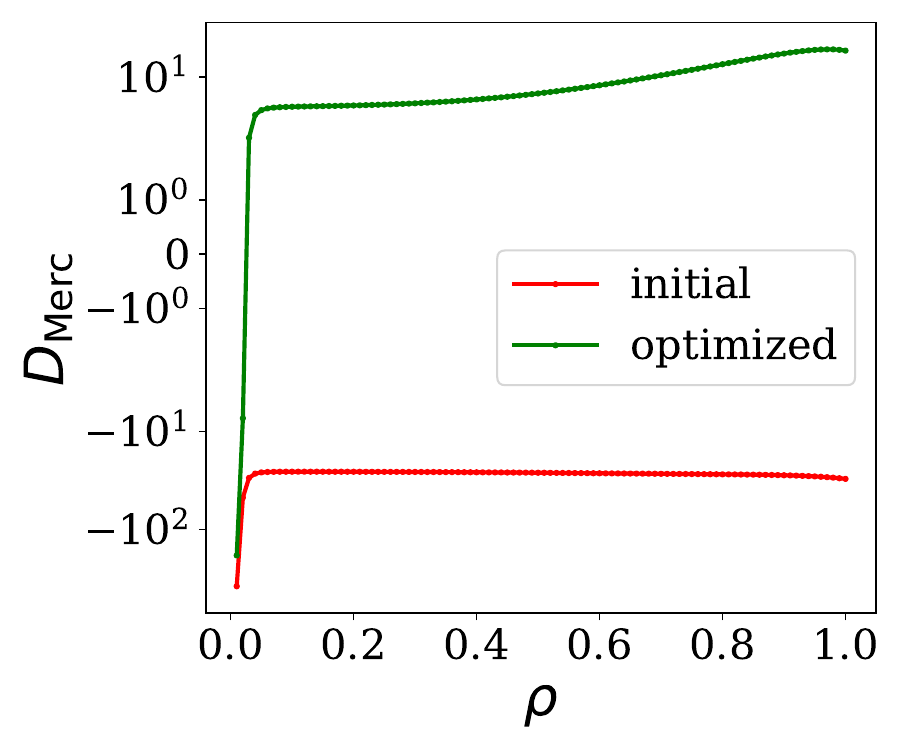}
        \caption{Mercier stability}
    \end{subfigure}
    \quad
    \begin{subfigure}[b]{0.32\textwidth}
        \centering
        \includegraphics[width=\textwidth]{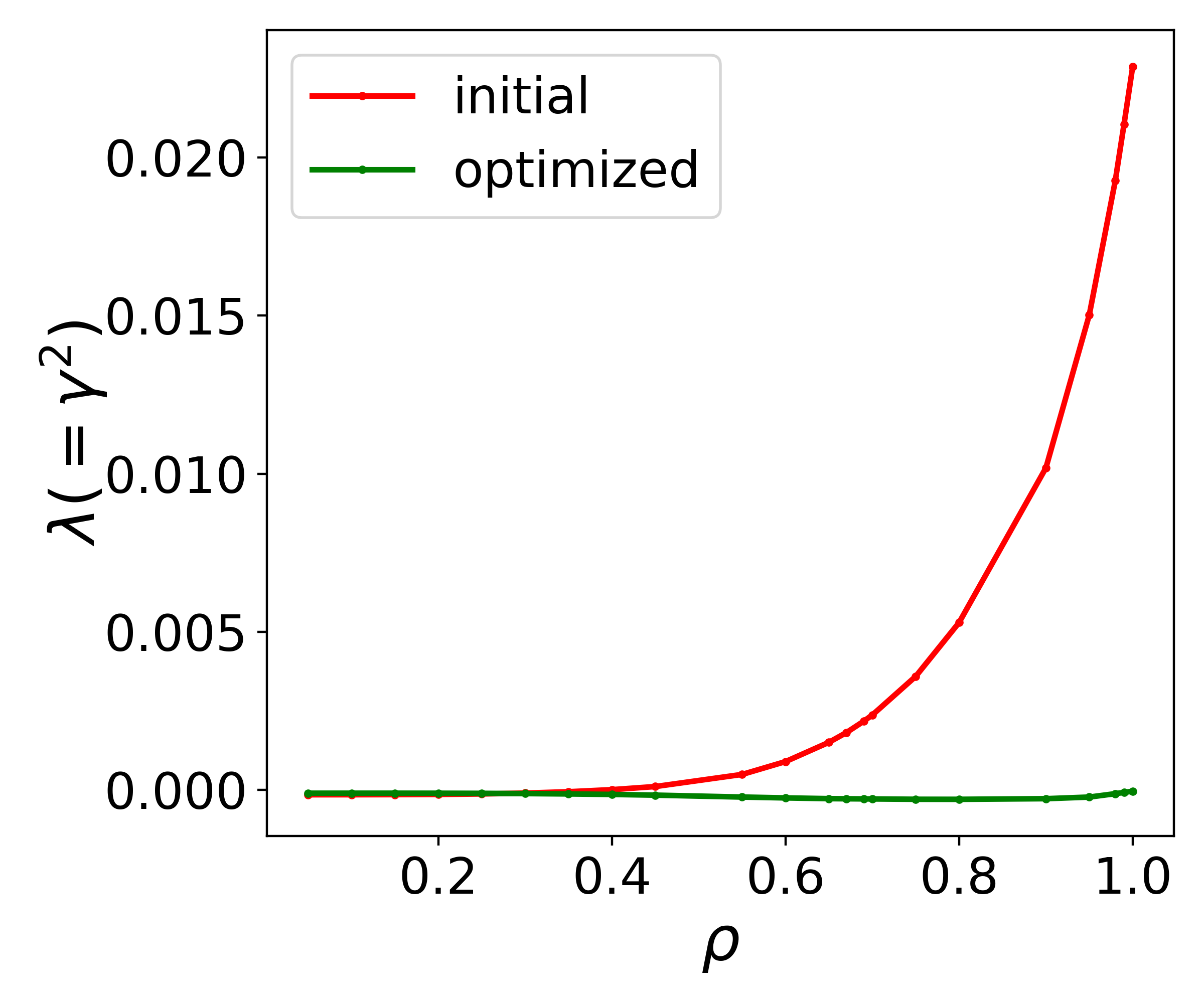}
        \caption{Ballooning stability}
    \end{subfigure}
    \begin{subfigure}[b]{0.32\textwidth}
        \centering
        \includegraphics[width=\textwidth, trim={0mm 0mm 0 0mm}, clip]{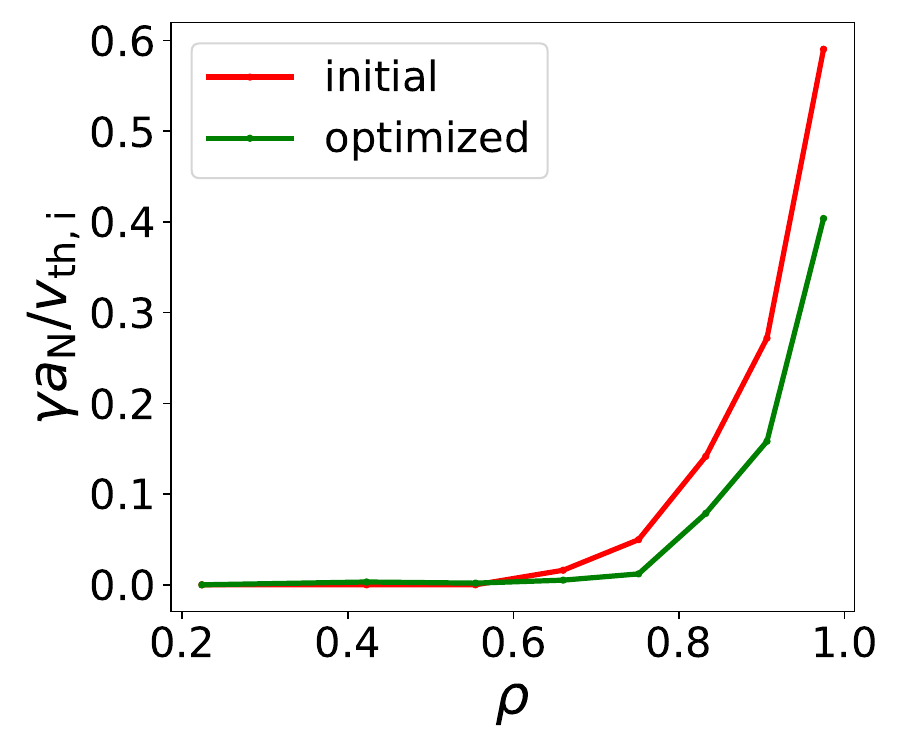}
        \caption{KBM growth rate}
    \end{subfigure}\\

    \clearpage
    \begin{subfigure}[b]{0.35\textwidth}
    \centering
        \includegraphics[width=\textwidth]{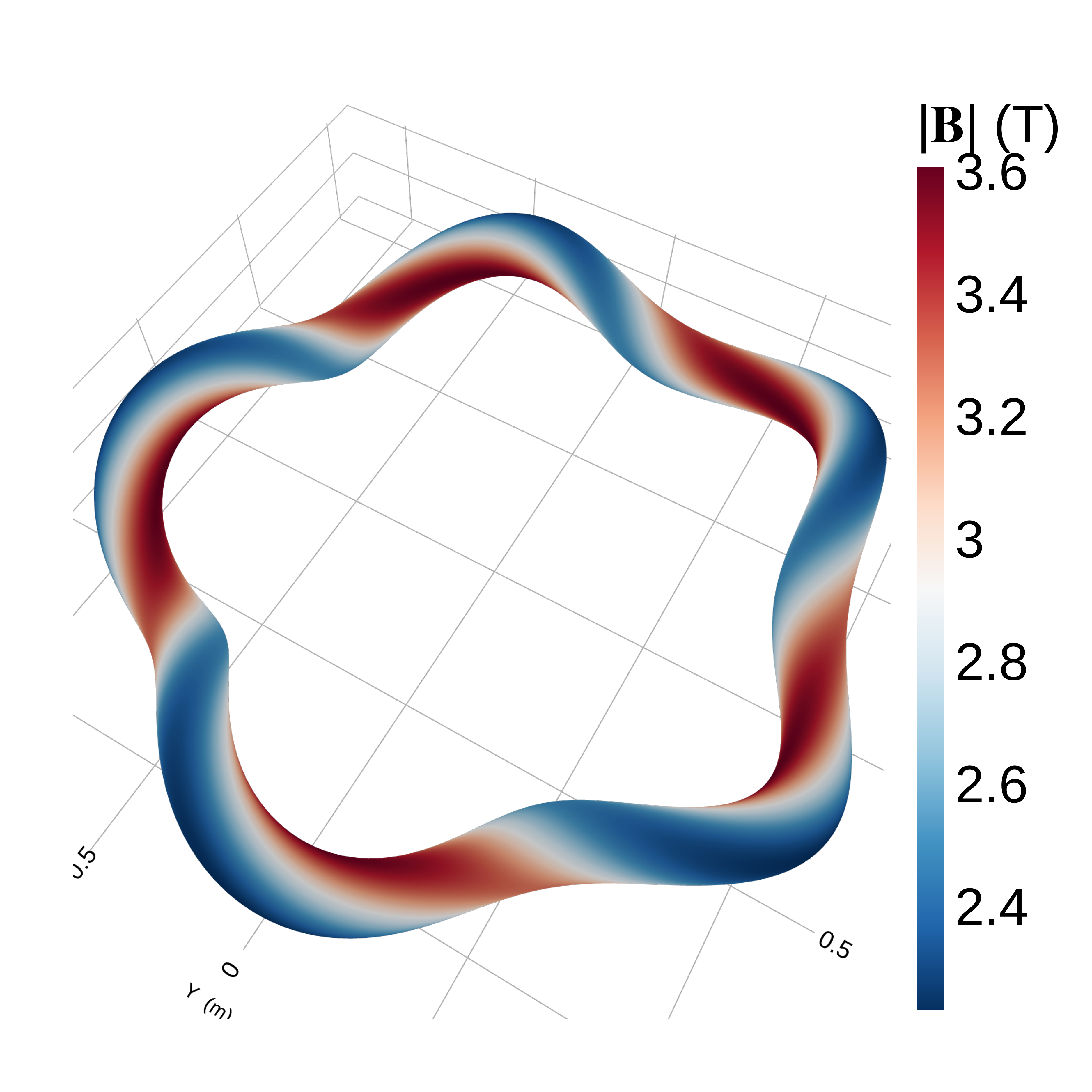}
        \caption{Initial $B$ on boundary}
    \end{subfigure}
    \qquad  \qquad
    \begin{subfigure}[b]{0.35\textwidth}
        \centering
        \includegraphics[width=\textwidth]{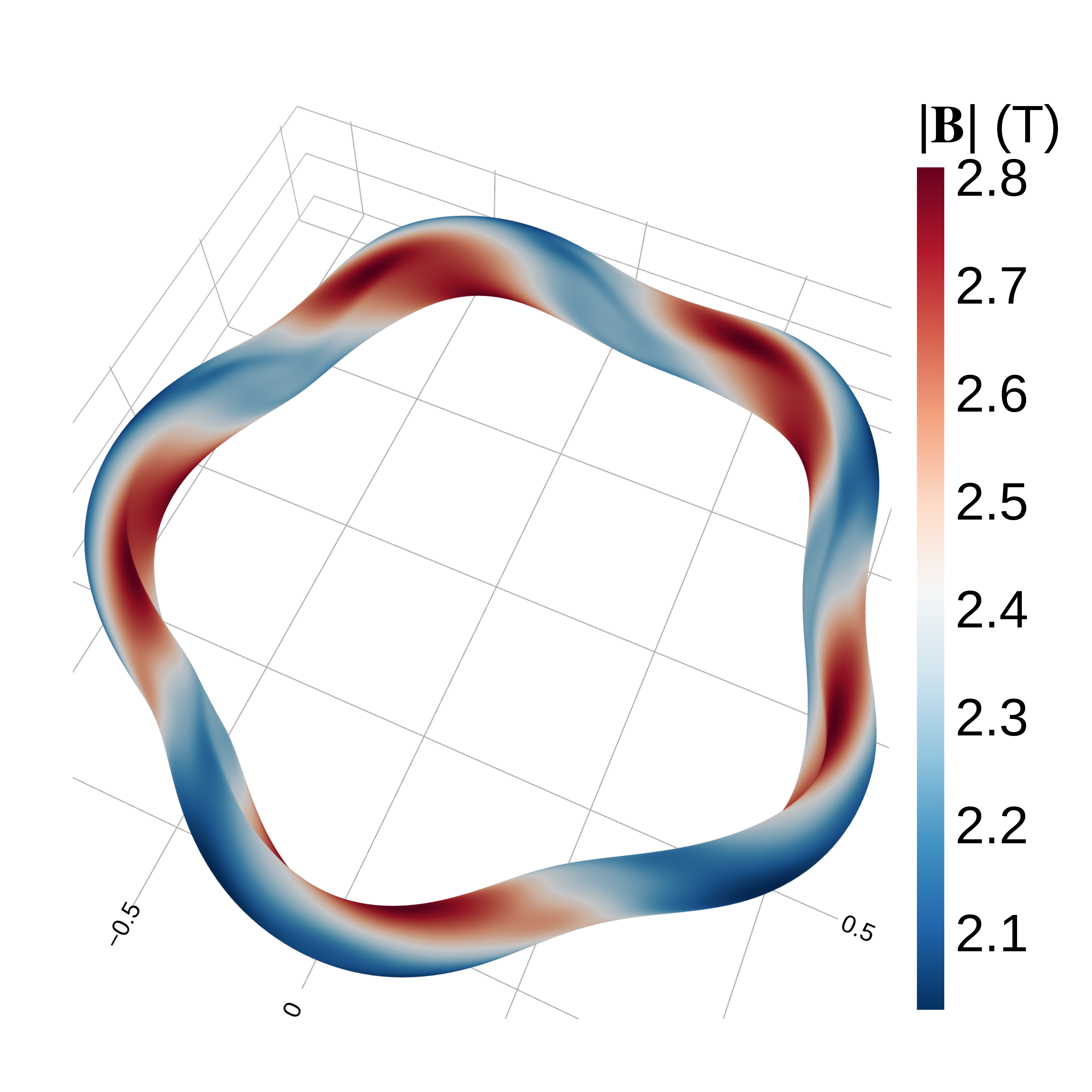}
        \caption{Optimized $B$ on boundary}
    \end{subfigure}
    \caption{Outputs from the OH stability optimization in~\texttt{DESC}. Figures $(a)$ and $(b)$ show the magnetic field strength $B$ on the boundary surface, respectively; $(c)$ is the optimized boundary cross-section at different toroidal angles for a single field period, figures $(d), (e), (f)$ show comparison of various instabilities between the initial and optimized equilibria and figures $(g), (h)$ illustrate the magnetic field strength on the plasma boundary.}
\label{fig:OH-outputs1}
\end{figure}
%
%
\begin{table}
\caption{Figures of merit of the initial and optimized OH equilibria}
\lineup
\begin{tabular*}{\textwidth}{@{}l*{15}{@{\extracolsep{0pt plus12pt}}l}}
\br                              
Equilibrium & Aspect ratio & \0$\langle \beta \rangle$ &\m OH error & \0$\Psi_{\mathrm{b}} (\mathrm{T-m^2})$ & \0 $I_{\mathrm{b}} (\mathrm{kA})$  & \0 $D_{\mathrm{mix}}(\rho=0.9)$ \cr 
\mr
initial   &\0\0 10.0   & 0.016 & \0\0\0 $0.32$ & \0\0 $0.079$ & \0\0 23.8 & \0\0 11.52\cr 
optimized &\0\0 9.37   & 0.025 & \0\0\0 $0.21$ & \0\0 $0.079$ & \0\0 120 & \0\0  0.23\cr 
\br
\end{tabular*}
\label{tab:OH-quantities}
\end{table}
Using DESC, we are able to successfully stabilize the equilibrium while improving the quality of omnigenity. However, we find that obtaining a stable OH is more difficult compared to the OP or OT for a high plasma beta. The optimized equilibria are also strongly ``bean''-shaped, which would affect coil design. Positive shear would also degrade turbulent transport compared to negative shear equilibria. 

The distance from ideal ballooning marginality significantly reduced the KBM growth rates and the mixing-length diffusion coefficient for the OP and OH cases, but only marginally for the OT case. This indicates that our hypothesis correlating the distance from marginality and the KBM growth rate is not always correct. To determine the effect of ideal ballooning stabilization on KBM stability, we perform an additional analysis in the next section.

\subsection{Neoclassical transport in the optimized configurations}
In addition to MHD and kinetic stability, it is important to have reduced neoclassical transport. Hence, we shall evaluate the  neoclassical transport due to the magnetic field ripple for the three optimized configurations. To do that, we calculate the effective ripple proxy $\epsilon_{\mathrm{eff}}^{3/2}$~\cite{nemov1999evaluation} using~\texttt{DESC}, which measures neoclassical transport in the $1/\nu$ collisionality regime. Details concerning the implementation of an effective ripple implementation in \texttt{DESC} can be found in Unalmis \etal~\cite{unalmis2024spectrally}.
\begin{figure}[!h]
    \centering
    \begin{subfigure}[b]{0.325\textwidth}
    \centering
        \includegraphics[width=\textwidth, trim={4mm 0mm 4mm 4mm}, clip]{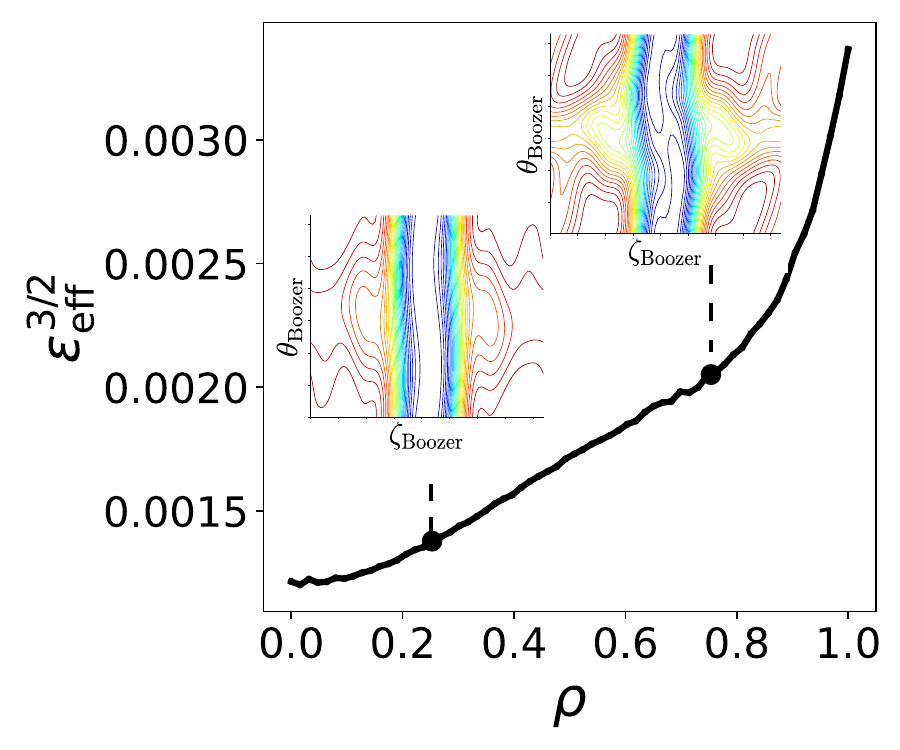}`
        \caption{OP}
    \end{subfigure}
    \begin{subfigure}[b]{0.325\textwidth}
        \centering
        \includegraphics[width=\textwidth, trim={4mm 0mm 4mm 4mm}, clip]{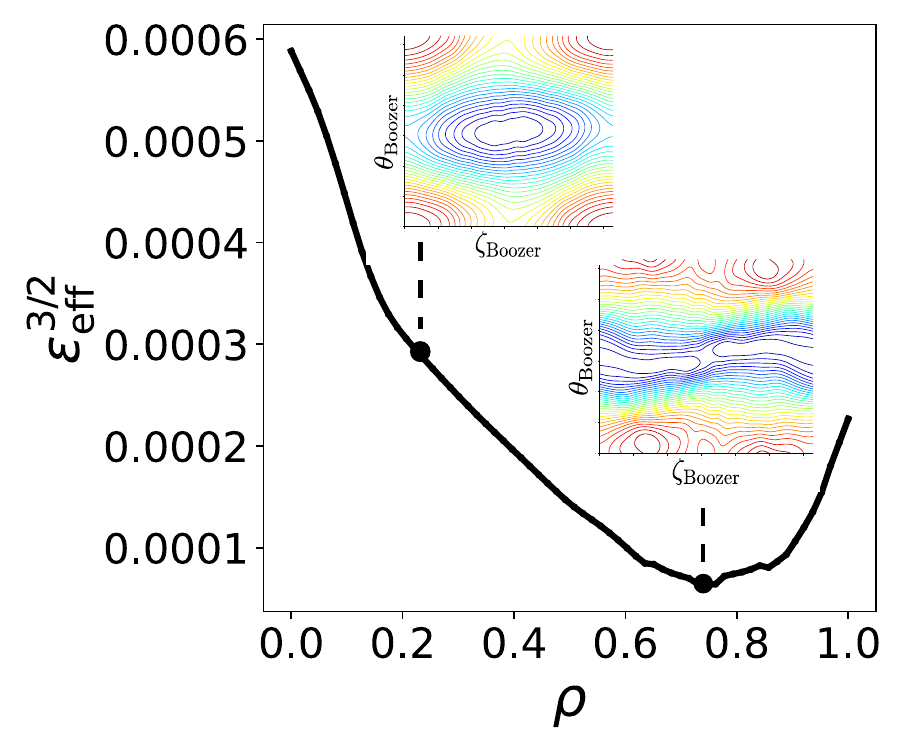}
        \caption{OT}
    \end{subfigure}
    \begin{subfigure}[b]{0.325\textwidth}
        \centering
        \includegraphics[width=\textwidth, trim={4mm 0mm 4mm 4mm}, clip]{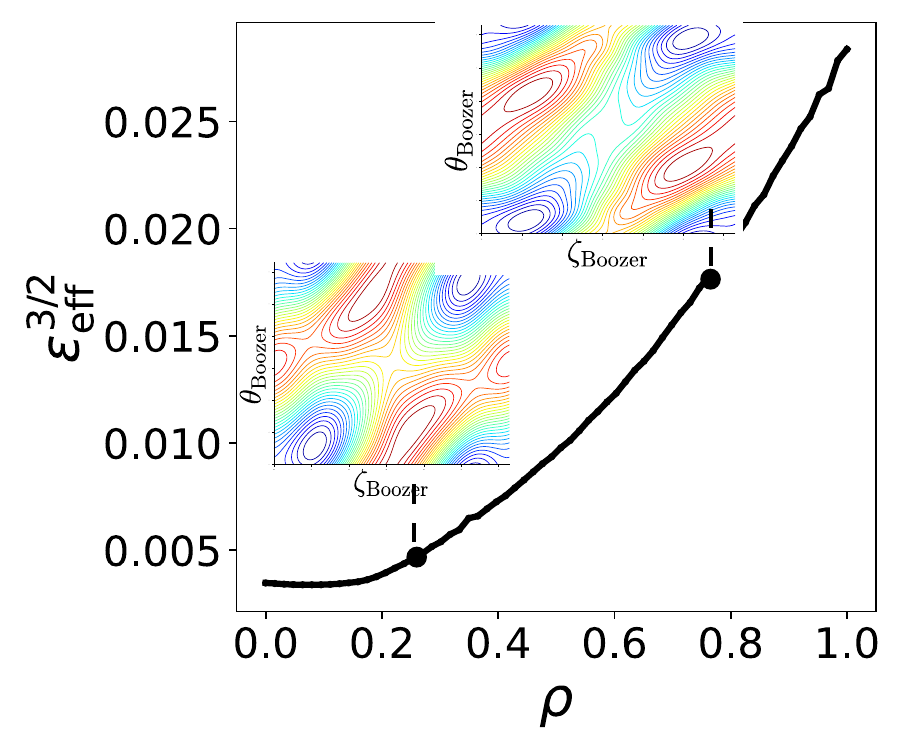}
        \caption{OH}
    \end{subfigure}
    \caption{Neoclassical transport coefficient $\epsilon_{\mathrm{eff}}^{3/2}$ calculated for the optimized omnigenous configurations along with the Boozer plots of optimized magnetic field at two radial locations $\rho = 0.25, 0.75$. The OT case has the lowest ripple whereas the OH case has the largest.}
    \label{fig:effective-ripple}
\end{figure}

From figure~\ref{fig:effective-ripple}, we can see that there is a strong correlation between the degree of omnigenity and effective ripple. As one can see, the contours for the OT case at $\rho = 0.25$ have more closed ``loops'' than the contours at $\rho = 0.75$, in the low-field region, resulting in a higher neoclassical transport. The same can be observed for the OP case where $\rho = 0.75$ has more closed loops than $\rho = 0.25$ in the high-field region. Indeed, according to the Cary-Shasharina prescription, Boozer plots with closed contours spanning the full toroidal, poloidal, or helical range are more omnigenous and lead to reduced neoclassical transport than plots with ``loops'' of the magnetic field. When compared with the optimization method used by~\cite{isaev1997plasma, mikhailov2002improved}, the optimized OP equilibrium has an $\epsilon_{\mathrm{eff}}^{3/2}$ that is more than twice as small while having a lower aspect ratio.

\section{$\hat{s}-\alpha_{\mathrm{MHD}}$ and sensitivity analysis}
\label{sec:s-alpha}
In this section, we analyze the ideal MHD and kinetic properties of the three optimized equilibria and determine when the distance from marginality proxy is suitable to reduce the KBM growth rates. To do so, we start by plotting the ideal ballooning growth rates in the $\hat{s}-\alpha_{\mathrm{MHD}}$ landscape for the three optimized equilibria at four different radii. To create a single $\hat{s}-\alpha_{\mathrm{MHD}}$ contour plot, we solve the ballooning equation for $N_{\hat{s}} = 24 \times N_{\alpha_{\mathrm{MHD}}} = 32$. For each solution, we use $\zeta = [-5\pi, 5\pi]$, $N = 1961$ points along a field line and scan $N_{\alpha} = 8$ field lines with $\alpha \in [0, \pi]$ and $N_{\zeta_0} = 9$ values of $\zeta_0 \in [-\pi/2, \pi/2]$.

\subsection{Optimized OP equilibrium}
\begin{figure}[!h]
    \centering
    \begin{subfigure}[b]{0.242\textwidth}
    \centering
        \includegraphics[width=\textwidth, trim={2mm 2mm 8.6mm 4mm}, clip]{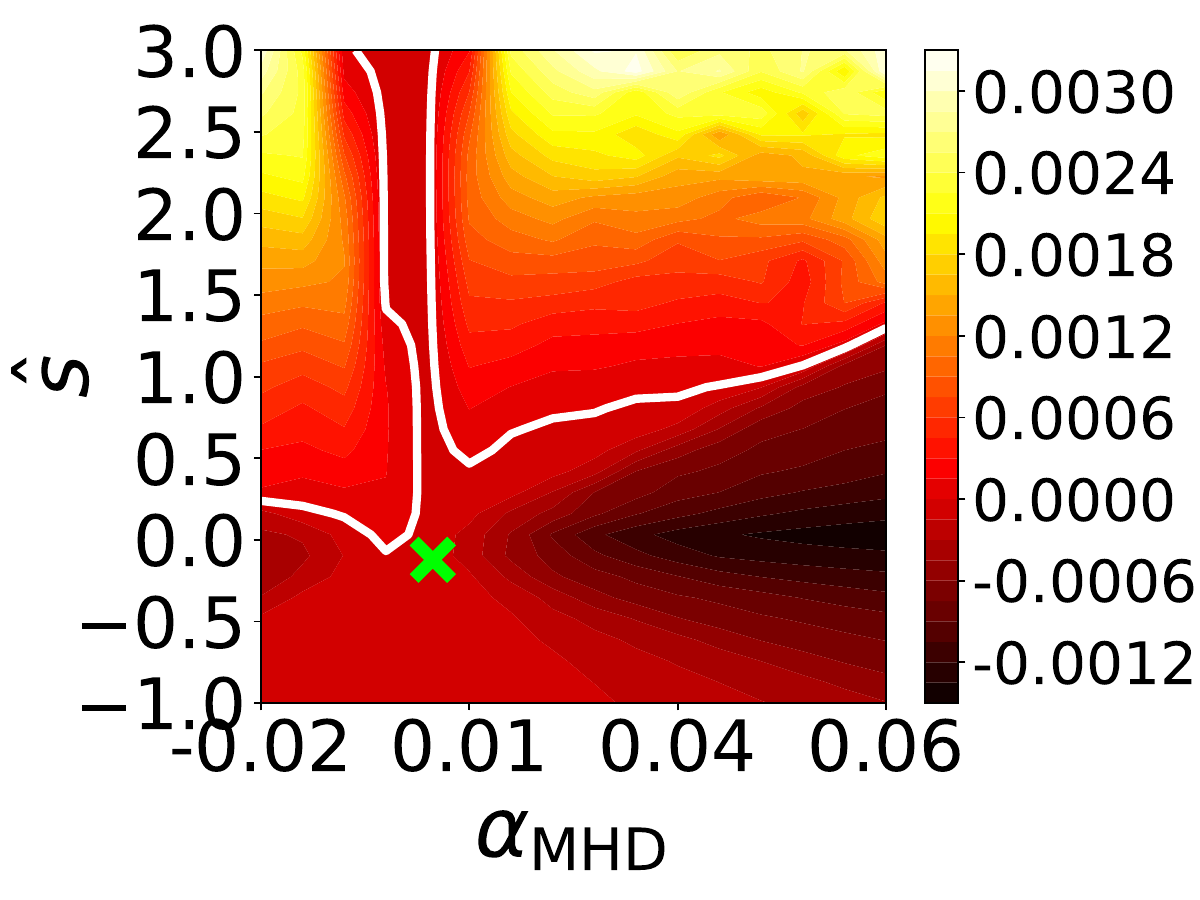}`
        \caption{$\rho=0.35$}
    \end{subfigure}
    \begin{subfigure}[b]{0.242\textwidth}
        \centering
        \includegraphics[width=\textwidth, trim={2mm 2mm 9.5mm 4mm}, clip]{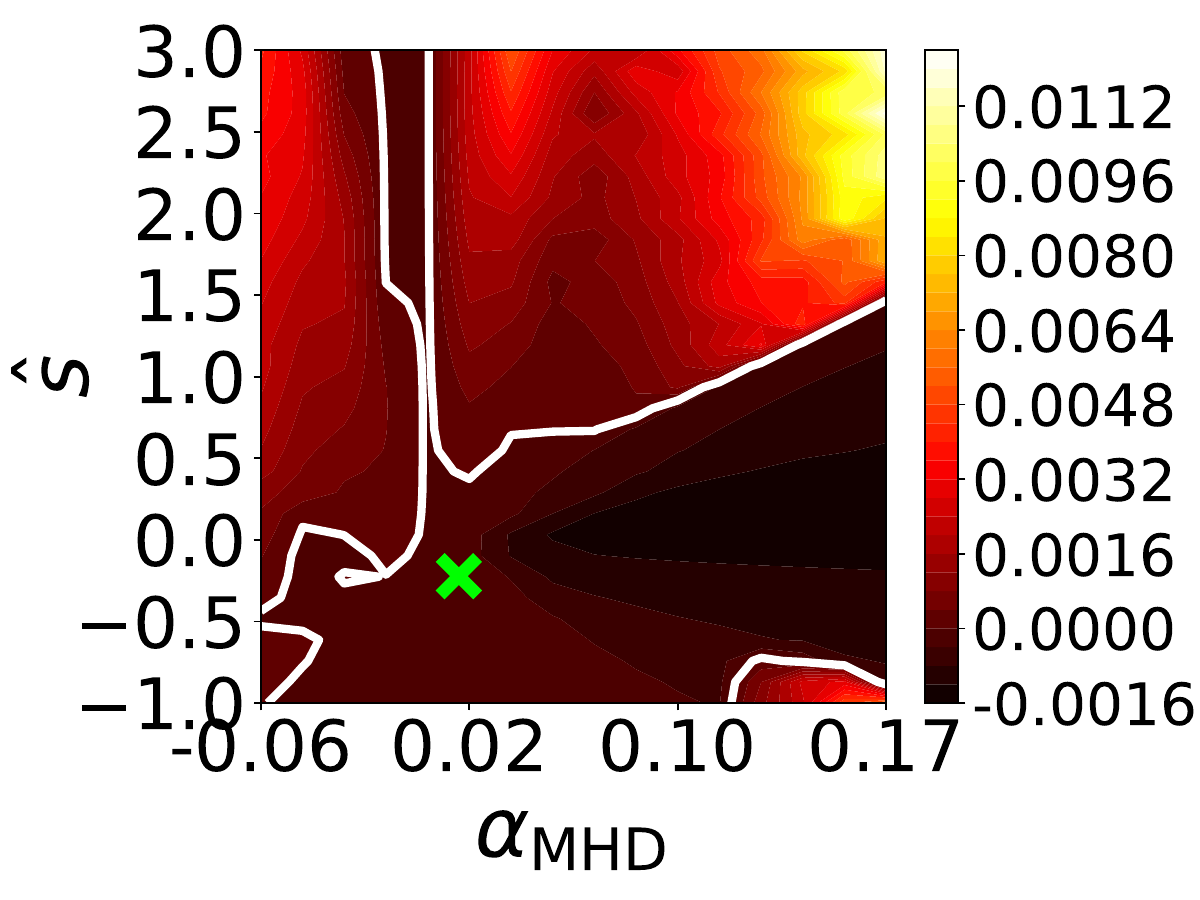}
        \caption{$\rho=0.65$}
    \end{subfigure}
    \begin{subfigure}[b]{0.242\textwidth}
        \centering
        \includegraphics[width=\textwidth, trim={2mm 2mm 8mm 4mm}, clip]{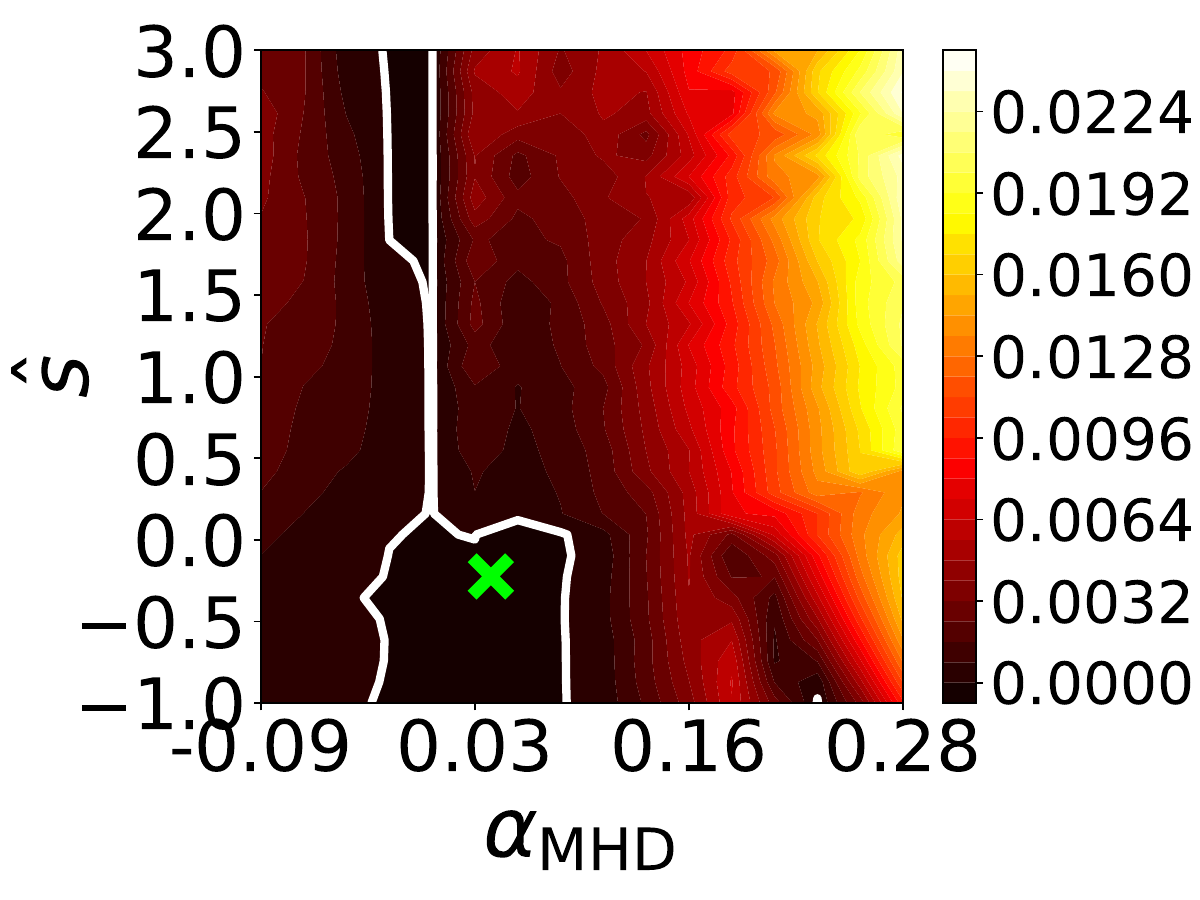}
        \caption{$\rho=0.8$}
    \end{subfigure}
    \begin{subfigure}[b]{0.242\textwidth}
        \centering
        \includegraphics[width=\textwidth, trim={2mm 2mm 8mm 4mm}, clip]{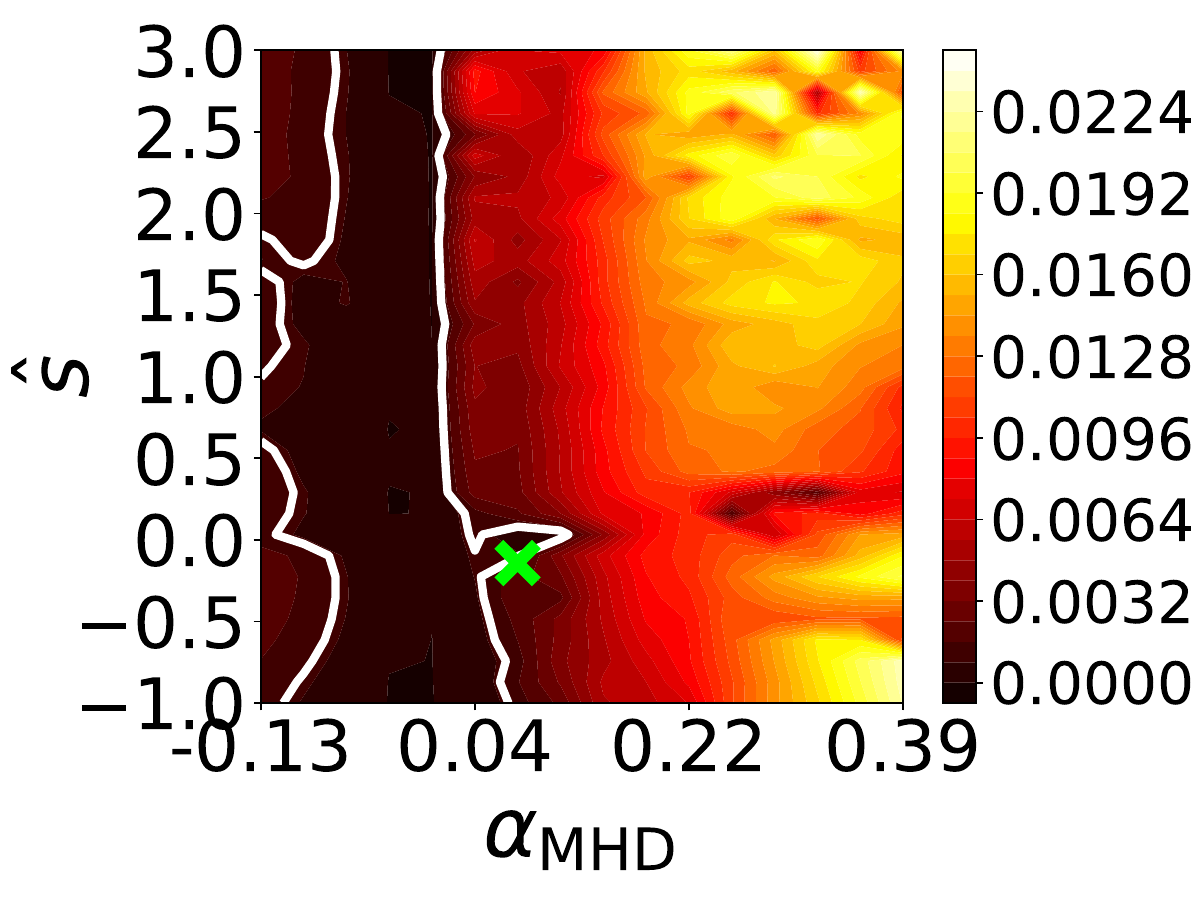}
        \caption{$\rho=0.97$}
    \end{subfigure}
    \caption{$\hat{s}-\alpha_{\mathrm{MHD}}$ scans of the maximum ballooning eigenvalue $\lambda_{\mathrm{max}}$ for the optimized OP configuration. The white line corresponds to the marginal stability curve. The nominal equilibrium (marked by a cross) always lies in the first stability region and access to second stability in the outer core, is inaccessible.}
\label{fig:s-alpha-OP-optimized}
\end{figure}

The $\hat{s}-\alpha_{\mathrm{MHD}}$ ideal-ballooning landscape for the OP equilibrium is shown in figure~\ref{fig:s-alpha-OP-optimized}.
 We demonstrate the existence of a second stability regime in the core region in poloidally omnigenous stellarators. However, as we move towards the outer core, second stability becomes inaccessible for this equilibrium. This suggests that despite the existence of this state, it might not be realistically achievable. When pressure rises from a vacuum equilibrium ($\alpha_{\mathrm{MHD}} = 0$) and the equilibrium enters the unstable region, KBM will drastically increase heat and particle transport, preventing the pressure profile from becoming steeper. Therefore, one must check accessibility by repeating this exercise, especially for finite-$\beta$ omnigenous stellarators such as the ones developed by Sanchez \etal~\cite{sanchez2023quasi} and Goodman \etal~\cite{goodman2024quasi}.

\subsection{Optimized OT equilibrium}
The $\hat{s}-\alpha_{\mathrm{MHD}}$ landscape for the OP equilibrium is shown in figure~\ref{fig:s-alpha-OP-optimized}. Similar to the OP case, we find that the OT equilibrium is situated in the second stability region, demonstrating the existence of a second stability regime in toroidally omnigenous stellarators. Moreover, for these equilibria, second stability is accessible and is not blocked by an unstable region. This means that one can achieve this state without crossing an unstable region --- a region of large turbulent transport resulting from unstable ideal and kinetic ballooning modes.
\begin{figure}
    \centering
    \begin{subfigure}[b]{0.243\textwidth}
    \centering
        \includegraphics[width=\textwidth, trim={2mm 2mm 8mm 4mm}, clip]{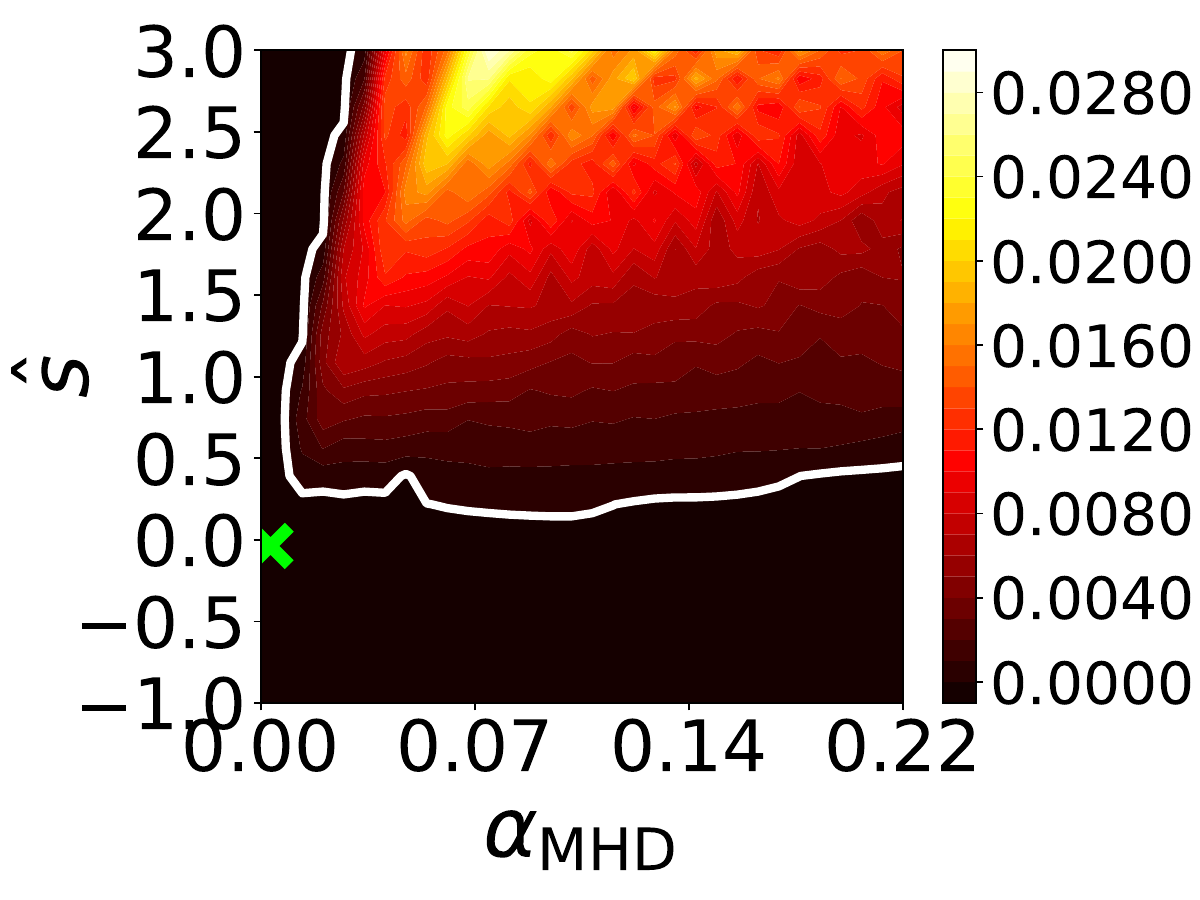}
        \caption{$\rho=0.35$}
    \end{subfigure}
    \begin{subfigure}[b]{0.243\textwidth}
        \centering
        \includegraphics[width=\textwidth, trim={2mm 2mm 8mm 4mm}, clip]{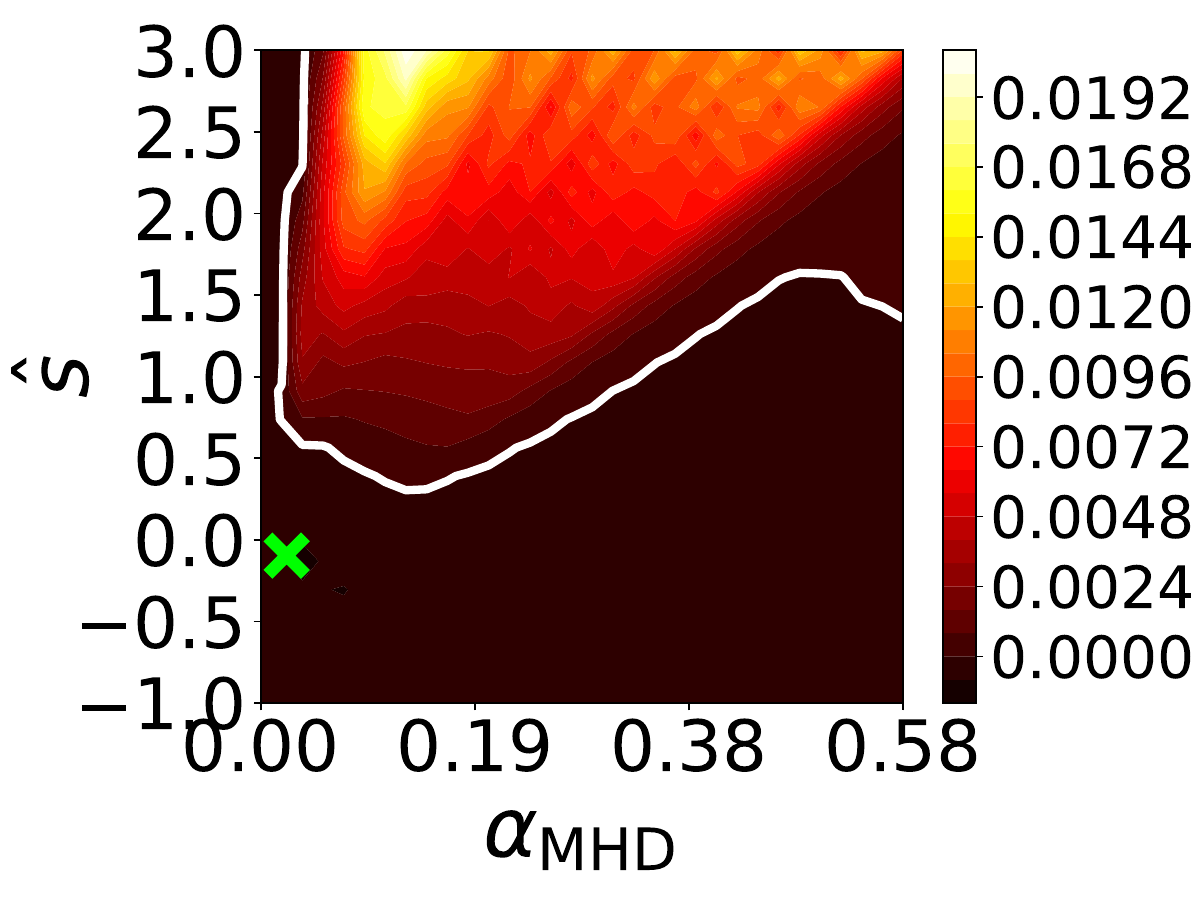}
        \caption{$\rho=0.65$}
    \end{subfigure}
    \begin{subfigure}[b]{0.243\textwidth}
        \centering
        \includegraphics[width=\textwidth, trim={2mm 2mm 8mm 4mm}, clip]{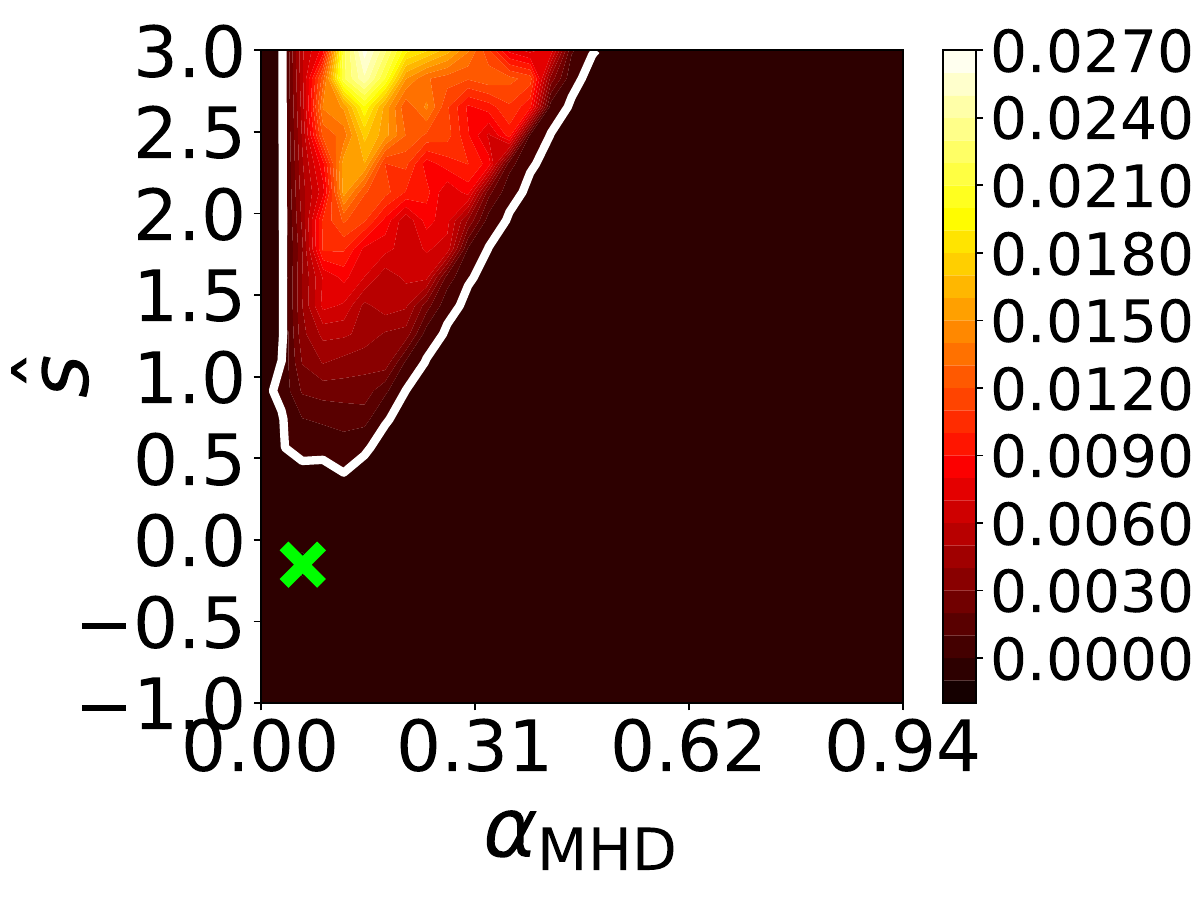}
        \caption{$\rho=0.8$}
    \end{subfigure}
    \begin{subfigure}[b]{0.243\textwidth}
        \centering
        \includegraphics[width=\textwidth, trim={2mm 2mm 8mm 4mm}, clip]{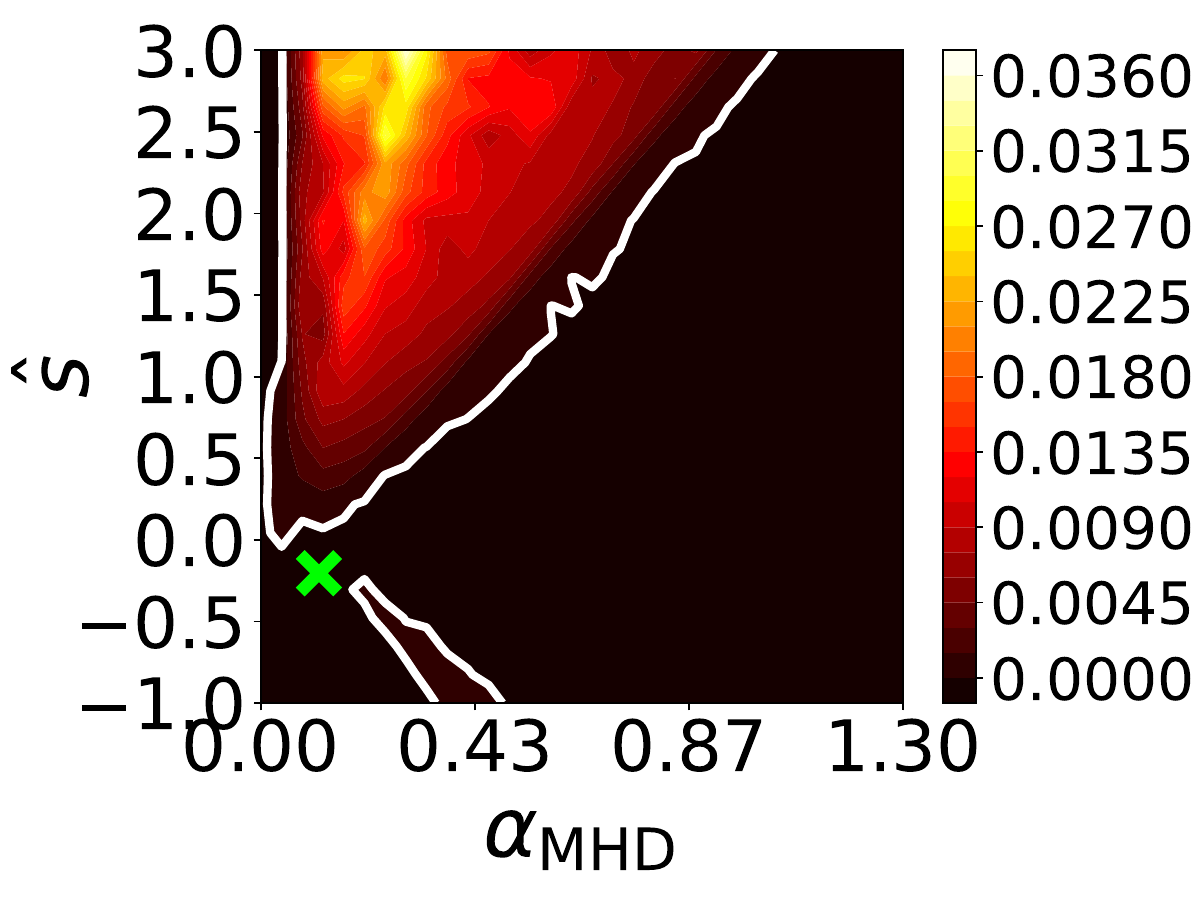}
        \caption{$\rho=0.97$}
    \end{subfigure}
    \caption{$\hat{s}-\alpha_{\mathrm{MHD}}$ scans of the maximum ballooning eigenvalue $\lambda_{\mathrm{max}}$ for the optimized OT configuration. The nominal equilibrium (marked by a cross) moves from the first to the second stable region as we move from the core to the edge. Second stability is fully accessible}
\label{fig:s-alpha-OT-optimized}
\end{figure}

\subsection{Optimized OH equilibrium}
\begin{figure}[!h]
    \centering
    \begin{subfigure}[b]{0.243\textwidth}
    \centering
        \includegraphics[width=\textwidth, trim={2mm 2mm 8mm 4mm}, clip]{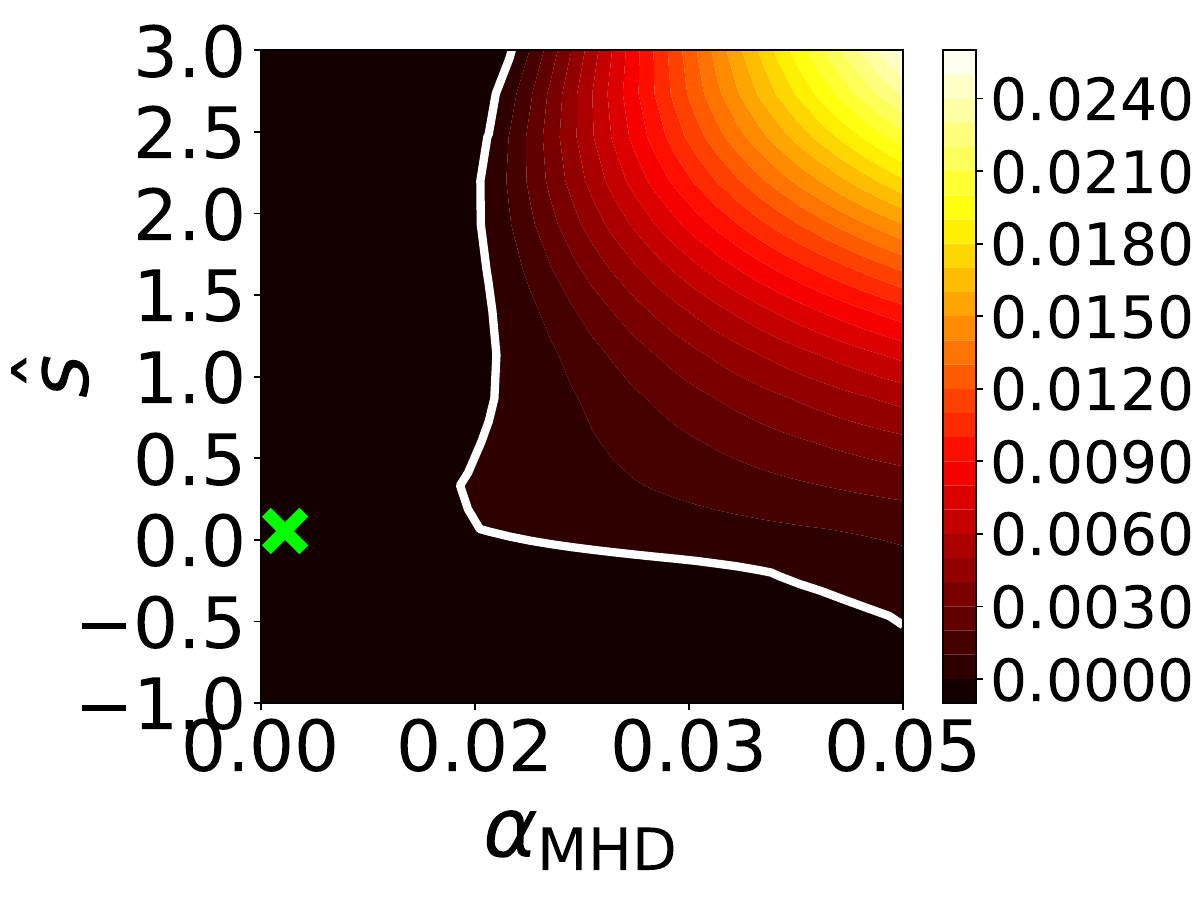}
        \caption{$\rho=0.35$}
    \end{subfigure}
    \begin{subfigure}[b]{0.243\textwidth}
        \centering
        \includegraphics[width=\textwidth, trim={2mm 2mm 8mm 4mm}, clip]{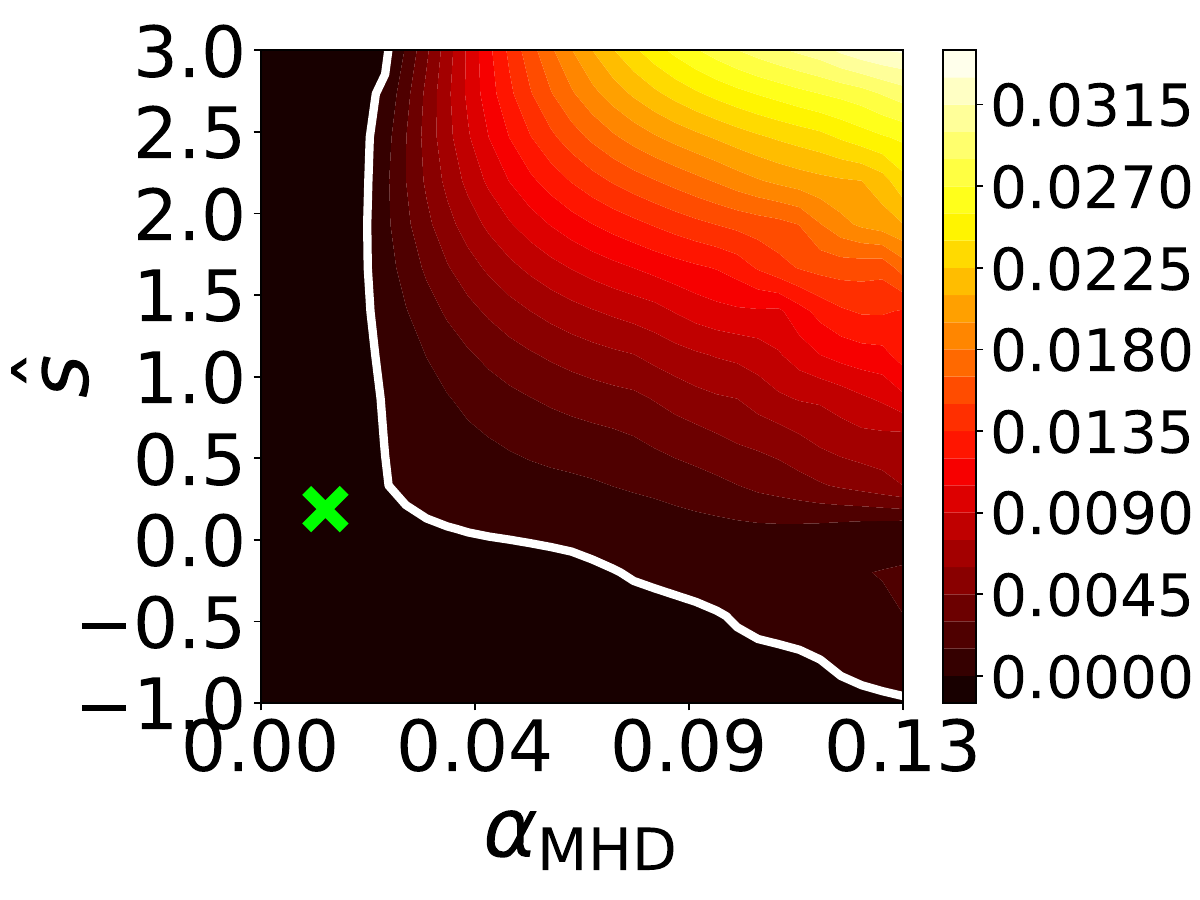}
        \caption{$\rho=0.65$}
    \end{subfigure}
    \begin{subfigure}[b]{0.243\textwidth}
        \centering
        \includegraphics[width=\textwidth, trim={2mm 2mm 8mm 4mm}, clip]{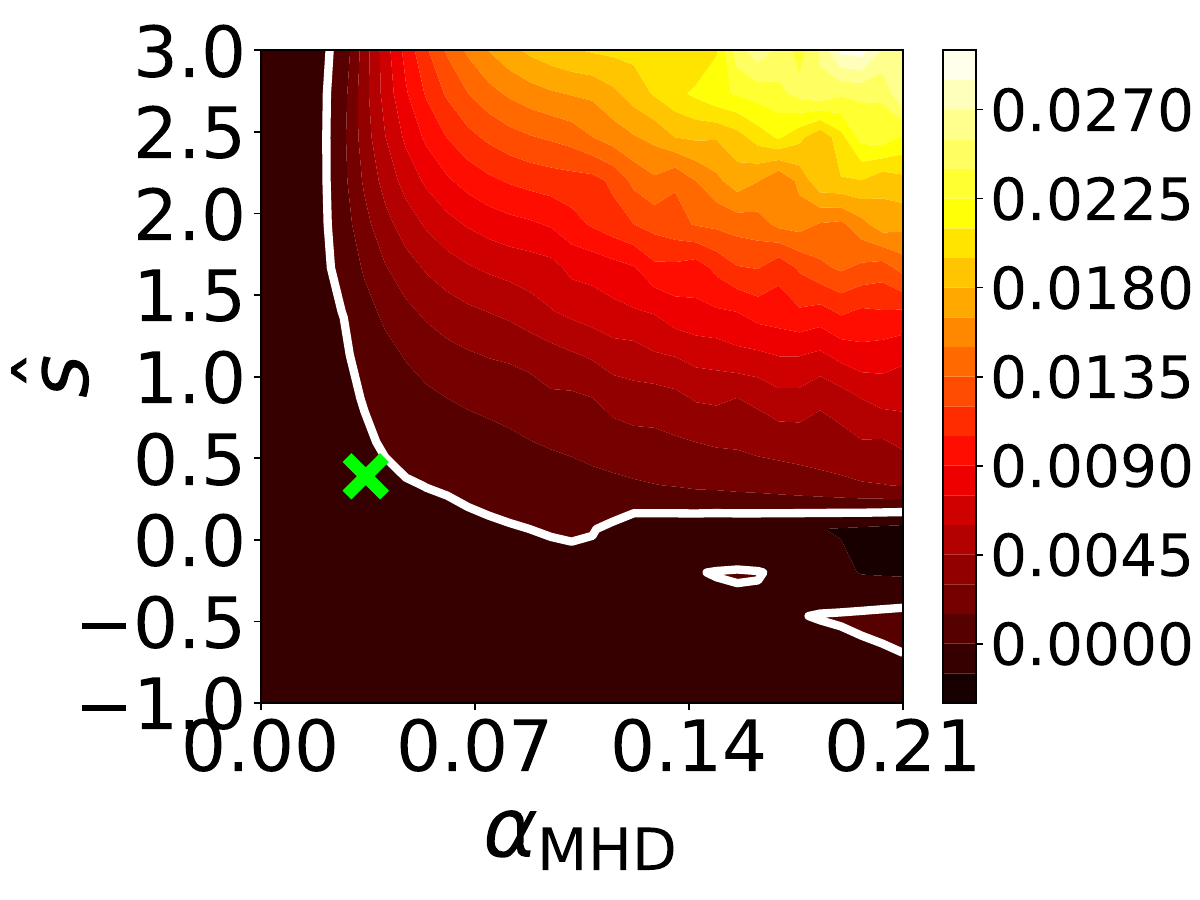}
        \caption{$\rho=0.8$}
    \end{subfigure}
    \begin{subfigure}[b]{0.243\textwidth}
        \centering
        \includegraphics[width=\textwidth, trim={2mm 2mm 8mm 4mm}, clip]{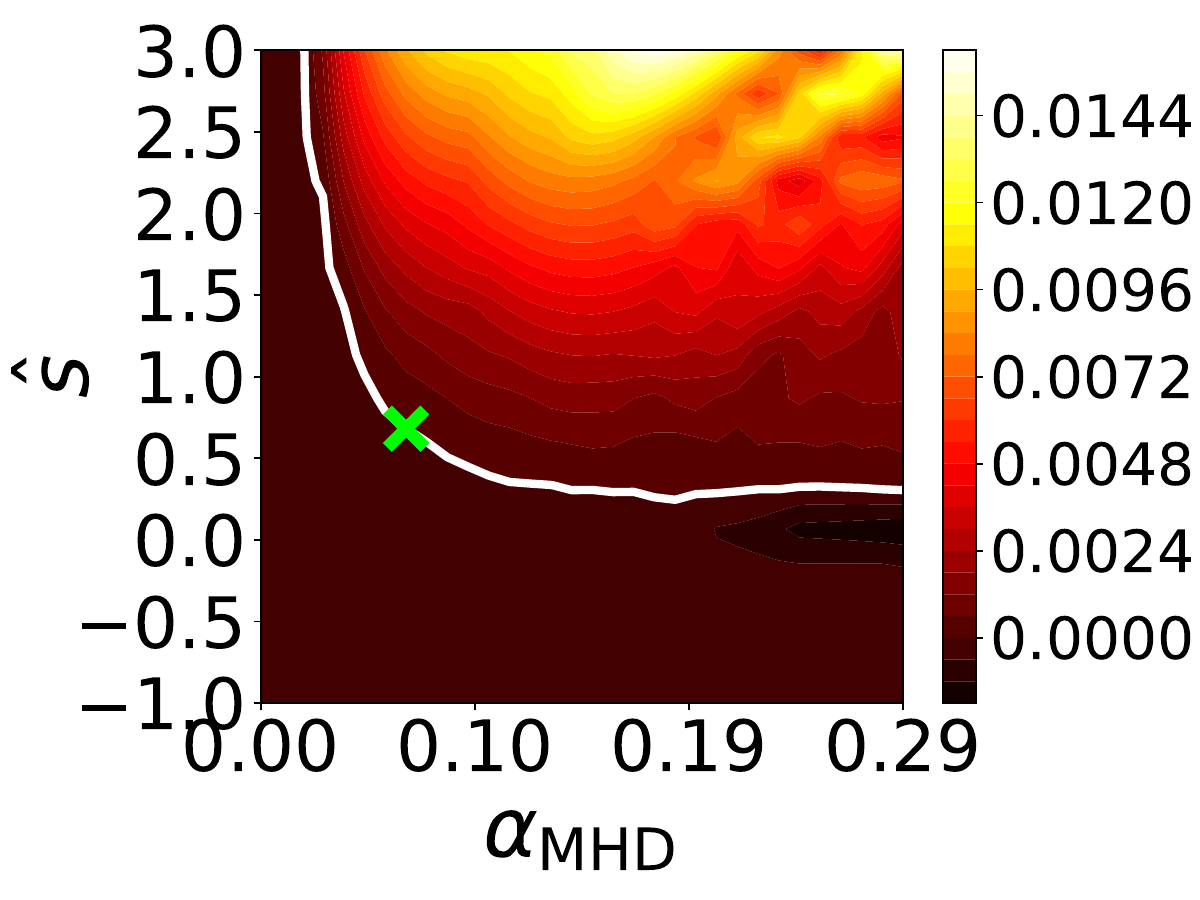}
        \caption{$\rho=0.97$}
    \end{subfigure}
    \caption{$\hat{s}-\alpha_{\mathrm{MHD}}$ scans of the maximum ballooning eigenvalue $\lambda_{\mathrm{max}}$ for the optimized OH configuration. The nominal equilibrium (marked by a cross) always lies in the first stability region. Second stability seems to exist near the edge but it inaccessible.}
\label{fig:s-alpha-OH-optimized}
\end{figure}

The $\hat{s}-\alpha_{\mathrm{MHD}}$ landscape for the OH equilibrium is shown in figure~\ref{fig:s-alpha-OH-optimized}. Unlike the OP and OT cases, we find that the OH equilibrium is situated in the first stability region.
Although a second stability region is present, the outer core equilibria cannot access it. It may be possible to access second stability for negative shear OH equilibria.

\subsection{When is the distance from marginality a good proxy for KBMs?}
\begin{figure}[!h]
    \centering
    \begin{subfigure}[b]{0.31\textwidth}
    \centering
        \includegraphics[width=\textwidth, trim={2mm 0mm 8mm 4mm}, clip]{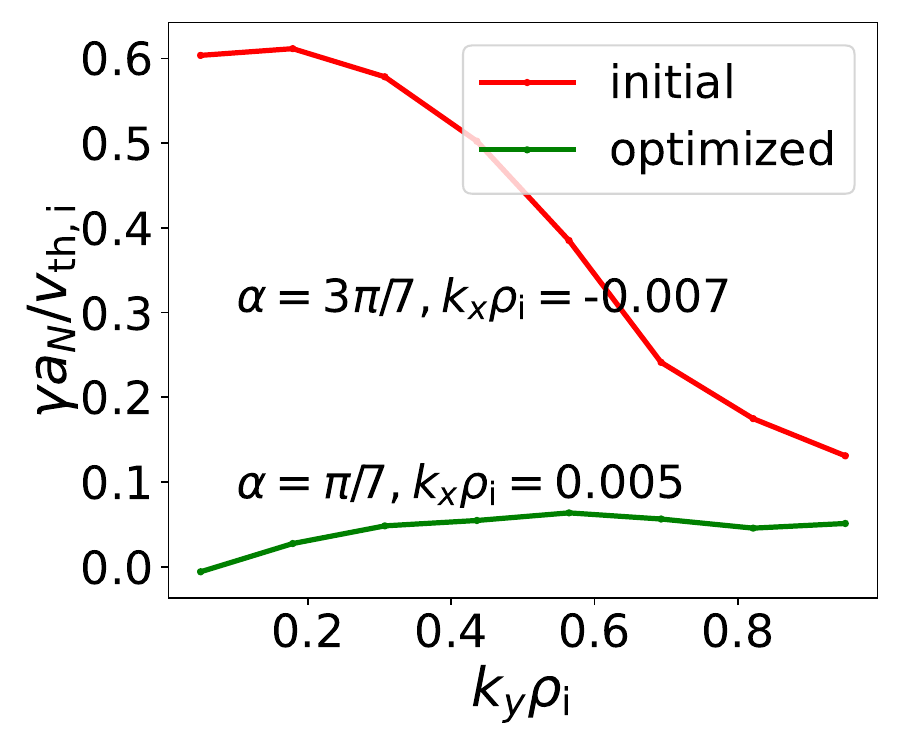}
        \caption{OP, $\rho=0.9$}
    \end{subfigure}\,
    \begin{subfigure}[b]{0.31\textwidth}
        \centering
        \includegraphics[width=\textwidth, trim={2mm 0mm 8mm 4mm}, clip]{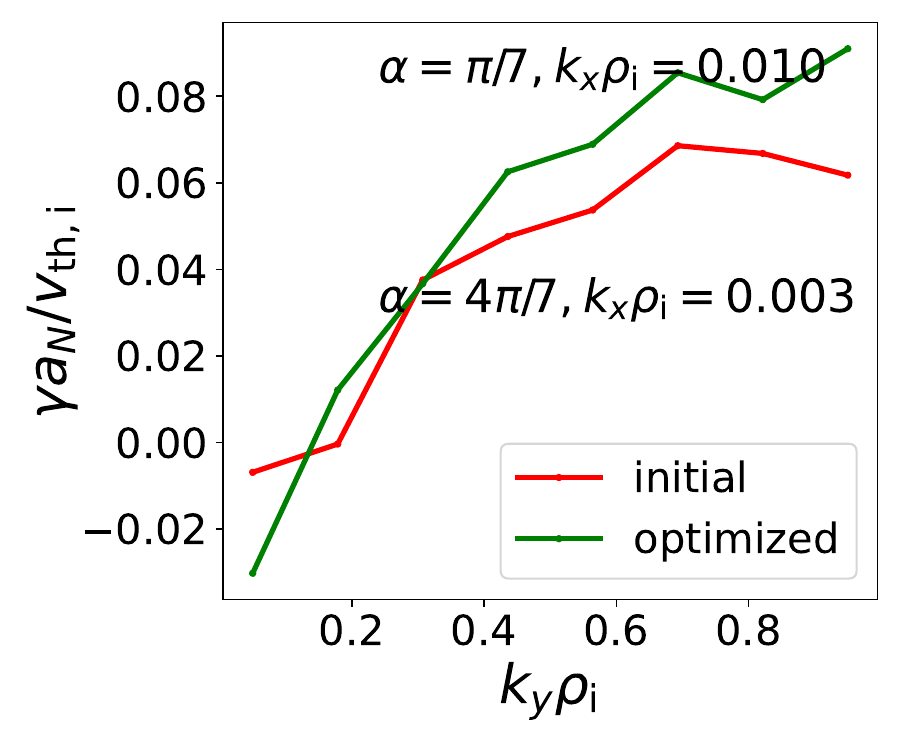}
        \caption{OT, $\rho=0.9$}
    \end{subfigure}\, 
    \begin{subfigure}[b]{0.31\textwidth}
        \centering
        \includegraphics[width=\textwidth, trim={2mm 0mm 8mm 4mm}, clip]{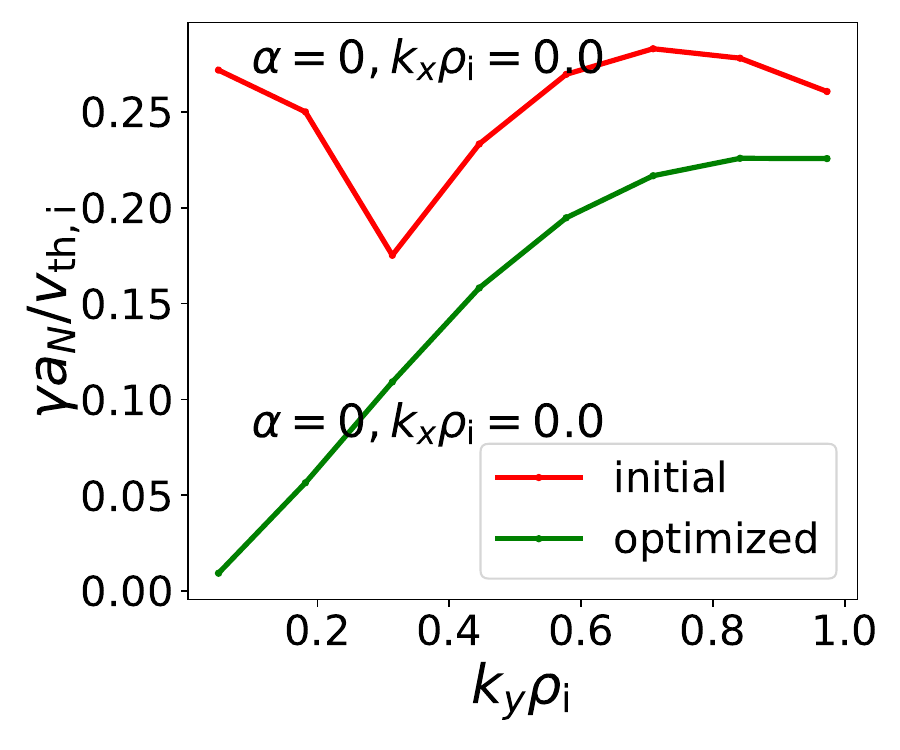}
        \caption{OH, $\rho=0.9$}
    \end{subfigure}
    \caption{Normalized gyrokinetic growth rate as a function of the binormal wavenumber $k_y \rho_{\mathrm{i}}$ for the fieldline $\alpha$ and radial wavenumber $k_x \rho_{\mathrm{i}}$ at which the maximum growth rate occurs for the three equilibria. Ideal-ballooning instability manifests as KBM in the long wavelength limit $k_y \rho_{\mathrm{i}} = 0$. Once the equilibrium is stabilized agaimst the ballooning mode, KBM peaks at a finite wavenumber $k_y \rho_{\mathrm{i}}$.
    }
\label{fig:gamma-vs-ky}
\end{figure}

For the OT equilibria, the KBM growth rates did not change significantly, even when we stabilized the ideal ballooning mode and increased the distance from marginality. In contrast, for the OP and OH case, the KBM growth rate decreased significantly. 
To better understand this effect, we plot the normalized KBM growth rate $\gamma a_{\mathrm{N}}/v_{th, \mathrm{i}}$ with respect to the normalized binormal wavenumber $k_y \rho_{\mathrm{i}}$ in figure~\ref{fig:gamma-vs-ky}
The OP and OH cases were ideal ballooning unstable for the initial equilibrium and stable for the optimized equilibrium. Hence, ideal ballooning stability also leads to KBM stabilization. Moreover, the wavenumber for the peak growth rate $k_y \rho_{\mathrm{i}}$ shifts from $k_y \rho_{\mathrm{i}} = 0$ (as assumed by the derivation in~\ref{app:GK-KBM}) to a finite value for these cases. This also explains the massive values of the mixing-length diffusion coefficients $D_{\mathrm{mix}}$ in the initial, ballooning-unstable OP and OH equilibria, as the growth rate peaks near the lowest wavenumber $k_x = 0, k_y = 0$. However, since the OT case was already stable against the ideal ballooning mode, we do not see a significant change in either the KBM growth rate or $D_{\mathrm{mix}}$.

Therefore, we observe that the distance from the ballooning marginality proxy is only effective in reducing the KBM growth rate when the equilibrium is ballooning unstable or close to marginal stability. In practice, it is difficult to increase the distance from marginality after the ballooning mode becomes stable due to the existence of the Alfv\`{e}n continuum or the optimizer getting stuck between the least unstable and second least unstable eigenvalues. A possible solution would be to include the first $M$ largest eigenvalues in the ballooning objective function and use the new objective
\begin{eqnarray}
    \fl f_{\mathrm{ball}}(\rho) =\sum_{l =1}^{M}\left[ w_0 \,  \mathrm{ReLU}(\lambda_{\mathrm{max}, l}(\rho) - \lambda_0) + w_1 \sum_{j=1}^{N_{\zeta_0}} \sum_{k = 1}^{N_{\alpha}} \mathrm{ReLU} (\lambda_{j, k, l}(\rho) - \lambda_0)\right].
    \label{eqn:modified-f_ball}
\end{eqnarray}
It is possible that increasing the distance from marginality further stabilizes the KBM but further analysis is beyond the scope of this paper.

\subsection{Robustness of the ideal ballooning stability to changes in the pressure and rotational transform profiles}
Understanding the resilience of ideal ballooning and the resulting kinetic ballooning stability to small changes in the pressure and rotational transform profiles is crucial. To this end, we calculate derivatives $\partial f_{\mathrm{ball}} (\rho)/\partial \hat{\bi{p}}, \hat{\bi{p}} = \{p_0, \iota_0, \iota_2\}$ subject to the constraint that the equilibrium remains in force balance. 
\begin{figure}
    \centering
    \begin{subfigure}[b]{0.41\textwidth}
    \centering
        \includegraphics[width=\textwidth, trim={2mm 0mm 4mm 4mm}, clip]{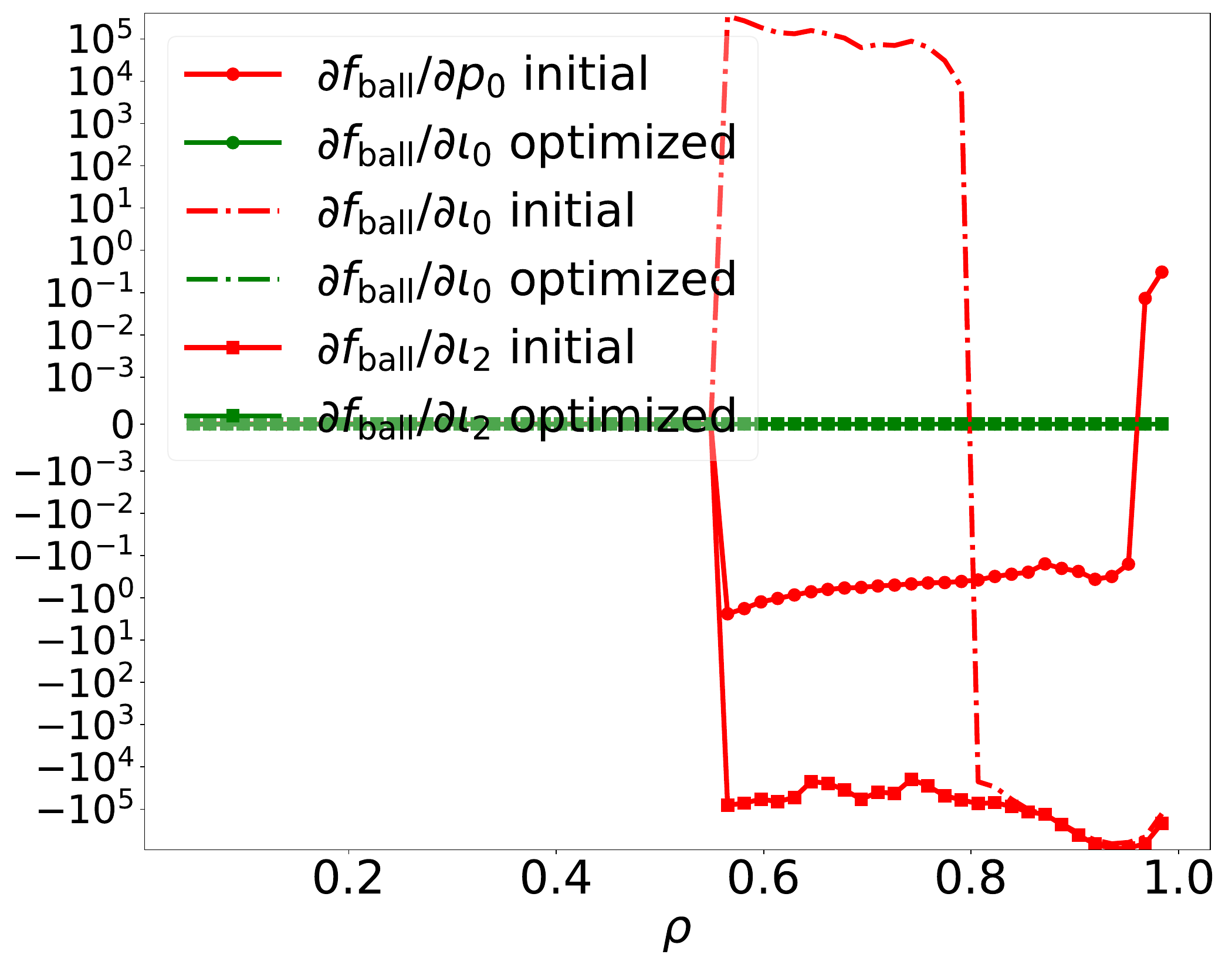}
        \caption{OP}
    \end{subfigure}\,
    \begin{subfigure}[b]{0.41\textwidth}
        \centering
        \includegraphics[width=\textwidth, trim={2mm 0mm 4mm 4mm}, clip]{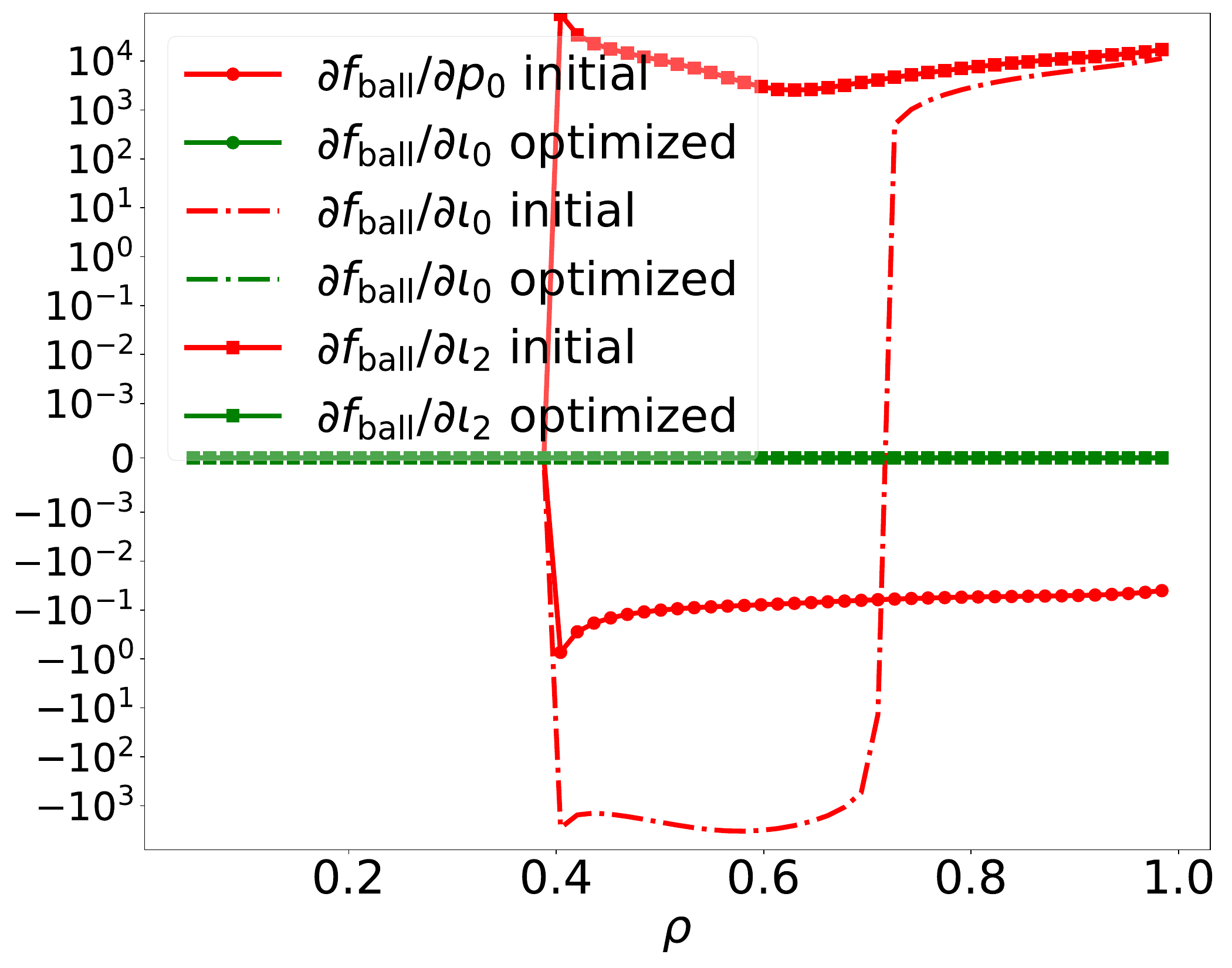}
        \caption{OH}
    \end{subfigure}\, 
    \caption{Sensitivity of the ballooning eigenfunction to the pressure and rotational transform profiles. For the OT case (not plotted), $\partial f_{\mathrm{ball}}/\partial \hat{\bi{p}} = 0$ for both the initial and optimized equilibria. Upon increasing the pressure, the unoptimized OP equilibrium becomes stable (except the outer core) whereas the unoptimized OH equilibrium becomes unstable. The discontinuous jump in the core is due to a derivative of the ReLU function in the $f_{\mathrm{ball}}$ objective.}
\label{fig:sensitivity}
\end{figure}
The derivatives $\partial f_{\mathrm{ball}}/\hat{\bi{p}} = 0$ for all the optimized equilibria, even though the profiles were fixed during optimization \footnote{By design, the derivatives $\partial f_{\mathrm{ball}}/\partial \bi{p} = 0$ for the optimized equilibria, as they correspond to a local minimium of the objective function. However, for parameters $\hat{\bi{p}}$ outside the set $\bi{p}$, in general partial derivatives will be non-zero}. The derivative $\partial f_{\mathrm{ball}}/\hat{\bi{p}} = 0$ implies that small variations in the global profiles still keep the ballooning eigenvalue below the threshold $\lambda < \lambda_0$, maintaining ballooning stability. This matches our observation from the $\hat{s}-\alpha_{\mathrm{MHD}}$ plots from figures~\ref{fig:s-alpha-OP-optimized}-~\ref{fig:s-alpha-OH-optimized}, as moving the local equilibrium (marked by a green cross) by a small amount will not affect the ideal ballooning, and consequently, kinetic ballooning stability of the optimized equilibrium. This indicates that variation of local gradients using the $\hat{s}-\alpha_{\mathrm{MHD}}$ method, performed in previous sub-section, leads to the same conclusion as small variations of the global profiles, that the optimized equilibria are resilient to small changes in the global profiles.

Note that it is straightforward to calculate the sensitivity of any other objective. Accurate values of the derivatives of all objective functions are easily accessible due to the automatic-differentiation-oriented design of~\texttt{DESC}.


\section{Summary and Conclusions}
 In this work, we developed omnigeneous finite-$\beta$ stellarators with improved kinetic stability using the GPU-accelerated~\texttt{DESC} optimizer.  

 In Section~\ref{sec:ideal-MHD}, we began by introducing the ideal MHD model along with the technique developed by Hegna and Nakajima used to vary a local equilibrium by changing the gradients $\hat{s}$ and $\alpha_{\mathrm{MHD}}$. In Section~\ref{sec:GK-W7X}, we introduce the gyrokinetic model and explain how it is related to the ideal ballooning equation. By analyzing a finite-$\beta$ W7-X equilibrium, we argue how distance from ideal-ballooning marginality can be used a prxy for KBM stability. In Section~\ref{sec:ballooning}, we presented the ideal-ballooning analysis along with benchmarks against~\texttt{COBRAVMEC}. In~\ref{sec:omnigenity}, we explain the prescription devised by Cary and Shasharina to obtain omnigenity with poloidal, helical, or toroidal helicity along with an example of this implementation. In section~\ref{sec:Results}, these objectives are then used to develop finite-$\beta$ stellarators with stability against Mercier, ideal ballooning, and as a consequence, kinetic ballooning modes. In section~\ref{sec:s-alpha}, we perform $\hat{s}-\alpha_{\mathrm{MHD}}$ and sensitivity analysis and explain the effectiveness of the ideal ballooning proxy for KBM stabilization. 

 The results are then analyzed by locally varying the optimized equilibria. Upon creating the $\hat{s}-\alpha_{\mathrm{MHD}}$ plot of the maximum ballooning eigenvalue $\lambda$. we found the existence of a region of second stability against the ideal ballooning mode in the toroidal (OT) and helical (OH) omnigenity cases and the poloidal(OP) omnigenity case in figure~\ref{fig:s-alpha-bad-OP}. The second stable region was only accessible for the toroidally omnigenous (OT) case. We found that the distance from ideal ballooning marginality is only a good proxy for KBMs if an equilibrium is ballooning unstable or close to marginally stable.

This work presents many avenues for future research. An important step would be to use the modified ballooning objective in~\eref{eqn:modified-f_ball}, increase the distance from marginality and further test the effectiveness of the distance-from-marginality proxy. Another key step would be to extend the stability optimization to ensure low-$n$ MHD stability. This is especially important for the OT and OH cases since they require a significant plasma current. This can be done with the help of new faster codes like~\texttt{DCON3D}~\cite{glasser2016direct,glasser2020dcon3d}.

To ensure optimization against KBMs, we can also couple~\texttt{DESC} with a fully electromagnetic solver such as~\texttt{GS2} and perform direct optimization of electromagnetic turbulence, as demonstrated in Gaur~\cite{gaur2023optimization}(pp. 110-114) or a linear adjoint gyrokinetic solver, based on the technique demonstrated by Acton~\cite{acton2024optimisation} \etal. This will allow us to directly reduce the growth rate of any unstable gyrokinetic mode. To reduce nonlinear heat and particle transport, one could also implement equilibrium-dependent proxies for nonlinear heat fluxes resulting from KBMs using improved quasilinear models as was done for the NCSX stellarator by Mynick \etal~\cite{mynick2010optimizing} or recently by Roberg-Clark \etal~\cite{roberg2022coarse}.

Since the $\hat{s}-\alpha_{\mathrm{MHD}}$ analysis is valid for any radially local calculation, we can apply it to study neoclassical transport, Mercier stability, and electrostatic (low $\beta$) turbulence. An $\hat{s}-\alpha_{\mathrm{MHD}}$ analysis of nonlinear heat and particle fluxes for these optimized configurations will be presented in a subsequent paper.\\

\textbf{Data availability}
The driver and post-processing scripts along with the optimized omnigenous equilibria and figures used in this paper are freely available in this \href{https://doi.org/10.5281/zenodo.13887566}{Zenodo repository}~\cite{gaur_2024_13887566}. The DESC code is open source and can be obtained, along with tutorials for omnigenity and ballooning stability from \href{https://github.com/PlasmaControl/DESC}{this GitHub repository}. Similarly, the GS2 code is open source and can be acquired from \href{https://bitbucket.org/gyrokinetics/gs2/src/master/}{this bitbucket repository}. In this work, we used a docker image of the GS2 source, which is freely available \href{https://hub.docker.com/r/rgaur104/gs2}{here}.

\textbf{Acknowledgements}
One of the authors, R. G. enjoyed fruitful discussions with W. Sengupta, X. Chu, M. Landreman, F. P. Diaz, I. G. Abel, M. Zarnstroff, M. Hardman, N. Barbour and B. Jang. 
This work is funded through the SciDAC program by the US Department of Energy,
Office of Fusion Energy Science, and Office of Advanced Scientific Computing Research
under contract No. DE-AC02-09CH11466, the Hidden Symmetries Grant from the Simons Foundation/SFARI (560651), 
and Microsoft Azure Cloud Computing mini-grant awarded by the Center for Statistics and Machine Learning (CSML), Princeton University. This research also used the computing resources of the Della, Stellar, and Traverse clusters at Princeton University.

\appendix
\section{The DESC optimizer suite}
\label{app:DESC-appendix}
DESC~\cite{dudt2020desc, conlin2023desc, panici2023desc, dudt2023desc} is an equilibrium solver and optimizer, designed primarily to generate and optimize stellarator equilibria. It solves the ideal MHD force balance equation~\eref{eqn:ideal-MHD-force-balance} as an inverse problem --- obtaining the cylindrical coordinates $(R, \phi, Z)$ for each point in a grid defined by a set of non-orthogonal curvilinear coordinates $(\psi, \theta, \phi)$. For a fixed boundary system, the user has to specify the pressure $p(\psi)$ and the enclosed toroidal current $I(\psi)$, or the rotational transform $\iota(\psi)$ along with the boundary shape $R_{\rm{b}}(\theta, \zeta), Z_{\rm{b}}(\theta, \zeta)$, and the enclosed toroidal flux by the boundary $\psi_{\mathrm{b}}$. Given these inputs, DESC solves~\eref{eqn:ideal-MHD-force-balance} to find the magnetic field $\bi{B}$ and the shape of the flux surfaces throughout the volume. DESC uses a $(\rho, \theta, \phi)$ coordinate system, where
\begin{eqnarray}
    \rho = \sqrt{\frac{\psi}{\psi_b}}, \\
    \theta = \theta_{\rm{PEST}} - \Lambda, \\
    \zeta = \phi,
\end{eqnarray}
The problem is defined in a cylindrical coordinate system $(R, \zeta, Z)$ by decomposing it into Fourier-Zernike spectral bases as shown below
\begin{equation}
    R(\rho,\theta,\zeta) = \sum_{m=-M,n=-N,l=0}^{M,N,L} R_{lmn} \mathcal{Z}_l^m (\rho,\theta) \mathcal{F}^n(\zeta)
\end{equation}
\begin{equation}
    \Lambda(\rho,\theta,\zeta) = \sum_{m=-M,n=-N,l=0}^{M,N,L} \Lambda_{lmn} \mathcal{Z}_l^m (\rho,\theta) \mathcal{F}^n(\zeta)
\end{equation}
\begin{equation}
Z(\rho,\theta,\zeta) = \sum_{m=-M,n=-N,l=0}^{M,N,L} Z_{lmn} \mathcal{Z}_l^m (\rho,\theta) \mathcal{F}^n(\zeta)
\end{equation}
where $l, m$, and $n$ are the radial, poloidal, and toroidal mode numbers whereas $L, M$, and $N$ define the largest values of $l, m$, and $n$, respectively. This defines the resolution of a~\texttt{DESC} equilibrium. The Zernike polynomials $\mathcal{Z}^{m}_l(\rho, \theta)$ are defined as 
\begin{equation}
    \mathcal{Z}_l^m(\rho,\theta) = \cases{
            \mathcal{R}_l^{m}(\rho) \cos(m\theta) & for $m\geq 0$ and $0\leq\rho\leq 1$\\
            \mathcal{R}_l^{|m|}(\rho) \sin(|m|\theta) & for $m < 0$ and $0\leq\rho\leq 1$
    },
    \label{zernike-poly}
\end{equation}
where 
\begin{equation}
    \mathcal{R}_l^{m} (\rho) = \sum_{s=0}^{(l-m)/2} \frac{(-1)^s(l-s)!}{s!\left( \frac{l+m}{2} - s\right)! \left( \frac{l-m}{2} - s\right)!} \rho^{l-2s} \quad \mbox{for } m\geq 0, 
\end{equation}
denotes the radial variation using shifted Jacobi polynomials. 

Using cylindrical coordinates and duality relations between cylindrical and curvilinear coordinates $(\psi, \theta, \zeta)$, we can calculate all terms in~\eref{eqn:ideal-MHD-force-balance}. Additionally, we impose the constraint $\bi{B}\cdot \bi{\nabla}\rho = 0$, enforcing the toroidal nestedness of all flux surfaces. This reduces the system from three to two coupled partial differential equations in three dimensions
\begin{eqnarray}
    F_\rho &=&  \frac{1}{\mu_0}\left[B^\zeta \left(\frac{\partial B_\rho}{\partial \zeta} - \frac{\partial B_\zeta}{\partial \rho}\right) - B^\theta \left(\frac{\partial B_\theta}{\partial \rho}- \frac{\partial B_\rho}{\partial \theta}\right) \right] -\frac{d p}{d \rho},\\
    F_{\mathrm{helical}} &=&  \frac{1}{\mu_0}\left(\frac{\partial B_{\zeta}}{\partial \theta} - \frac{\partial B_\theta}{\partial \zeta}\right),
\end{eqnarray}
where a subscript $x$ denotes the covariant component $F_{x} = (\bi{F}\cdot \bi{e}^{x})$ and superscript denotes the contravariant component $F^{x} = (\bi{F}\cdot \bi{e}_x)$, so the helical component $F_{\mathrm{helical}} =  \bi{F} \cdot (B^\zeta \bi{e}^\theta - B^\theta \bi{e}^\zeta)$.
Using this formalism, combined with the nestedness assumption,~\texttt{DESC} minimizes the two components of the force error $\bi{F}$, which gives us the desired equilibrium.

Using the spectral representation, it becomes possible to accurately express any continuous optimizable quantity as a combination of the components of $\bi{B}$ and the input profiles $p(\psi)$ and $\iota(\psi)$ (or current $I(\psi)$). This capability enables us to optimize multiple objectives simultaneously, rather than solely minimizing the force balance error.
In addition, by combining this spectral solver with the GPU-accelerated JAX~\cite{jax2018github} package, we can efficiently and accurately compute the gradients of any quantity using automatic differentiation. This enables us to perform rapid optimization using gradient-based methods, as demonstrated in the main body of this paper.

\section{Local variation of a 3D equilibrium}
\label{app:s-alpha-appendix}
In this section, we will briefly explain the method of local variation of gradients, first developed by Greene and Chance~\cite{greene1981second} for tokamaks and later by Hegna and Nakajima~\cite{hegna-nakajima} for stellarators. Using this method, one can vary the gradients $p^{'}(\psi)$ and $\iota^{'}(\psi)$ on a flux surface while locally satisfying the ideal MHD force balance. By varying these gradients, and consequently the local equilibrium around a flux surface, we can obtain a deeper understanding of the characteristics of tokamaks and stellarators in relation to small-scale instabilities. We use this analysis to vary the normalized pressure gradient $\alpha_{\mathrm{MHD}}$, and global shear $\hat{s}$ and to calculate the growth rates against the ideal ballooning and the kinetic ballooning instabilities. 

We start by writing the magnetic field in Boozer coordinates
\begin{equation}
    \bi{B} = \bi{\nabla} \psi \times \bi{\nabla} (\theta_{\mathrm{B}} - \iota \zeta_{\mathrm{B}})
    \label{eqn:covariant-B}
\end{equation}
and equivalently defined as    
\begin{equation}
   \bi{B} =  G(\psi)\, \bi{\nabla} \zeta_{\mathrm{B}}  + I(\psi)\, \bi{\nabla} \theta_{\mathrm{B}} +\hat{\beta}\,  \bi{\nabla} \psi,
   \label{eqn:contravariant-B}
\end{equation}
where $G(\psi)$ is the enclosed poloidal current, $I(\psi)$ is the enclosed toroidal current and $\hat{\beta}$ is related to the Pfirsch-Schl\"{u}ter current. We also define the Jacobian
\begin{equation}
    \sqrt{g}_{\mathrm{B}} = [(\bi{\nabla}\psi \times \bi{\nabla} \theta_{\mathrm{B}})\cdot \bi{\nabla}\zeta_{\mathrm{B}}]^{-1} = \frac{G + \iota I}{B^2}
   \label{eqn:Boozer-jacobian}
\end{equation}
To ensure both the definitions~\eref{eqn:covariant-B} and~\eref{eqn:contravariant-B} of the magnetic field are consistent, we must ensure
\begin{equation}
    G = \frac{g_{\zeta_{\mathrm{B}} \zeta_{\mathrm{B}}} + \iota g_{\zeta_{\mathrm{B}} \theta_{\mathrm{B}}}}{\sqrt{g}_B}
    \label{eqn:constraint-1}
\end{equation}
\begin{equation}
    I = \frac{g_{\zeta_{\mathrm{B}} \theta_{\mathrm{B}}} + \iota g_{\theta_{\mathrm{B}} \theta_{\mathrm{B}}} }{\sqrt{g}_B}
    \label{eqn:constraint-2}
\end{equation}
\begin{equation}
    \hat{\beta} = \frac{ g_{ \zeta_{\mathrm{B}} \psi} + \iota g_{\theta_{\mathrm{B}} \psi}}{\sqrt{g}_B}
    \label{eqn:constraint-3}
\end{equation}
where the metric tensor element $g_{a b} = (\partial \bi{x}/\partial a)\cdot(\partial \bi{x}/\partial b)$. Substituting~\eref{eqn:contravariant-B} into~\eref{eqn:ideal-MHD-force-balance}, and separating the radial component, we get
\begin{equation}
    -G^{'} - \iota I^{'} + \left(\frac{\partial}{\partial \zeta_{\mathrm{B}}} + \iota \frac{\partial }{\partial \theta_{\mathrm{B}}}\right)\hat{\beta} = p^{'}\sqrt{g}_{\mathrm{B}}.
    \label{eqn:ideal-MHD-force-balance-Boozer}
\end{equation}
Applying the surface averaging operator $\oiint = (1/2\pi)^2 \int d\theta_{\mathrm{B}} \int d \zeta_{\mathrm{B}}$ to~\eref{eqn:ideal-MHD-force-balance-Boozer}, we get the surface-averaged force balance equation
\begin{eqnarray}
    -G^{'} - \iota I^{'} = p^{'} V^{'}.
    \label{eqn:FSA_ideal-MHD-force-balance-Boozer}
\end{eqnarray}
Subtracting~\eref{eqn:FSA_ideal-MHD-force-balance-Boozer} from~\eref{eqn:ideal-MHD-force-balance-Boozer}, we get an equation for the variable $\hat{\beta}$
\begin{equation}
    \left(\frac{\partial}{\partial \zeta_{\mathrm{B}}} + \iota \frac{\partial }{\partial \theta_{\mathrm{B}}}\right)\hat{\beta} = p^{'}(\sqrt{g}_{\mathrm{B}} - V^{'}).
    \label{eqn:non-FSA_ideal-MHD-force-balance-Boozer}
\end{equation}
Variables such as $G^{'}, I^{'}$, and $\hat{\beta}$ will change when we locally vary the pressure gradient to maintain force balance. A final quantity of interest is the Pfirsch-Schl\"{u}ter current
\begin{eqnarray}
    \lambda^{*} = \frac{1}{p^{'}V^{'}} \left(\frac{\bi{J}\cdot\bi{B}}{B^2} - \oiint \frac{\bi{J}\cdot\bi{B}}{B^2}\right) =  (\bi{B}\times \bi{\nabla}\psi)\cdot \bi{\nabla}\hat{\beta}.
    \label{eqn:Pfirsch-Schluter}
\end{eqnarray}
With all the important expressions, we proceed to demonstrate the method of variation of a local equilibrium. We expand all flux functions in a local coordinate $y =(\psi-\psi_0)/\mu$ around the flux surface labelled $\psi_0$
\begin{eqnarray}
    X = X^{(0)}(\psi) + \mu X^{(1)}(y) + \mu^2 X^{(2)}(y) + \ldots, \quad X = \{p, \iota, G, I\}
    \label{eqn:Hegna-Nakajima_series_flux_functions}
    \label{eqn:Hegna-Nakajima_series_flux_functions}
\end{eqnarray}
where $\mu \ll 1$. This expansion allows us to vary the gradients of the flux functions by the same order as $p(\psi_0)$ while varying the equilibrium values by order $\mu$. 
This also changes the position of a points $\bi{x}$ on a flux surface
\begin{equation}
    \bi{x} = \bi{x}^{(0)}(\psi, \theta_{\mathrm{B}}, \zeta_{\mathrm{B}}) + \mu  \bi{x}^{(1)}(y, \theta_{\mathrm{B}}, \zeta_{\mathrm{B}}) + \ldots.
\end{equation}
and the variable
\begin{equation}
    \hat{\beta} = \hat{\beta}^{(0)}(\psi, \theta_{\mathrm{B}}, \zeta_{\mathrm{B}}) + \mu\,  \hat{\beta}^{(1)}(y, \theta_{\mathrm{B}}, \zeta_{\mathrm{B}}) + \ldots.
    \label{eqn:beta-series}
\end{equation}
Varying the local gradient would require calculating the gradient of the quantities in the series expansion --- the only gradients that will be comparable to the lowest-order gradients will be in radial gradients of the form
$\partial /\partial y$. Therefore, solving for these gradients will fully define the new equilibrium. 

First, we substitute~\eref{eqn:beta-series} and into~\eref{eqn:non-FSA_ideal-MHD-force-balance-Boozer} to get
\begin{equation}
    \left(\frac{\partial}{\partial \zeta_{\mathrm{B}}} + \iota \frac{\partial }{\partial \theta_{\mathrm{B}}}\right)\hat{\beta}^{(1)} = {p^{(1)}}^{'}(\sqrt{g}_{\mathrm{B}} - V^{'}).
\end{equation}
Any general three-dimensional vector can be decomposed using the basis vectors $(\bi{B}, \nabla \psi, \bi{B}\times \nabla \psi)$. To find the change in the basis vectors upon locally deforming an equilibrium, we solve for  
\begin{eqnarray}
    \frac{\partial \bi{x}}{\partial y} = C \bi{B} + M \frac{\bi{\nabla}\psi}{|\bi{\nabla}\psi|^2}  + D \frac{\bi{B}\times \bi{\nabla}\psi}{B^2},
    \label{eqn:partial_x_partial_y}
\end{eqnarray}
where $C, M$, and $B$ are unknowns. Using~\eref{eqn:contravariant-B},~\eref{eqn:beta-series}, and~\eref{eqn:constraint-3} gives us $C = \hat{\beta}^{(1)}/B^2$. Similarly, enforcing that the jacobian $\sqrt{g}_{\mathrm{B}}$ remains fixed after varying the local pressure and current gradients, we must set $M = 0$. The final and most important quantity in this analysis, the cross-field variation $D$, can be obtained by substituting~\eref{eqn:partial_x_partial_y} into constraints~\eref{eqn:constraint-1}-\eref{eqn:constraint-2}. After some algebraic manipulations, using~\eref{eqn:non-FSA_ideal-MHD-force-balance-Boozer} and the surface averaging operator $\oiint$ , we obtain
\begin{equation}
   \eqalign{\left(\frac{\partial}{\partial \zeta_{\mathrm{B}}} + \iota \frac{\partial }{\partial \theta_{\mathrm{B}}}\right)D &= {\iota^{(1)}}^{'}\frac{1}{\oiint 1/g^{\psi \psi}}\left(\frac{1}{g^{\psi \psi}} - \oiint \frac{1}{g^{\psi \psi}}\right)\\
    &- {p^{(1)}}^{'} \frac{V^{'}(G + \iota I)}{\oint 1/g^{\psi \psi}}\left(\frac{\lambda^{*}}{g^{\psi\psi}}\oiint\frac{1}{g^{\psi \psi}} - \frac{1}{g^{\psi \psi}} \oiint \frac{\lambda^{*}}{g^{\psi\psi}}\right),}
\end{equation}
where $g^{\psi \psi} = | \bi{\nabla} \psi|^2$, and $\lambda^{*}$ is the Pfirsch-Schl\"{u}ter  current, obtained from~\eref{eqn:Pfirsch-Schluter}. Finally, we can write the integrated local shear 
\begin{eqnarray}
    L = \frac{\bi{\nabla}\alpha \cdot \bi{\nabla}\psi}{|\bi{\nabla}\psi|^2} =  {\iota^{(1)}}^{'} (\zeta - \zeta_0) +  D +  L_0
\end{eqnarray}
where $L_0 = ((\bi{\nabla}\alpha \cdot \bi{\nabla}\psi)/|\bi{\nabla}\psi|^2)_{0}$ is the integrated local shear of the original equilibrium without any local variation. This process differs from the calculation presented in Hegna and Nakajima but should be mathematically equivalent.
Given the variation in the pressure gradient ${p^{(1)}}^{'}$ and the rotational transform gradient ${\iota^{(1)}}^{'}$, this completely defines the new local equilibrium. Once the local equilibrium is defined, we calculate the geometric quantities required to solve~\eref{eqn:ideal-ballooning-equation} and~\eref{eqn:electrostatic-gyrokinetic-normal-mode}-\eref{eqn:Perpedicular-Ampere's-Law-Fourier} and understand how they would behave if the pressure gradient or magnetic shear were varied. Hudson and Hegna~\cite{hegna-hudson2002ideal}, and Hegna and Nakajima~\cite{hegna-nakajima} have used this technique to analyze the Mercier and ideal ballooning marginal stability of stellarators. We use the same technique to calculate the ideal ballooning and KBM gyrokinetic growth rates.

In practice, we calculate the coefficients $\hat{\beta}, \lambda^{*}, \sqrt{g}_{\mathrm{B}}$ on each flux surface by evaluating the Fourier-Zernike coefficients at a fixed $\rho$. This gives us the Boozer Jacobian $\sqrt{g}_{\mathrm{B}}$ and its Fourier components, from which we can calculate $\hat{\beta}$ and then $\lambda^{*}$ in Fourier space. Upon conversion of $\lambda^{*}$ to real space, we perform flux surface averaging to calculate the cross-field term $D$ and subsequently the integrated local shear $L$. This gives us all the necessary terms needed to calculate the new set of geometry coefficients for the varied equilibrium.

\begin{figure}
    \centering
    \begin{subfigure}[b]{0.36\textwidth}
    \centering
        \includegraphics[width=\textwidth, trim={2mm 0mm 8mm 8mm}, clip]{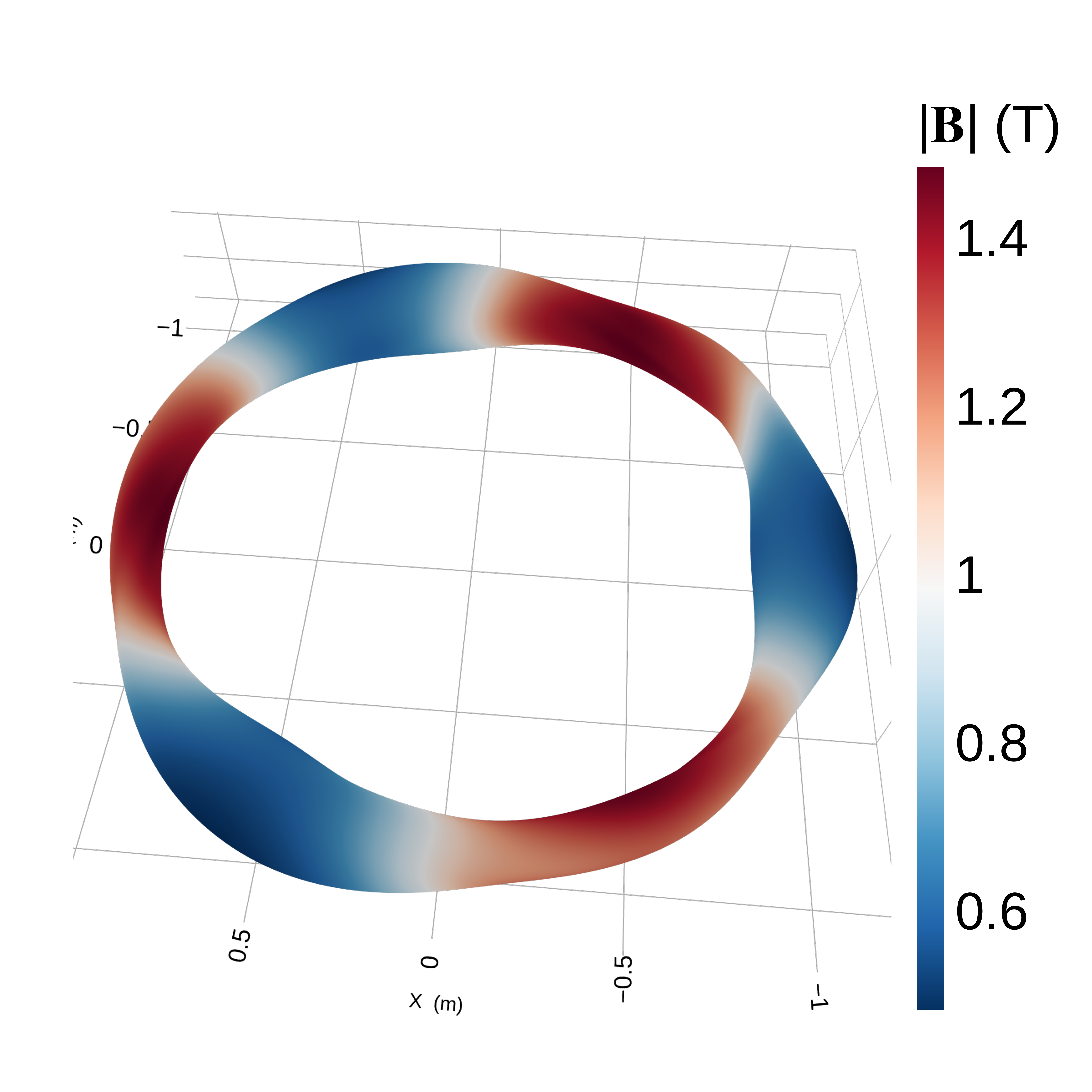}
        \caption{Optimized B $(\rho = 1.0)$}
    \end{subfigure}\quad
    \begin{subfigure}[b]{0.3\textwidth}
        \centering
        \includegraphics[width=\textwidth, trim={2mm 0mm 8mm 4mm}, clip]{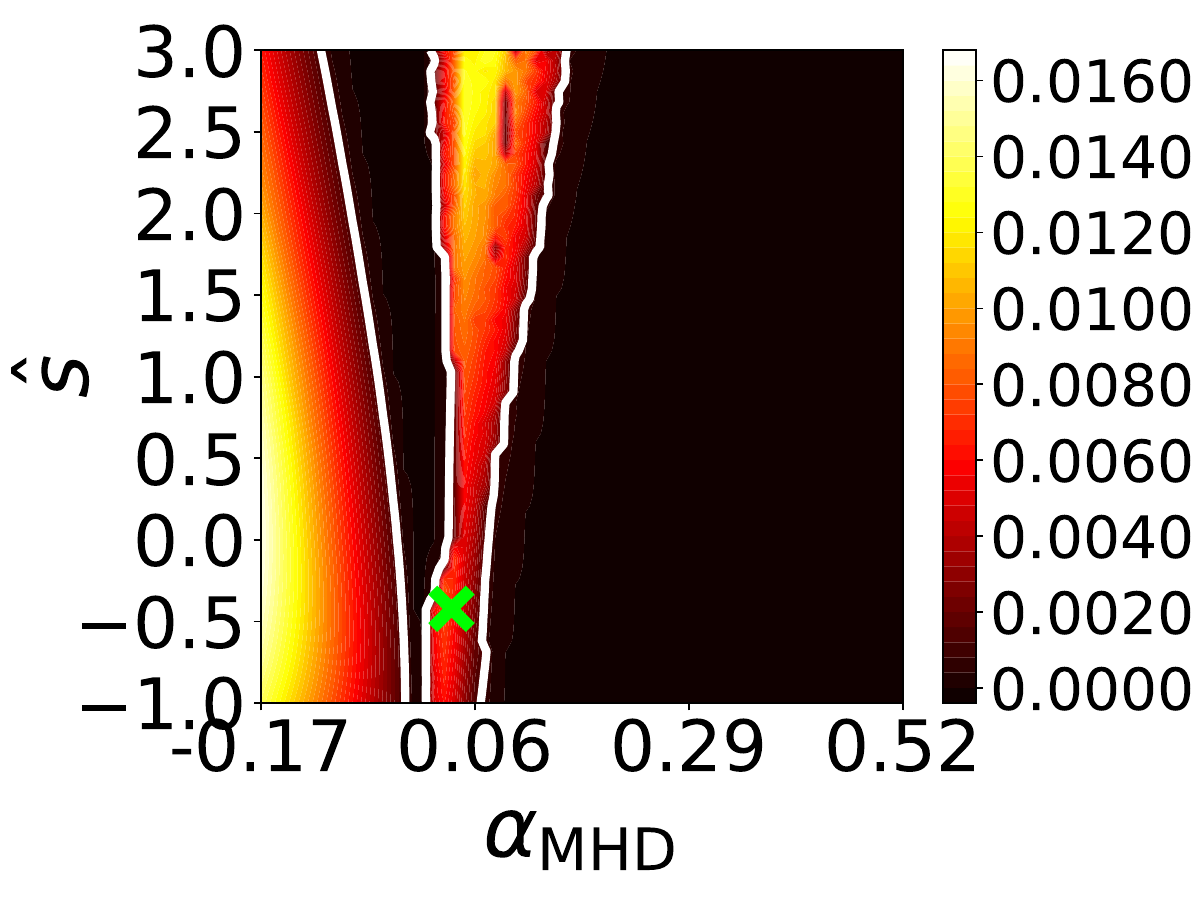}\\[4mm]
        \caption{$\lambda_{\mathrm{max}}$, $\rho=0.9$}
    \end{subfigure}
    \caption{Magnetic field strength of an unstable OP equilibrium and $\hat{s}-\alpha_{\mathrm{MHD}}$ contour plot of the maximum ideal-ballooning eigenvalue $\lambda_{\mathrm{max}}$. The equilibrium is recalculated for each value of $\hat{s}$ and $\alpha_{\mathrm{MHD}}$ and the ideal-ballooning equation is solved to create figure(b). At large pressure gradients $\alpha_{\mathrm{MHD}}$, the equilibrium becomes stable again, demonstrating the existence of second-stability in poloidally omnigenous stellarators.}
    \label{fig:s-alpha-bad-OP}
\end{figure}

As an example, we present $\hat{s}-\alpha_{\mathrm{MHD}}$ of the ballooning eigenvalue for a finite-$\beta$ OP equilibrium, demonstrating the existence of second stability in figure~\ref{fig:s-alpha-bad-OP}. Note that some features of this equilibrium are qualitatively similar to the high-$\beta$ QuasiPoloidal (QP) equilibrium presented in Ware \textit{et al}~\cite{ware2002high}, specifically the existence to second-stability.

\section{Simplified linear gyrokinetic model in the intermediate frequency range}\label{app:GK-KBM}
In this Appendix, we simplify the linearized gyrokinetic model and show how, subject to a set of assumptions, reduces to the ideal ballooning equation. We start by assuming that the frequency of interest $\omega$ is faster than the ion transit frequency, but slower than the electron transit frequency
\begin{equation}
    \frac{v_{\mathrm{th, i}}}{a_{\rm{N}}} \ll \epsilon\, \omega \ll  \epsilon^2 \frac{v_{\mathrm{th, e}}}{a_{\rm{N}}}, \quad \epsilon \equiv \sqrt{\frac{m_{\mathrm{e}}}{m_{\mathrm{i}}}} \ll 1,
    \label{eqn:intermediate-frequency-ordering}
\end{equation}
where $v_{\mathrm{th,i}} = \sqrt{2 T_{\mathrm{i}}/m_{\mathrm{i}}}, v_{\mathrm{th,e}} = \sqrt{2 T_{\mathrm{e}}/m_{\mathrm{e}}}$ are the thermal speeds of the ions and electrons, respectively. This separation in time scales is a consequence of the difference in the masses of ions and electrons. 

In this limit, we can simplify the electron gyrokinetic equation
\begin{eqnarray}
    \fl i \left(\omega - \omega_{Ds}\right)h_{\mathrm{e}} &+ (\bi{b}\cdot \bi{\nabla}\zeta)  w_{\parallel} \frac{\partial h_{\mathrm{e}}}{\partial \zeta} \nonumber \\
    &= (\omega - {\omega}^{T}_{*,\mathrm{e}}) \Bigg[J_0\left(\frac{k_{\perp}w_{\perp}}{\Omega_{\mathrm{e}}}\right) \left(\varphi -\frac{w_{\parallel}\delta\! A_{\parallel}}{c}\right) +
J_1\left(\frac{k_{\perp}w_{\perp}}{\Omega_{\mathrm{e}}}\right) \frac{w_{\perp}}{k_{\perp}} \frac{\delta \! B_{\parallel}}{c}\Bigg] F_{0\mathrm{e}},
\label{eqn:Appendix-electron-GK-equation}
\end{eqnarray}
to lowest order as
\begin{eqnarray}
     w_{\parallel}(\bi{b}\cdot \bi{\nabla}\zeta) \frac{\partial h_{\mathrm{e}}}{\partial \zeta}= w_{\parallel} (\omega - {\omega}^{T}_{*,\mathrm{e}}) F_{0s} \frac{\delta\! A_{\parallel}}{c},
\label{eqn:AppendixA-electron-GK-equation-lowest-order}
\end{eqnarray}
which can be integrated to give us 
\begin{equation}
    h_{\mathrm{e}}(\zeta, E, \hat{\lambda}; \sigma) = \left(1 - \frac{\omega^{T}_{*,e}}{\omega}\right) \hat{\delta\! A_{\parallel}} F_{0\mathrm{e}} + c_0(\hat{\lambda}),
\label{eqn:AppendixA-electron-GK-equation-lowest-order-integrated}
\end{equation}
where $c_0(\hat{\lambda})$ is a constant of integration and we have defined the line-integrated vector potential
\begin{equation}
    \hat{\delta\! A_{\parallel}}(\zeta; \sigma) = \frac{i \omega}{c} \int_{- \sigma \infty}^{\zeta}\frac{d\zeta}{\bi{b}\cdot\bi{\nabla}\zeta} \delta\! A_{\parallel},
\end{equation}
where $\sigma = v_{\parallel}/|v_{\parallel}|$ can either be $+1$ or $-1$ depending on the streaming direction of the particle with respect to the background magnetic field. Since all the variables that govern the passing species ($\hat{\lambda} \leq 1/B_{\mathrm{max}}$) satisfy the Dirichlet boundary condition $\lim_{\zeta \rightarrow \pm \infty} \delta \! \hat{A}_{\parallel} =\lim_{\zeta \rightarrow \pm \infty} h_{\mathrm{e}} = 0$, $c_0(\hat{\lambda}) = 0 \ \forall \ \hat{\lambda} \leq 1/B_{\mathrm{max}}$. For trapped particles, we must further solve~\eref{eqn:Appendix-electron-GK-equation} to find $c_0(\hat{\lambda}) \neq 0$.

To calculate the trapped electron response to the fields, we rewrite~\eref{eqn:Appendix-electron-GK-equation} after multiplying both sides with an integrating factor $\exp\left[-i\int d\theta (\omega- \omega_{D\mathrm{s}})/(|w_{\parallel}|\, \bi{b}\cdot\bi{\nabla}\theta)\right]$
\begin{eqnarray}
    & \fl  w_{\parallel}(\bi{b}\cdot\bi{\nabla}\zeta)\frac{\partial}{\partial \zeta}\left\{\left[h_{\mathrm{e}}  - \left(1 - \frac{\omega^{T}_{*,e}}{\omega}\right) F_{0\mathrm{e}}\delta\!\hat{A}\right] e^{-i\int d\zeta (\omega- \omega_{D\mathrm{e}})/(|w_{\parallel}|\, \bi{b}\cdot\bi{\nabla}\zeta)}\right\} e^{i\int d\zeta (\omega- \omega_{D\mathrm{e}})/(|w_{\parallel}|\, \bi{b}\cdot\bi{\nabla}\zeta)} \nonumber \\
    &=  (\omega - {\omega}^{T}_{*,\mathrm{e}})
    \left[ \varphi - \left(1 - \frac{\omega_{D\mathrm{e}}}{\omega}\right)\delta\!\hat{A}_{\parallel}  + 
\frac{w_{\perp}^2}{\Omega_s} \frac{\delta B_{\parallel}}{c}\right] F_{0s},
\label{eqn:AppendixA-electron-GK-equation-intfactor}
\end{eqnarray}
and apply the bounce-averaging operation
\begin{equation}
    \overline{X}  = \oint \frac{d\zeta}{(\bi{b}\cdot \bi{\nabla}\zeta)} \frac{X}{|w_{\parallel}|} \Bigg/\oint \frac{d\zeta}{(\bi{b}\cdot \bi{\nabla}\zeta)} \frac{1}{|w_{\parallel}|},
\label{eqn:AppendixA-bounce-average}
\end{equation}
to~\eref{eqn:AppendixA-electron-GK-equation-intfactor}, use integration by parts on the left side of~\eref{eqn:AppendixA-electron-GK-equation-intfactor}, and then use~\eref{eqn:AppendixA-electron-GK-equation-lowest-order-integrated}, to find $c_0(\hat{\lambda})$ for trapped electrons. This gives us the complete trapped electron response
\begin{equation}
    \fl h^{(t)}_{k_{\perp},\mathrm{e}} = \left(1 - \frac{\omega^{T}_{*,e}}{\omega}\right) F_{0\mathrm{e}}\delta\!\hat{A}_{\parallel} + \frac{(\omega - {\omega}^T_{*,\mathrm{e}})}{\left(\omega - \langle \omega_{Ds}\rangle \right)} \overline{\left[\varphi_{k_{\perp}} - \left(1 - \frac{\omega_{D\mathrm{e}}}{\omega}\right)\delta\!\hat{A}_{\parallel} + \frac{w_{\perp}^2}{2\, \Omega_{\mathrm{e}}} \frac{\delta B_{\parallel}}{c}\right]} F_{0s}
\label{eqn:AppendixA-eq-4}
\end{equation}
Note that, to lowest order we have used $J_0(k_{\perp}w_{\perp}/\Omega_{\mathrm{e}}) =  1, J_{1}(k_{\perp}w_{\perp}/\Omega_{\mathrm{e}}) = k_{\perp}w_{\perp}/(2\, \Omega_{\mathrm{e}})$ due to the intermediate frequency ordering in~\eref{eqn:intermediate-frequency-ordering}. The total electron response is the sum of the trapped ($1/B_{\rm{max}} < \hat{\lambda} \leq 1/B_{\rm{min}}$) and passing ($\hat{\lambda} \leq 1/B_{\rm{max}}$) distributions.

Next, we repeat the same process for ions. However, the trapped and passing frequency of ions is slower than the frequency of interest $\omega$, so the ion gyrokinetic equation to lowest order becomes
\begin{equation}
    h^{(0)}_{k_{\perp}, \mathrm{i}}  = \frac{(\omega - {\omega}^T_{*,\mathrm{i}})}{(\omega - \omega_{D\mathrm{i}})} \Bigg[J_0\left(\frac{k_{\perp}w_{\perp}}{\Omega_{\mathrm{i}}}\right) \varphi_{k_{\perp}}  +J_1\left(\frac{k_{\perp}w_{\perp}}{\Omega_{\mathrm{i}}}\right) \frac{w_{\perp}}{k_{\perp}} \frac{\delta B_{\parallel}}{c}\Bigg] F_{0\mathrm{i}},
\label{eqn:Appendix-eq-4}
\end{equation}
Using the expressions for the distribution functions $h$, we can now calculate their moments and plug those moments into the Maxwell's equations. We calculate the velocity integral in $(E, \hat{\lambda}, \vartheta)$ space
\begin{equation}
    \int d^3\bi{v} = 2\pi\, \mathlarger{\sum_{\sigma}} \int_{0}^{\infty} dE \sqrt{E} \int_{0}^{1/B}\frac{d\hat{\lambda} B}{\sqrt{1-\hat{\lambda} B}},
\end{equation}
where $\sum_{\sigma} h(\sigma) = [h(\sigma=1) + h(\sigma=-1)]$ takes into account both streaming directions of particles.
We now have all the information to write the Maxwell's equations. Substituting the gyrokinetic distribution functions into the quasineutrality equation~\eref{eqn:Poisson's-equation-Fourier}, we obtain
\begin{eqnarray}
 (2 - Q) \varphi + Q^{'}\tilde{\delta\! B_{\parallel}} + \int_{\mathrm{Tr}} d^3\bi{v} \left(\frac{\omega - \omega_{*, \rm{e}}^T}{\omega - \langle \omega_{D, \mathrm{e}} \rangle}\right)\overline{X_{\rm{e}}} = \left(1 - \frac{\omega_{*, \mathrm{e}}}{\omega}\right) \hat{\psi}_{\parallel}, 
 \label{eqn:quasineutrality-1}
\end{eqnarray}
where $\hat{\psi}_{\parallel} = \sum_{\sigma} \delta\!\hat{A}_{\parallel} = \left[\delta\!\hat{A}_{\parallel}(\sigma=1) + \delta\!\hat{A}_{\parallel}(\sigma=-1)\right]$ and $\int_{(t)}$ implies integration in $\hat{\lambda}$ for trapped particles only, \textit{i.e.}, over a truncated domain $1/B_{\mathrm{min}} \leq \hat{\lambda} \leq 1/B_{\mathrm{min}}$
\begin{eqnarray}
    X_s = \varphi - \left(1 - \frac{\omega_{\rm{D}s}}{\omega} \right) \hat{\psi}_{\parallel} + J_{1s} \frac{w_{\perp}}{k_{\perp}} \frac{\delta B}{c},\\
    Q = \int d^3\bi{v} \left(\frac{\omega  - \omega_{*, i}^T}{\omega - \omega_{D\mathrm{i}}}\right) J_0^2 F_{0\mathrm{i}}, \\
    Q^{'} = \int d^3\bi{v} \left(\frac{\omega  - \omega_{*, i}^T}{\omega - \omega_{D\mathrm{i}}}\right) \frac{d J_0^2}{db} F_{0\mathrm{i}}, \quad b = \frac{(k_{\perp}\rho_{\mathrm{i}})^2}{2},
\end{eqnarray}
and a typo has been corrected from equation $(3.31)$ in Tang \etal to get~\eref{eqn:quasineutrality-1}. Next, we write Ampere's law governing parallel fluctuations of the magnetic field strength
\begin{eqnarray}
 \tilde{\delta\! B_{\parallel}} &= \frac{4 \pi n_0 T_i}{B^2}\Bigg\{ Q^{'}\varphi - R\, \tilde{\delta\! B_{\parallel}} + \left[1 - \frac{\omega_{\mathrm{e,*}}}{\omega}(1+\eta_e)\right] \hat{\psi}_{\parallel} \nonumber \\
  &+ \frac{1}{n_0}\int_{(t)} d^3\bi{v} F_{\mathrm{e}0} \left(\frac{\omega - \omega_{*, \mathrm{e}}^T}{\omega - \langle \omega_{D, \mathrm{e}} \rangle}\right) \overline{X_{\mathrm{e}}} v_{\perp}^2\Bigg\},
  \label{eqn:parallel-fluctuations-Ampere}
\end{eqnarray}
where 
\begin{eqnarray}
    R = \int d^3\bi{v} \left(\frac{\omega  - \omega_{*, i}^T}{\omega - \omega_{D\mathrm{i}}}\right) v_{\perp}^2 J_1^2 F_{0\mathrm{i}}.
    \label{eqn:R-integral}
\end{eqnarray}
Note that the power of $v_{\perp}$ has been corrected in~\eref{eqn:R-integral} from Tang \etal. The final equation comes from substituting the simplified gyrokinetic distribution function in the equation $(3.34)$ in Tang \etal, governing the parallel current $j_{\parallel}$ 
\begin{eqnarray}
  \fl  \bi{B}\cdot\bi{\nabla} \zeta \frac{\partial}{\partial \zeta} \left[ \frac{k_{\perp}^2}{B^2} \bi{B}\cdot \bi{\nabla}\zeta \frac{\partial \hat{\psi}_{\parallel}}{\partial \zeta} \right] =& -4\pi n_i m_i \omega^2   \left\{ \left[Q - \left(1 - \frac{\omega_{*e}}{\omega}\right)\right]\varphi - \left[1 - \frac{\omega_{*e}}{\omega}(1+ \eta_{\mathrm{e}})\right] \left(\frac{\omega_{\kappa} + \omega_B}{\omega}\right) \hat{\psi}_{\parallel} \right\} \nonumber \\ 
  & + \int_{(t)} d^3\bi{v} \left[\frac{F_{0s}}{n_0} \frac{\omega_{\mathrm{De}}}{\omega} \left(\frac{\omega-\omega^T_{*}}{\omega - \langle \omega_{\mathrm{De}} \rangle} \right) \overline{X_{\mathrm{e}}} \right]
    \label{eqn:Appendix-j_parallel-equation}
\end{eqnarray}
where we have corrected an omission and a sign error from Tang \etal on the right side of~\eref{eqn:Appendix-j_parallel-equation}. The model now comprises three coupled integro-differential equations~\eref{eqn:quasineutrality-1},~\eref{eqn:parallel-fluctuations-Ampere} , and~\eref{eqn:Appendix-j_parallel-equation}.

It is possible to solve the intermediate-frequency electromagnetic gyrokinetic model numerically as an eigenvalue problem. However, this is beyond the scope of this paper and in this work we shall make a set of additional assumptions and impose subsidiary orderings similar to the ones used by Aleynikova
\etal~\cite{aleynikova2017quantitative, aleynikova2018kinetic} to remove trapped particle effects, assume steep density and temperature gradients, finite-$\beta$, and long-wavelength modes to reduce the model further. Applying the following orderings
\begin{equation}
   \fl k_{\perp} \rho_i \sim \delta^{1/2}, A \sim 1/\delta^2, \beta \sim \delta, \omega_d/\omega \sim \delta,  a_N/L_{n} \sim a_N/L_{n} \sim 1/\delta, \delta \gg \epsilon
\end{equation}
we can drop all the trapped particle integrals because the integral over pitch angle has the factor $1/\sqrt(1-\lambda B) \sim 1/\sqrt{A} \sim \delta$. This ordering reduces the three coupled differential equations to three decoupled equations
\begin{eqnarray}
    \varphi = \hat{\psi}_{\parallel},\\
    \delta B_{\parallel} = \frac{\beta_\mathrm{i}}{2} \hat{\psi}_{\parallel}, \\
    (\bi{B}\cdot \nabla \zeta)\frac{\partial}{\partial \zeta} \frac{(k_{\perp} \rho_i)^2}{2} (\bi{B}\cdot \nabla \zeta)\frac{\partial \hat{\psi}_{\parallel}}{\partial \zeta}  = -4 \pi n_i m_i \omega^2 K \hat{\psi}_{\parallel},
    \label{eqn:C22}
\end{eqnarray}
where
\begin{eqnarray}
K &= \left\{ \left[ Q - \left( 1 - \frac{\omega_{*\rm{e}}}{\omega}\right)\right]\left[ \upsilon_{\rm{0e}} \left(1 + \frac{\beta_{\rm{i}}}{2}R\right) - \upsilon_{1\rm{e}} \tau Q^{'}
\frac{\beta_{\rm{i}}}{2}\right]\right. \nonumber \\
& \left. - \frac{\beta_{\rm{i}}}{2}(Q^{'} + \upsilon_{1\rm{e}})\left[\upsilon_{0\rm{e}} Q^{'}  + \upsilon_{1\rm{e}}(2-\tau_{\rm{e}} Q) \right] \right\}\\
& \left[ (2 -\tau_{\rm{e}} Q)\left( 1 + \frac{\beta_{\rm{i}}}{2}R\right) + {Q^{'}}^2\frac{\beta_i}{2}\right]^{-1} - \upsilon_{1e}\frac{\omega_{\kappa} + \omega_B}{\omega} \nonumber,
\label{eqn:K-appendix}
\end{eqnarray}
$\upsilon_{ls} = (1 - \frac{\omega_{*s}}{\omega}(1+ l\eta_s)),\, \omega_{*s} = k_y \rho_{s} \frac{w_{\rm{th},s}}{L_{n,s}} B, \beta_{\mathrm{i}} = \frac{8\pi p_{\rm{i}}}{B^2}$.
This effectively requires solving a single ODE.
We simplify $Q, Q^{'}$,
\begin{equation}
\fl
\eqalign{
    Q = \left(1 - \frac{\omega_{*, j}}{\omega}\right) + \left(\frac{\omega_B + \omega_{\kappa}}{\omega} - \frac{(k_{\perp}\rho_i)^2}{2}\right)\left(1 - \frac{\omega_{*}}{\omega} (1+\eta_{i, j})\right) + \mathcal{O}(\delta^2) \\
    Q^{'} = -\left[\left(1 - \frac{\omega_{*,j}}{\omega}(1 + \eta)\right)\right] - \left[\frac{2 \omega_{B} + \omega_{\kappa}}{\omega}  + \frac{3}{4}\frac{(k_{\perp}\rho_{\rm{i}})^2}{2} \right]\left[\left(1 - \frac{\omega_{*,j}}{\omega}(1 +2\eta)\right)\right] + \mathcal{O}(\delta^2)}
\end{equation}
substitute into~\eref{eqn:K-appendix} and simplify
\begin{eqnarray}
K = 2 \frac{\omega_{*, \mathrm{i}}}{\omega^2}(1+\eta) \omega_{\kappa} \hat{\psi}_{\parallel}  +  b \hat{\psi}_{\parallel}.
\end{eqnarray}
Finally, substituting $K$ in~\eref{eqn:C22}, and cancelling the factor of $k_y^2/2$ from both sides, and normalizing everything else with $a_{\mathrm{N}}$ and $B_{\mathrm{N}}$, we recover the infinite-$n$ ideal ballooning equation
\begin{equation}
\eqalign{
   \fl (\bi{b}\cdot \nabla_{\rm{N}} \zeta) \frac{\partial }{\partial \zeta} \frac{|\nabla_{\rm{N}} \alpha|^2}{(B/B_{\rm{N}})} (\bi{b}\cdot \nabla_{\rm{N}} \zeta) \frac{\partial \hat{\psi}_{\parallel}}{\partial \zeta}\! +\! \frac{2}{B^2} \frac{dp}{d\psi_{\rm{N}}} [\bi{b}\times (\bi{b}\cdot\bi{\nabla}\bi{b}) \cdot \bi{\nabla}_{\rm{N}}\alpha] \hat{\psi}_{\parallel} = -\frac{a_{\rm{N}}^2 \omega^2}{v_{A}^2} \frac{|\nabla_{\rm{N}} \alpha|^2}{(B/B_{\rm{N}})^3} \hat{\psi}_{\parallel}
   }
\end{equation}
Details such as identities and algebra used to obtain these equations are presented in the supplementary notes.

\section{Definition of various objectives and figures of merit} \label{app:objectives-FoMs-defn}
In this appendix, we will explain the definition of various objective functions and figures of merit used by us in the main body of this paper.
\subsection{Objective functions}
We start by defining the curvature objective
\begin{eqnarray}
    f_{\mathrm{curv}} = \mathrm{ReLU}(\max(\kappa_{2, \rho} - \kappa_{2, \rho, \mathrm{bound1}}, \kappa_{2, \rho} - \kappa_{2, \rho, \mathrm{bound2}}))
\end{eqnarray}
where 
\[
\centering
\kappa_{2, \rho} = \min\left\{x:\det\left[\begin{array}{cc}
    L_{\mathrm{sff}, \rho} - x E & M_{\mathrm{sff}, \rho} - x F \\
    M_{\mathrm{sff}, \rho} - x F & N_{\mathrm{sff}, \rho} - x G 
\end{array}\right] = 0\right\}
\]
is the second principal curvature on a given point on a flux surface, and 
\begin{eqnarray}
    E = \frac{\partial \bi{x}}{\partial \zeta} \cdot \frac{\partial \bi{x}}{\partial \zeta},\quad F = \frac{\partial \bi{x}}{\partial \theta} \cdot \frac{\partial \bi{x}}{\partial \zeta},\quad  G = \frac{\partial \bi{x}}{\partial \theta} \cdot \frac{\partial \bi{x}}{\partial \theta}, \\
    L_{\mathrm{sff}, \rho} = \frac{\partial^2 \bi{x}}{\partial \theta^2} \cdot \frac{\bi{\nabla}\rho}{|\bi{\nabla}\rho|}, \quad   M_{\mathrm{sff}, \rho} = \frac{\partial^2 \bi{x}}{\partial \theta \partial \zeta} \cdot \frac{\bi{\nabla}\rho}{|\bi{\nabla}\rho|}, \quad N_{\mathrm{sff}, \rho} = \frac{\partial^2 \bi{x}}{\partial \zeta^2} \cdot \frac{\bi{\nabla}\rho}{|\bi{\nabla}\rho|}
\end{eqnarray}
are the metric coefficients corresponding to the first and second fundamental forms. The user-specified values $\kappa_{2, \mathrm{bound1}}, \kappa_{2, \mathrm{bound2}}$ define the limits on the curvature. Using this objective, we impose a maximum and minimum value on $\kappa_{2, \rho}$ that helps the optimizer avoid plasma boundaries with sharp convex or concave curvatures.

To maintain a shape that is practically feasible using coils, we also put bounds on the elongation of the boundary, by using the elongation objective
\begin{eqnarray}
    f_{\mathrm{elongation}} = \mathrm{ReLU}(\max{e - e_{\mathrm{bound1}}, e - e_{\mathrm{bound2}}})
\end{eqnarray}
where 
\begin{eqnarray}
 e =\frac{1}{2 \pi}\int \hat{e}\,  d\zeta, \quad \hat{e} = \frac{a_{\mathrm{major}}}{a_{\mathrm{minor}}}
\end{eqnarray} 
is the averaged elongation of a flux surface averaged over the cylindrical toroidal angle $\zeta$, $\hat{e}$ is the elongation of a toroidal cross-section, and $e_{\mathrm{bound}}$ are the bounds imposed on the elongation. To calculate $\hat{e}$, we first calculate $a_{\rm{major}}$ using the area $\hat{A}$ and perimeter $\hat{P}$ of a toroidal cross-section and invert Ramanujan's approximation for the perimeter of an ellipse~\cite{Villarino2006,Ramanujan1914}
\begin{eqnarray}
\fl a_{\rm{major}} = \frac{\sqrt{3} \left(\sqrt{8\pi \hat{A} + \hat{P}^2} + \sqrt{\left|2\sqrt{3} \hat{P} \sqrt{8\pi \hat{A} + \hat{P}^2} - 40\pi \hat{A} + 4\hat{P}^2\right|}\right) + 3\hat{P}}{12\pi}
\end{eqnarray}
Using the major radius $a_{\rm{major}}$, the area $\hat{A}$, and assuming the section has an elliptical shape, we get $a_{\rm{minor}} = \hat{A}/(\pi a_{\rm{minor}})$ which gives us the elongation $\hat{e} = a_{\rm{major}}/a_{\rm{minor}}$ and the average elongation $e = \int \hat{e}\,  d\zeta /(2 \pi)$.

Finally, we define the Mercier (or interchange)~\cite{mercier1964equilibrium, carreras1988low} stability objective
\begin{eqnarray}
    f_{\mathrm{D_{Merc}}} = (D_{\mathrm{Merc}} - D_{\mathrm{Merc}0}),
\end{eqnarray}
where
\begin{equation}
    D_\mathrm{Merc}  = -\frac{p^{'}}{\iota^{'}}V^{\dagger \dagger} - \frac{1}{4} > 0,
\end{equation}
where the quantity $V^{\dagger \dagger}$ is related to the magnetic well~\cite{greene1997brief}. A $D_{\mathrm{Merc}} > 0$ corresponds to stability against interchange modes, whereas $D_{\mathrm{Merc}} < 0$ implies  ideal interchange instability.  Note that for negative shear equilibria, $D_{\mathrm{Merc}} > 0$ implies a positive (favorable) magnetic well $V^{\dagger \dagger} > 0$, which is the case for OP and OT cases presented in this paper.

\subsection{Figures of Merit}
In this section, we define the figure of merits. We start with the volume-averaged $\beta$
\begin{eqnarray}
    \langle \beta \rangle = \frac{1}{\int_{0}^{1} \oint \oint \sqrt{g}\, d\rho d\theta d\zeta} {\int_{0}^{1} \oint \oint \sqrt{g}\, d\rho d\theta d\zeta \, \frac{2 \mu_0 p}{B^2}}.
\end{eqnarray}
Next, we define another important quantity of interest, the toroidal magnetic flux enclosed by the plasma boundary
\begin{eqnarray}
    \Psi_{\mathrm{b}} = \int_{0}^{1} \oint \oint \sqrt{g}\, d\rho d\theta d\zeta\,  \bi{B}\cdot\bi{\nabla}\zeta.  
\end{eqnarray}
and finally, the total enclosed toroidal plasma current represented by the contour integral
\begin{eqnarray}
I_{\mathrm{b}} = \frac{1}{2\pi \mu_0} {\oint d\theta \, \bi{B} \cdot \frac{\partial \bi{x}}{\partial \theta}},
\end{eqnarray}
where all the elements in the integrand and the integrals are calculated on the boundary.

\section{Characterizing gyrokinetic modes}\label{app:mode-filter}
In this Appendix, we will explain how to characterize and separate different unstable modes from the output of an initial value gyrokinetic solver such as~\texttt{GS2}. We will also provide details of the~\texttt{GS2} runs used to calculate the maximum KBM growth rates in Section~\ref{sec:Results}.
\subsection{Separating KBMs from other microinstabilities}
To obtain the maximum growth rate, we require a method to separate Kinetic Ballooning Modes (KBM) from other gyrokinetic instabilities such as the Ion Temperature Gradient (ITG) mode, Trapped Electron Mode (TEM), Microtearing mode (MTM), and the Electron Temperature Gradient (ETG) mode. 
To this end, we use a set of classification rules with which we can find the most unstable KBMs. To define these rules, we first define the quasilinear particle flux and heat flux
\begin{eqnarray}
    \Gamma_s = \Big\langle \int d^3 v\, h_{s} \bi{V}_{E}\cdot\frac{\bi{\nabla}\psi}{|\bi{\nabla}\psi|} \Big\rangle_{\psi} =  \frac{a_{\mathrm{N}}}{L_{ns}} D_s, \\
    q_s = \Big\langle \int d^3 v\,  h_{s} \bi{V}_{E}\cdot\frac{\bi{\nabla}\psi}{|\bi{\nabla}\psi|} \Big\rangle_{\psi} = \frac{a_{\mathrm{N}}}{L_{Ts}} \chi_s + \frac{3}{2} \Gamma_s,
\end{eqnarray}
where the $\bi{E}\times\bi{B}$ drift $\bi{V}_{E}$ is defined in~\eref{eqn:E-cross-B-drift-velocity}, the gradient scale lengths $a_{\mathrm{N}}/L_{\mathrm{n}s}, a_{\mathrm{N}}/L_{\mathrm{T}s}$ are defined in~\eref{eqn:tprim-fprim-definitions}, and $\chi_s$ and $D_s$ are the heat and particle diffusivities, respectively. The flux surface averaging operator $\langle \rangle_{\psi}$ is defined in~\eref{eqn:FSA}. We also define the parity of the gyrokinetic parallel vector potential $\delta \! A_{\parallel}$
\begin{eqnarray}
    \mathcal{P} = 1 - \frac{|\int d\theta \delta\! A_{\parallel}|}{\int d\theta |\delta\! A_{\parallel}|}.
\end{eqnarray}
Using these classifying variables and the prescription provided in Table 1 of Parisi \textit{et al.}~\cite{parisi2024stability, kotschenreuther2019fingerprints}, we create table~\ref{tab:Characterization-table} to separate KBM from the rest of the modes in the output produced by~\texttt{GS2}.

\begin{table}[h]
\caption{Characterizing various gyrokinetic modes}
\lineup
\begin{tabular*}{\textwidth}{@{}l*{15}{@{\extracolsep{0pt plus12pt}}l}}
\br                              
Mode &\0$\chi_{\mathrm{i}}/\chi_{\mathrm{e}}$ & \m $D_{\mathrm{e}}/\chi_{\mathrm{e}}$ & \0$D_{\mathrm{i}}/\chi_{\mathrm{i}}$& \0 $\mathcal{P}(A_{\parallel})$& $\partial \gamma/\partial \beta$   \cr 
\mr
KBM      &\0 $\sim 1$  & \0 $\sim 1$ & \0\0 $\sim 1$ & \0 $ 1$ & \0 $> 0$ \cr
TEM      &\0 $\sim 1$  & \0 $\sim 1$ & \0\0 ---      & \0 $ 1$ & \0 $< 0$ \cr 
ITG      &\0 $\gg 1 $  & \0 ---      & \0\0 $\ll 1$  & \0 $ 1$ & \0 $< 0$ \cr 
ETG      &\0 $\ll 1 $  & \0 $\ll 1$  & \0\0 ---      & \0 $ 1$ & \0 $< 0$ \cr
MTM      &\0 $\ll 1 $  & \0 $\ll 1$  & \0\0 ---      & \0 $< 1$ & \0 $> 0$ \cr
EM-ETG   &\0 $\ll 1 $  & \0 $\ll 1$  & \0\0 ---      & \0 $ 1$ & \0 $> 0$ \cr
\br
\end{tabular*}
\label{tab:Characterization-table}
\end{table}

\subsection{Details of~\texttt{GS2} runs}
\begin{figure}
    \centering
    \includegraphics[width=\textwidth, trim={2mm 2mm 8mm 0mm}, clip]{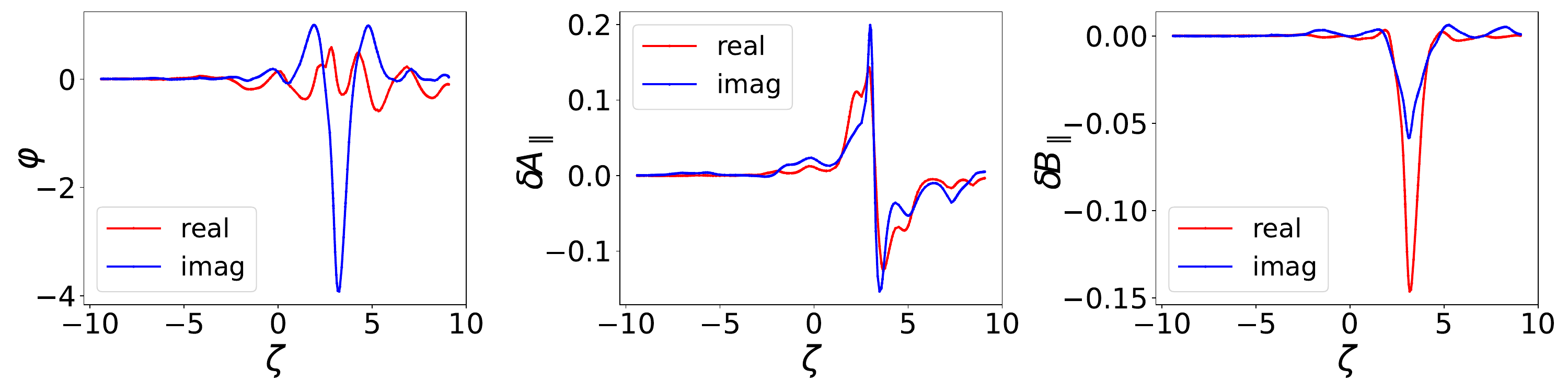}
    \caption{The set of eigenfunctions for the most unstable mode for the the initial OT equilibrium at $\rho = 1.0$. The peak growth rate occurs at $\alpha = 0, k_x=1.2, k_y = 0.4$. Note the odd parity of the $\delta\! A_{\parallel}$, signature of a KBM.}
    \label{fig:KBM-eigenfunctions}
\end{figure}

On each flux surface a~\texttt{GS2} simulation scans $N_{\alpha} = 8$ fieldlines. For each fieldline, we scan $N_{k_y} = 8$ values of the binormal wavenumber linearly spaced between the $k_y \in [0.05,  1.2]$ and $N_{k_x} = 12$ values of the radial wavenumbers linearly spaced between $k_x \in [-\pi/(2\hat{s} k_y), \pi/(2\hat{s} k_y)]$.
 The number of grid points along a flux tube is determined by the pitch angle resolution $N_{\hat{\lambda}}$ for which we choose $N_{\hat{\lambda}} \geq 45$. The pitch angle resolution is sensitive to the type of omnigenity. In general OP equilibria require a higher $N_{\hat{\lambda}}$ and OT equilibria require the lowest $N_{\hat{\lambda}}$. The resolution of the velocity space in GS2 is $\mathrm{ngauss} = 6, \mathrm{negrid} = 12$. 
Upon imposing all the criteria from table~\label{tab:Characterization-table} we can separate out the KBMs from the rest of the unstable modes. A typical set of eigenfunctions corresponding to the KBM are provided in figure~\ref{fig:KBM-eigenfunctions}.

\pagebreak

\section*{References}
\bibliographystyle{ieeetr}
\bibliography{main}

\end{document}